\documentclass[epj]{svjour}
\pdfoutput=1
\usepackage{graphicx}
\usepackage{amssymb}
\usepackage{amsmath}
\usepackage{hyperref}

\begin{document}

\title{Eigenfunction structure and scaling of two interacting particles 
in the one-dimensional Anderson model} 

\author{
Klaus~M.~Frahm
}
\institute{
Laboratoire de Physique Th\'eorique, Universit\'e de Toulouse, CNRS, UPS, France
}

\titlerunning{Eigenfunction structure and scaling of two interacting particles}
\authorrunning{K.~M.~Frahm}

\abstract{The localization properties of eigenfunctions 
for two interacting particles 
in the one-dimensional Anderson model are studied for system sizes 
up to $N=5000$ sites corresponding to a Hilbert space of dimension 
$\approx 10^7$ using the Green function Arnoldi method. 
The eigenfunction structure is illustrated 
in position, momentum and energy representation, the latter corresponding 
to an expansion in non-interacting product eigenfunctions. 
Different types of localization lengths are computed for 
parameter ranges in system size, disorder and interaction strengths 
inaccessible until now. We confirm that one-parameter scaling theory 
can be successfully applied 
provided that the condition of $N$ being significantly larger than 
the one-particle localization length $L_1$ is verified. 
The enhancement effect of the two-particle localization length $L_2$ 
behaving as $L_2\sim L_1^2$ is clearly confirmed for a certain quite large 
interval of optimal interactions strengths. Further new results for the 
interaction dependence in a very large interval, an energy value outside the 
band center, and different interaction ranges are obtained. 
}

\date{Received 25 February 2016 / Received in final form 14 March 2016 \\
Published online 2 May 2016, 
Eur. Phys. J. B (2016) 89: 115, DOI: \href{http://dx.doi.org/10.1140/epjb/e2016-70114-7}{10.1140/epjb/e2016-70114-7}}

\maketitle

\section{Introduction}
\label{sec1}

The phenomenon of Anderson localization of quantum eigenstates 
of non-interacting particles moving in a random potential \cite{lee1} 
is well-understood from powerful numerical tools combined with 
scaling theory \cite{mackinnon1,mackinnon2,kramer1} and also 
by analytical approaches such as the supersymmetric non-linear $\sigma$-model 
\cite{efetov1,guhr1} or the Fokker-Planck approach for the transfer matrix 
\cite{been_rev1} which have been shown to be equivalent \cite{brouwer1} 
for the 
particular case of quasi one-dimen\-sional geometries (with many transverse 
channels) where localization even persists for arbitrarily small disorder. 
Also the exact one-dimensional Anderson model 
(with only one transverse channel) of non-interacting particles is 
well-understood with a localization length $L_1=105t^2/W^2$ \cite{kramer1} 
in the band center expressed in terms of the width $W$ of the distribution 
of the disorder potential and the hopping matrix element $t$. 

Dorokhov \cite{dorokhov1} 
considered two interacting particles (TIP) in one dimension with 
an attractive long range potential which he mapped to the transfer 
matrix Fokker-Planck 
approach using some approximations and assumptions about the statistics 
and correlations of the effective disorder potential. He found a 
strong enhancement of the two-particle localization length $L_2$ 
if compared to the localization length $L_1$ of 
one particle without interaction. Shepelyansky \cite{dlstip} considered 
two particles in the one-dimen\-sional Anderson model coupled by a repulsive 
or attractive local Hubbard interaction of strength $U$, 
a type of systems which are 
potentially accessible by experiments on cold atoms similar as 
in \cite{bloch}. It came as a surprise when he found for this case 
the enhancement 
$L_2/L_1\sim (U/t)^2 L_1$ using an assumption of random phases of the 
one-particle wave functions inside the localized domain and mapping the 
initial model to a different random band matrix with preferential basis 
for which he numerically extracted an analytic expression of the 
localization length. The enhancement effect was soon confirmed \cite{imry} 
by a different argument based on the Thouless scaling block picture 
\cite{thouless1} and by direct numerical computations 
using a finite size transfer matrix \cite{frahm1} and exact diagonalization 
\cite{weinmann}. 

At the same time also the understanding of the random band matrix 
used in \cite{dlstip} was substantially improved \cite{jacquod1} 
and even analytically solved by mapping it onto the one-dimensional 
supersymmetric $\sigma$-model \cite{fyodorov1,frahm2}. This provided 
analytical expressions \cite{jacquod1,fyodorov1,frahm2} 
for the localization length, the inverse participation ratio (IPR) 
and established a Breit-Wigner regime. The latter is characterized 
by an energy scale $\Gamma$, called the Breit-Wigner width, corresponding 
to the inverse 
life time of an unperturbed eigenstate in absence of interaction 
and the energy interval over which these unperturbed states are mixed 
by the interaction. 

The enhancement effect also appears in related models of two interacting 
kicked rotors for which it is possible to determine directly the 
quantum time-evolution \cite{dlstip,borgonovi1,borgonovi1a} or a bag model 
\cite{dlstip,frahm1,halfpap1} corresponding 
to an attractive long range interaction and for which the standard 
transfer matrix method is well suited.
An efficient method to calculate the two-particle Green function 
projected onto the subspace of states with both particles on the same site 
was introduced by von Oppen et al. \cite{vonoppen} who 
proposed the scaling relation $L_2/L_1\approx 0.5+0.054\,|U|\,L_1$ 
(for $t=1$) with a linear dependence on $|U|$ contradicting the quadratic 
behavior predicted in \cite{dlstip}. 

An explanation 
of this was given by Jacquod et al. \cite{jacquod2} who 
calculated analytically to all orders in the interaction the Breit-Wigner 
width for the limit of vanishing disorder suggesting the modified behavior 
$L_2\sim L_1^2 |U|/\sqrt{t^2+(U/4)^2}$. The physical picture behind this 
result is that for weak disorder the behavior of the one-particle 
wave functions is essentially ballistic inside the localization domain. 
This provides 
well correlated phases of plane waves due to a rather well defined 
quasi-momentum thus modifying the initial estimation of 
the Breit-Wigner width or typical interaction matrix elements used in 
\cite{dlstip,imry} \cite{jacquod1,shep2} obtained from the random phase 
approximation (or an ergodic hypothesis). 

The effect of well defined quasi-momenta was also independently 
investigated in detail by Ponomarev et al. \cite{ponomarev1} 
who showed by analytical arguments 
that the interaction matrix elements have a long tail 
distribution with maximum values corresponding to approximate momentum 
conservation with uncertainty $\sim 1/L_1$. 
Based on this they proposed and studied a modified random 
matrix model confirming the enhancement but suggesting 
a power law $L_2/L_1\sim L_1^\gamma$ with $0.56\le\gamma\le 0.95$ 
dependent on a certain model parameter. 
A numerical study of the interaction matrix elements confirmed 
the long tail distribution \cite{roemer4}.

It is worth mentioning that in \cite{frahm3,frahm3a} 
a sophisticated random matrix 
model was proposed and solved by the supermatrix non-linear $\sigma$-model 
which works for arbitrary space dimension and takes properly both particle 
coordinates (relative and center of mass coordinate) into account. Later 
a different random matrix model for quasi-one-dimensional geometries 
(with many transverse channels already for non-interacting particles) was 
introduced and investigated by the $\sigma$-model \cite{richert1,richert1a}. 
Other work was concerned with the role of the level 
statistics \cite{wein2}, level curvatures \cite{akkermans} or with the 
fractal structure of the interaction 
matrix elements \cite{waintal1,waintal1a,waintal1b}. 
The arguments of a claim that the effect completely vanishes in the limit 
of infinite system size \cite{roemer1,roemer1a} were shown to be specific for 
a certain intermediate disorder value where the modest enhancement effect (of 
a factor of $\sim 2$) can only be measured by the Green function approach 
but not by the finite size transfer matrix method \cite{frahm4,song2}. 

Further numerical work \cite{song2,song1,leadbeater1,frahm5} based on 
different methods to compute the Green function confirmed the enhancement 
but they were mostly limited to system sizes $N$ between 200 and 300 
\cite{song2,song1,leadbeater1}. Up to now only reference \cite{frahm5} 
has considered large systems sizes up to $N=1400$ 
for disorder values down to $W=1$ with $L_1=105$ using finite size 
extrapolation to determine the infinite size localization lengths. 
Based on these numerical data combined with an extension of the 
analytical calculation of the Breit-Wigner width of \cite{jacquod2} the 
approximate expression $L_2/L_1\approx 0.5+0.074\,L_1|U|/(1+|U|)$ was 
suggested \cite{frahm5}.
In \cite{song1,leadbeater1} the method of finite size scaling, 
which is in principle a more powerful tool than the finite size 
extrapolation, was applied to disorder values down 
to $W=0.5$ (or even $W=0.4$ in \cite{leadbeater1}). However, at this 
disorder value all considered system sizes ($N\le 200$ \cite{song1} 
or $N\le 251$ \cite{leadbeater1}) are below $L_1=420$ and therefore 
clearly outside 
the range of validity of one-parameter scaling theory requiring that 
$N$ is significantly larger than all other typical length scales 
in the system \cite{mackinnon1,mackinnon2,kramer1}, 
especially $L_1$ which somehow plays the role of the mean free path for TIP 
(see our discussion below 
in Appendix~\ref{appC} for more details on this point). 
In view of this the results of \cite{song1,leadbeater1} obtained by 
finite size scaling for small 
disorder values appear to be invalid and therefore there are no published 
reliable numerical data available for $W<1$. 

The situation concerning numerical calculations of exact eigenfunctions 
such as in \cite{weinmann} is similar with results only available for 
system sizes up to a few hundred sites. 
New claims \cite{flach2011,flach2014} disputing the existence or size of the 
enhancement effect have recently 
surfaced based on numerical data for eigenfunctions 
with limited parameters in system size ($N\le 234$) and disorder ($W\ge 2$). 

Recently a very powerful new method to compute exact eigenfunctions of 
TIP in one-dimensional systems, the Green function Arnoldi method, 
was developed and applied in the context of TIP in a quasi-periodic 
potential for system sizes up to $N\approx 10^5$ \cite{frahm6} (see 
references therein for the physics and history of this model). 
This confirmed and extended previous results 
for smaller systems \cite{flach2012} for this model about 
eigenstates being completely delocalized over the full system size for 
certain particular values of energy and interaction strength 
even though $L_1\approx 4.5$ for the chosen parameters. 

In this work we apply this method to TIP for the disordered case 
of the one-dimensional Anderson model and we will present results 
for exact eigenfunctions up to systems sizes $N=5000$ for 
some individual samples and $N=2000$ for a systematic study 
of disorder values down to $W=0.75$ and interaction values up to $U=12$ 
for different disorder realizations. 
Furthermore we also employ the projected Green function method 
\cite{vonoppen} implemented very efficiently in \cite{frahm5} together 
with a new optimization allowing to treat many different 
interaction values simultaneously without additional effort. 
Here we use system sizes up to 
$N=2000$, disorder values down to $W=0.5$ (with at least 7 data points 
respecting the condition $N\ge 1.5 L_1$) and a very large range 
of positive and negative interaction values covering 6 orders of magnitude. 
In most cases two energy values $E=0$ (in the band center) and $E=1$ 
(outside the band center) are considered. Our results for both methods 
clearly confirm a scaling of the type $L_2/L_1\sim L_1+$const. 
for a rather wide range of optimal interaction values. 

In Section~\ref{sec2} we introduce the model and remind the basic ideas 
of the two numerical methods. In Section~\ref{sec3} and \ref{sec4} 
results for eigenfunctions computed by the Green function Arnoldi method 
are presented. Section~\ref{sec3} discusses some of their general 
properties and introduces three types of inverse participation ratios 
while Section~\ref{sec4} provides results of finite size 
scaling for them. Section~\ref{sec5} presents results of finite size 
scaling for the Green function localization length. 
Section~\ref{sec6} discusses the internal structure of TIP 
eigenfunctions inside the localization domain in energy representation 
while Section~\ref{sec7} provides the discussion of the main results. 
Appendix~\ref{appA} 
describes some details of our particular implementation of the scaling 
procedure, Appendix~\ref{appB} provides a separate discussion 
of the different two-particle localization lengths at vanishing 
interaction, and Appendix~\ref{appC} discusses various scenarios 
of finite size scaling using insufficient system sizes and 
establishes that data in the regime $N<L_1$ clearly do not obey 
one-parameter scaling.

\section{Model and numerical methods}
\label{sec2}

The Hamiltonian of the TIP 1d-disorder problem is given by 
\begin{equation}
\label{ham_tot}
H=h^{(1)}+h^{(2)}+\hat U
\end{equation}
where 
\begin{equation}
\label{eq_h1}
h^{(j)}=-\sum_{<x,y>} |x\!>_j\,<\!y|_j+\sum_x V(x)\,|x\!>_j\,<\!x|_j
\end{equation}
is the one-particle Hamiltonian of the particle $j$ corresponding 
to the 1d-Anderson model with hopping matrix element $t=-1$ between 
nearest neighbor sites $x$ and $y$ and $V(x)$, uniformly 
distributed in $[-W/2,\,W/2]$ and uncorrelated for different values 
of $x$, is the random disorder potential with $W$ being the disorder 
parameter. 
We consider systems of finite size $N$ with sites $x=0,\,1,\,\ldots,\,N-1$. 
The interaction operator in (\ref{ham_tot}) can be written as $\hat U=UP$ 
with the projector 
\begin{equation}
\label{P_operator}
P=\sum_{|x_1-x_2|<U_R} |x_1,x_2\!><\!x_1,x_2|
\end{equation}
on sites with $|x_1-x_2|<U_R$ and the notation 
$|x_1,x_2\!>=|x_1\!>_1|x_2\!>_2$ for the two particle basis states 
in position representation. The number $U$ is the overall 
interaction strength and $U_R$ is the interaction 
range where $U_R=1$ corresponds to the case of the Hubbard on-site interaction. 
In Sections~\ref{sec3} and \ref{sec4}, where we study eigenfunction properties 
of the 
TIP Hamiltonian (\ref{ham_tot}), we assume periodic boundary conditions 
for the hopping matrix elements and also the interaction, i.~e. the condition 
$|x_1-x_2|<U_R$ in (\ref{P_operator}) is understood to be true also if 
$N-|x_1-x_2|<U_R$. For the case $U_R>1$ 
the second condition corresponds to a situation where one particle is 
close to one boundary and the other one to the other boundary. In this work 
we only consider the case of a uniform interaction strength $U$ 
for distances smaller then $U_R$. In Section~\ref{sec5} where we 
study the localization length determined by the exponential decay of 
the two-particle Green function we will limit ourselves to 
the Hubbard interaction case ($U_R=1$) and use open boundary conditions. 

The eigenfunctions of the one-particle Hamiltonian (\ref{eq_h1}) 
are exponentially localized for $W>0$ with 
a localization length at the band center $\epsilon=0$ given by $L_1=105/W^2$ 
\cite{kramer1}, an expression which is ideally valid for small disorder values 
but even for $W=7$ the error is only $\sim 7$\% (and $3.5$\% for $W=5$). 
Throughout this work 
we will use this expression of $L_1$ as a disorder dependent 
reference length scale representing the one-particle localization length 
and regularly express other length scales, especially 
two-particle localization lengths, in units of $L_1$.

In the following of this section 
we remind the basic ideas of the two numerical methods, 
the {\em Green function Arnoldi method} \cite{frahm6} 
to compute a certain number of eigenfunctions close to a given energy 
value $E$ and the {\em projected Green function method} \cite{vonoppen} 
implemented efficiently in \cite{frahm5} to determine the exponential 
decay of the two-particle Green function along the diagonal of doubly 
occupied sites. A reader not interested in the details of these methods 
may skip the remainder of this section and directly continue with Section 
\ref{sec3} where the first eigenfunction results are discussed. 

The Hilbert space associated to the TIP Hamiltonian (\ref{ham_tot}) 
is of dimension $N_2=N(N-1)/2+sN\approx N^2/2$ with $s=1$ for bosons or $s=0$
for fermions. Therefore a direct numerical computation of all eigenfunctions 
is only feasible for relatively small systems with sizes up to $N=200$-$300$. 
Due to the sparse matrix structure one can try to apply the Arnoldi-Method 
\cite{stewart,frahmulam} (or the Lanczos method which is similar in spirit for 
hermitian matrices). The basic idea of this method is to choose some 
normalized initial vector and then to apply an iterative scheme of matrix 
multiplication and orthogonalization to construct 
an orthonormal basis on a {\em Krylov space} of modest {\em Arnoldi 
dimension} $n_A$ with typically $1\ll n_A\ll N_2$. 
During this scheme one obtains a representation matrix 
of $H$ on this Krylov space and it is necessary to neglect a last 
coupling element to the next vector of index $n_A+1$ 
\cite{stewart,frahmulam} which introduces a mathematical approximation. 
It turns out that the largest eigenvalues of the 
rather small representation matrix of size $n_A\times n_A$ are typically 
very good approximation to the largest eigenvalues of $H$. Furthermore 
it is also possible to compute the corresponding eigenvectors of $H$ by 
first calculating the eigenvectors of the representation matrix and 
transforming them to the full eigenvectors of $H$ using the orthonormal basis 
of the Krylov space. 

The method allows for much larger matrix sizes but in its most simple variant 
it has the flaw that it concentrates on the largest eigenvalues of $H$ 
(in modulus). To obtain some accurate eigenvalues close to 
some given energy $E$ in the middle of the spectrum of a large sparse 
matrix one may use a different quite complicated variant called 
the implicitly restarted Arnoldi method \cite{stewart} where the initial 
vector is iteratively modified/improved by a subtle procedure based on 
implicit QR-steps. Even though this method 
can be applied to the Hamiltonian (\ref{ham_tot}) for system sizes up to 
$N=700$-$1000$ (with a considerable effort) we did not use it. 

Instead we used a still more efficient method, the Green function 
Arnoldi method, which exploits more clearly the particular TIP-structure 
of (\ref{ham_tot}). The details of this method are given in 
\cite{frahm6} where it was applied to a Hamiltonian similar to 
(\ref{ham_tot}) but with a quasi-periodic one-particle potential $V(x)$. 
Here, we will only remind its main ideas. The key for this method 
is the efficient numerical evaluation of the matrix vector product 
of the Green function or resolvent $G=(E-H)^{-1}$ applied to some arbitrary 
vector of the Hilbert space using the following formula~:
\begin{equation}
\label{eq_Green}
G=G_0+G_0 ({\bf 1}-\hat U\bar G_0)^{-1} \hat U G_0
\end{equation}
where $G_0=(E-H_0)^{-1}$, $H_0=h^{(1)}+h^{(2)}$ is the TIP Hamiltonian 
in absence of interaction, and $\bar G_0=PG_0P$ with $P$ being the projector 
(\ref{P_operator}). This expression is exact and details of its demonstration 
can be found in \cite{frahm6}. 

Let us denote by $\phi_\nu(x)$ 
the eigenfunctions of the one-particle Hamiltonian (\ref{eq_h1}) with 
eigenvalues $\epsilon_\nu$ which we use to 
construct a basis of the two-particle Hilbert space using 
product states $|\phi_{\nu_1}\,\phi_{\nu_2}\!>$ which 
are also eigenvectors of $H_0$ with eigenvalues 
$\epsilon_{\nu_1}+\epsilon_{\nu_2}$. Let $|\psi\!>$ be some arbitrary 
vector of the TIP Hilbert space which can be expanded either in position 
representation~:
\begin{equation}
\label{eq_pos_rep}
|\psi\!>=\sum_{x_1,\,x_2} \psi(x_1,x_2)\,|x_1,x_2\!>
\end{equation}
or in energy representation~:
\begin{equation}
\label{eq_en_rep}
|\psi\!>=\sum_{\nu_1,\,\nu_2} \chi_{\nu_1,\,\nu_2}\,
|\phi_{\nu_1}\,\phi_{\nu_2}\!>
\end{equation}\
in terms of {\em non-interacting product eigenfunctions}. 
In (\ref{eq_pos_rep}) 
and (\ref{eq_en_rep}) we use for simplicity a non-symmetrized representation 
of two-particle states but the wave functions must of course satisfy the 
(anti-)symmetry for bosons (fermions) with respect to exchange 
of $x_1$ with $x_2$ or of $\nu_1$ with $\nu_2$. 
In the following we omit the details 
due to the complications of (anti-)symmetrization of the two-particle states 
but of course such details need to be dealt with care and precision 
in a concrete implementation of the method. 

Let us assume we 
know a vector $|\psi\!>$ in energy representation, i.~e. the vector of 
coefficients $\chi_{\nu_1,\,\nu_2}$ is known. First we compute 
$G_0|\psi\!>$ (with $N^2$ operations) and then we transform the 
resulting vector into position representation (\ref{eq_pos_rep}) which 
can be done with $2N^3$ operations by first applying the orthogonal 
transformation (corresponding to $|\phi_{\nu}\!>\to |x\!>$) 
to the first particle for all 
values of the second particle index and then transforming the second 
particle for all values of the first particle index. Once this is done we can 
efficiently apply the factor $({\bf 1}-\hat U\bar G_0)^{-1} \hat U$ 
(with $U_R^2N^2$ operations) 
to the resulting vector where the matrix inverse is done once in advance 
(for some fixed value of the energy $E$) and concerns only a matrix 
of effective size $U_RN\times U_RN$ (due to the projector in 
$\bar G_0=PG_0P$). Then the vector is transformed back in energy 
representation, also with $2N^3$ operations, and in total this gives 
an efficient algorithm to compute $G|\psi\!>$ with the help of 
(\ref{eq_Green}). 

It remains to clarify how to calculate $\bar G_0$ efficiently which is 
possible by \cite{frahm5,frahm6}~:
\begin{eqnarray}
\nonumber
<\!x_1,x_2|&G_0&|y_1,y_2\!>=\\
\label{eq_proj_resolv000}
&=&\sum_{\nu,\mu} \frac{\phi_\nu(x_1)\,\phi_\mu(x_2)\,
\phi_\mu(y_2)\,\phi_\nu(y_1)}{E-\epsilon_\nu-\epsilon_\mu}\\
\label{eq_proj_resolv0}
&=&\sum_\nu \phi_\nu(x_1)\,g(E-\epsilon_\nu;x_2,y_2)\,
\phi_\nu(y_1),\\
\label{eq_onepart_green}
g(E;x,y)&=&\sum_\mu\frac{\phi_\mu(x)\,\phi_\mu(y)}
{E-\epsilon_\mu}=<\!x|(E-h)^{-1}|y\!>
\end{eqnarray}
where $g(E;x,y)$ is the one-particle Green function.
To determine $\bar G_0$ 
one needs to compute $U_R^2\,N^2$ matrix elements of 
$G_0$. A naive application of (\ref{eq_proj_resolv000}) 
would require ${\cal O}(U_R^2\,N^4)$ operations but using 
(\ref{eq_proj_resolv0}) the effort can be reduced 
to ${\cal O}(U_R^2\,N^3)$ operations since the one-particle Green function, 
as inverse of a tri-diagonal matrix, can be computed by ${\cal O}(N^2)$ 
operations (for each value of $E-\epsilon_\nu$). 

At this point we mention that the eigenfunctions $\phi_\nu(x)$ can be 
computed efficiently and 
with great accuracy even for the case $|\phi_\nu(x)|\ll 10^{-15}$ 
by inverse vector iteration \cite{schwartz1} for tridiagonal 
matrices, and both expressions (\ref{eq_proj_resolv000}) and 
(\ref{eq_proj_resolv0}) provide correct exponential tails far below 
$10^{-15}$ for the matrix elements of $\bar G_0$ far away from its 
diagonal. 

In summary this provides a method to compute each product 
$G|\psi\!>$ with 
${\cal O}(N^3)$ operations (as long as the value of the Green 
function energy $E$ is not changed) and the initial preparation steps 
to compute $\bar G_0$ and the further matrix 
inverse in (\ref{eq_Green}) for a given value of $E$ 
in advance cost ${\cal O}(U_R^3\,N^3)$ operations \cite{frahm6}. 
Using this algorithm one can implement the simple variant of the 
Arnoldi method to compute the eigenvectors of $G=(E-H)^{-1}$ with largest 
eigenvalues $g_j$ which correspond exactly to the eigenvectors of $H$ with 
eigenvalues $E_j=E-g_j^{-1}$ closest to some fixed energy value $E$ 
which can be arbitrarily chosen in advance. 
Finally, the quality of obtained eigenvectors $|\psi_j\!>$ is tested by a 
completely independent computation of the energy variance~: 
\begin{equation}
\label{deltaE2}
\delta^2 E_j=<\!\psi_j|\,(E_j-H)^2\,|\psi_j\!>
\end{equation}
and only eigenvectors with $\delta^2 E_j<10^{-8}$ are accepted.
For reasonable values of the Arnoldi dimension $n_A$, e.~g. between 
$100$ and $2000$, the method 
selects typically $2/3$ of the $n_A$ initial eigenvectors, those 
whose eigenvalues are closest to $E$. The non-selected eigenvectors 
are either purely artificial due to the mathematical approximation 
of the Arnoldi method or their quality is too low because the corresponding 
eigenvalue is too far away from $E$. It turns out that 
most of the selected eigenvectors have actually a much better quality 
with typical values of $\delta^2 E_j$ below $10^{-20}$ and only very few 
eigenvalues close to the energy borders provided by the method 
correspond to values of $\delta^2 E_j$ close to $10^{-8}$. 

Using this method with $n_A=1000$ we have been able to compute 
about $\sim 650$ eigenvectors for system sizes up to $N=5000$ for a 
few individual samples and up to $N=2000$-$3000$ for a systematic study 
with several different parameter values (for $W$, $U$, $E$, etc.) and 10 
disorder realizations for each case. In \cite{frahm6} it was 
even possible to consider values up to $N\sim 10^4$ using a possible 
reduction of the Hilbert space dimension in energy representation 
by removing non-interacting product eigenstates where both particles 
are localized so far away such that their numerical contribution in 
(\ref{eq_Green}) is below $10^{-17}$ which typically happens at particle 
distances larger then $40L_1$ (with $L_1\approx 4.5$ in \cite{frahm6}). 
However, here with larger values of $L_1$ this optimization has at best 
only a modest effect and concerns a small fraction of non-interacting 
product eigenstates with particularly small values of $L_1$ (due to 
statistical fluctuations or one particle energies close to the band edge). 
For too small ratios $N/L_1$, in particular for small disorder, there 
is even no optimization effect at all. 

In the next two sections we will discuss different properties of eigenstates 
of the TIP Hamiltonian (\ref{ham_tot}) computed by the Green function 
Arnoldi method for the case of periodic boundary conditions. 

Another method to study the localization length is to compute directly 
the two-particle Green function $G=(E-H)^{-1}$ or more precisely the 
projected Green function $\bar G=PGP$ using the expression 
\cite{vonoppen} 
\begin{equation}
\label{eq_green_project}
\bar G=\bar G_0({\bf 1}-\hat U\bar G_0)^{-1}
\end{equation}
which can also be directly derived from (\ref{eq_Green}). 
Let us assume for simplicity the boson case with the Hubbard interaction 
($U_R=1$). Then from $\bar G$ one gets access to all Green function 
matrix elements of the type $<\!x,x|\,G\,|y,y\!>$ describing the 
propagation amplitudes between configurations with both particles on 
the same position. Assuming an exponential decay between $x$ and $y$ 
one can define a two-particle localization length $L_2$ (see Section~\ref{sec5}
for details) and this quantity has been used in various works by 
different methods to compute the projected Green function, either 
by the decimation method \cite{leadbeater1}, the recursive Green function 
method \cite{song2,song1}, or a direct application of (\ref{eq_green_project}) 
combined with the expression (\ref{eq_proj_resolv0}) to determine 
efficiently $\bar G_0$  \cite{frahm5}. One should mention 
that both decimation and recursive Green function method are of 
complexity ${\cal O}(N^4)$ and in \cite{song2,song1,leadbeater1} 
only system sizes up to $\sim 300$ were considered while 
the method based on (\ref{eq_green_project}) and (\ref{eq_proj_resolv0}) 
is of complexity ${\cal O}(N^3)$ (for $U_R=1$) and has allowed to study 
system sizes up to $N=1000$ (or even $N=1400$ for a few data points) 
in \cite{frahm5}. 

The difference of the algorithmic complexity can be understood by the 
fact that decimation and recursive Green function method can be applied to 
generic 2d-tight binding models with arbitrary potential configurations in 
two dimensions while 
the method based on (\ref{eq_green_project}) and (\ref{eq_proj_resolv0}) 
exploits very efficiently the particular TIP structure of disorder 
and interaction potential. 

Actually (\ref{eq_green_project}) allows for further optimizations 
if one considers simultaneously several values for the interaction 
strength $U$. In this case the quite expensive computation of $\bar G_0$ 
by  (\ref{eq_proj_resolv0}) needs to be done only once providing 
a considerable reduction of the computational effort. 

If one limits the number of needed values of $x$ and $y$, e.~g. with 
a few values of $x$ close to one boundary and of $y$ to the other boundary, 
one can apply an even better optimization to (\ref{eq_green_project}) 
by diagonalizing the symmetric matrix $\bar G_0$ 
which provides its normalized eigenvectors $O_j(x)$ and corresponding 
eigenvalues $\lambda_j$. 
Then the computation of individual matrix elements 
\begin{equation}
\label{eq_green_project3}
<\!x,x|\,G\,|y,y\!>=\sum_j O_j(x)\frac{\lambda_j}{1-U\lambda_j} O_j(y)
\end{equation}
is only of complexity ${\cal O}(N)$. Therefore the simultaneous computation 
of (\ref{eq_green_project3}) for many different interaction values 
($\sim 250$) and a modest number of $x$ and $y$ values is nearly 
free of charge if compared to the 
diagonalization of $\bar G_0$ or the matrix inverse in 
(\ref{eq_green_project}) both with complexity ${\cal O}(N^3)$.

This method has however a certain numerical shortcoming if applied 
to the case where $N/L_1$ is so large that the exponential decay 
of one-particle eigenfunctions (and therefore also of the matrix 
elements of $\bar G_0$ far away from the diagonal) 
leads to values below $\sim 10^{-15}$. Normally 
when using directly (\ref{eq_green_project}) the direct matrix inverse 
produces correct exponential tails well below 
$10^{-15}$ if done properly by a stable implementation of Gauss algorithm. 
However, when computing the eigenvectors $O_j(x)$ of the full matrix 
$\bar G_0$ with complexity ${\cal O}(N^3)$ the obtained eigenvectors
are not reliable for the exponential tails below $10^{-15}$. 
Therefore we have used the very efficient variant 
(\ref{eq_green_project3}) only for the case $N\le 20L_1$ 
(i.~e. $e^{-N/L_1}\ge 2\times 10^{-9}$) which is most important 
for the small disorder values where many disorder realizations are needed. 
The few particular cases with $N>20L_1$ for rather strong disorder values, 
which require less disorder realizations, were treated in a more stable way 
using directly (\ref{eq_green_project}) 
but still with the optimization of a single computation of $\bar G_0$ 
for the simultaneous calculation for many different $U$-values. 

We have compared both variants and verified that for $N\le 20L_1$ 
the numerical errors induced by (\ref{eq_green_project3}) 
are several orders of magnitude below the statistical errors arising 
from different disorder realizations. Furthermore, we 
have also numerically verified the general validity of 
(\ref{eq_green_project}) and (\ref{eq_proj_resolv0}) for a few cases 
with sufficiently small system sizes $N$ by directly 
comparing with $G$ computed from a full matrix inverse of $E-H$ 
(on the full two-particle Hilbert space of dimension $N_2\approx N^2/2$). 

In Section~\ref{sec5} and Appendix~\ref{appC} 
we present an extensive discussion of the results 
for the localization length obtained from the Green function, also in 
relation with previous work \cite{song2,song1,leadbeater1,frahm5} 
and concerning details for the precise definition of 
$L_2$ in terms of the matrix elements of $\bar G$ and the finite size 
scaling. 
For this method we limit our studies to the Hubbard interaction case $U_R=1$ 
and for obvious reasons we consider open boundary conditions (instead of 
periodic boundary conditions used 
for the eigenfunction computations in Sections~\ref{sec3} and \ref{sec4}). 

\section{Eigenfunction structure}

\label{sec3}

Let $|\psi\!>$ be an eigenvector of the TIP Hamiltonian ({\ref{ham_tot}) 
with $\psi(x_1,x_2)$ and $\chi_{\nu_1,\nu_2}$ being the corresponding 
wavefunctions in position representation (\ref{eq_pos_rep}) 
or energy representation (\ref{eq_en_rep}). 
To characterize the localization properties of such eigenstates we 
will use three variants of the inverse participation ratio 
(IPR) and one further length scale for the relative distance 
between the two particles. The first IPR type length scale is given by:
\begin{equation}
\label{eq_ipr_x}
\xi_x=\left(\sum_x \rho_1^2(x)\right)^{-1}
\quad,\quad \rho_1(x)=\sum_{x_2} |\psi(x,x_2)|^2
\end{equation}
where $\rho_1(x)$ is the one-particle density corresponding to the 
probability to find a particle at position $x$ and obeying the 
normalization $\sum_x\rho_1(x)=1$. The quantity $\xi_x$ is the 
inverse participation ratio in position representation 
and corresponds roughly to the number of sites $x$ contributing in 
$\rho_1(x)$. In a similar way we introduce also an inverse participation 
ratio for the center of mass by:
\begin{eqnarray}
\label{eq_CM_x}
\xi_{\rm CM}&=&\left(\sum_{s=0}^{2N-1} \rho_{\rm CM}^2(s)\right)^{-1}\\
\label{eq_CM_dens}
\rho_{\rm CM}(s)&=&\sum_{-N/2<r\le N/2} |\psi(X_1(s,r),X_2(s,r))|^2
\end{eqnarray}
where $s$ corresponds to twice the center of mass $x_1+x_2$ and 
$r$ to the relative coordinate $x_1-x_2$ between the two particles. 
However, the exact translation from $(x_1,x_2)$ to $(s,r)$ is somewhat 
tricky due to the periodic boundary conditions. The sum over $s$ 
in (\ref{eq_CM_x}) runs over all integer values $0\le s<2N$ 
while the sum over $r$ in (\ref{eq_CM_dens}) runs over 
all integer (or half-integer) values for the case of even (odd) $s$ 
such that $-N/2<r\le N/2$. The mapping functions $X_1$ and $X_2$ are given by
\begin{eqnarray}
\label{eq_map_func1}
X_1(s,r)&=&\Bigl((s+r)/2+N\Big) \mod N,\\
\label{eq_map_func2}
X_2(s,r)&=&\Big((s-r)/2+N\Big) \mod N
\end{eqnarray}
using the integer modulo operation 
to map the final values to the set $\{0,\,\ldots,\,N-1\}$. 
Using these mapping functions 
in (\ref{eq_CM_dens}) implies that we take the difference $x_1-x_2$ 
with respect to the periodic boundary conditions, e.~g. if $x_1$ is close 
to one boundary and $x_2$ to the other boundary we map $x_2$ 
behind the first boundary by adding or removing $N$ and compute the 
difference $r=x_1-x_2$ after the mapping such that $r$ always 
obeys $-N/2<r\le N/2$. In other words the square domain for $(x_1,x_2)$ 
is mapped to a rectangle with its longer side parallel to the diagonal 
and its constant width orthogonal to the diagonal such that points outside 
the initial square have been mapped from the square to the rectangle 
by periodic boundary conditions. 
We choose for $s$ twice the center of mass in order to assure integer values 
for this quantity. 
Therefore for a typical cigar-shape state, rather strongly delocalized in $s$ 
along the diagonal but stronger localized in $r$ orthogonal to the diagonal, 
we expect roughly $\xi_{\rm CM}\approx 2\,\xi_x$ but this relation 
does not need to hold for other shapes of localized states. 

We furthermore introduce the average particle distance by the 
expectation value~:
\begin{equation}
\label{eq_drel}
d_{\rm rel}=
\sum_{s=0}^{2N-1}\sum_{-N/2<r\le N/2} |r|\, |\psi(X_1(s,r),X_2(s,r))|^2
\end{equation}
and due to the use of the mapping functions (\ref{eq_map_func1}) 
and (\ref{eq_map_func2}) this quantity also takes into account the periodic 
boundary conditions when measuring the distance between the two particles. 

Finally we introduce the inverse participation ratio in energy representation 
by 
\begin{equation}
\label{eq_ipr_E}
\xi_E=\left(
s\sum_{\nu}|\chi_{\nu,\nu}|^4+
4\sum_{\nu_1>\nu_2} |\chi_{\nu_1,\nu_2}|^4
\right)^{-1}
\end{equation}
with $s=1$ for bosons and $s=0$ for fermions, and 
using the wave function in energy representation 
(\ref{eq_en_rep}). 
The factor $4$ for $\nu_1>\nu_2$ is due to the modified coefficient 
$\sqrt{2}\,\chi_{\nu_1,\nu_2}$ 
associated to the (anti-)symmetrized basis states 
$(|\phi_{\nu_1}\,\phi_{\nu_2}\!>\pm |\phi_{\nu_2}\,\phi_{\nu_1}\!>)/\sqrt{2}$
when rewriting (\ref{eq_en_rep}) in its (anti-)symmetrized form. 
The quantity $\xi_E$ 
essentially measures the number of non-interacting product eigenstates (of 
$H_0$) which contribute to the state $|\psi\!>$. Note that $\xi_E$ is 
defined in terms of a two-particle density while $\xi_x$ uses 
a one-particle density. 

As explained in the last section we employ the Green function Arnoldi 
method for various system sizes $N$, disorder strengths $W$ with 
the Arnoldi dimension $n_A=1000$ to compute about $\sim 650$ eigenstates 
with energies either close to $E=0$ or $E=1$. In this section we limit 
ourselves to $U=2$ and the case of the Hubbard interaction $U_R=1$. 

\begin{figure}
\begin{center}
\includegraphics[width=0.47\textwidth]{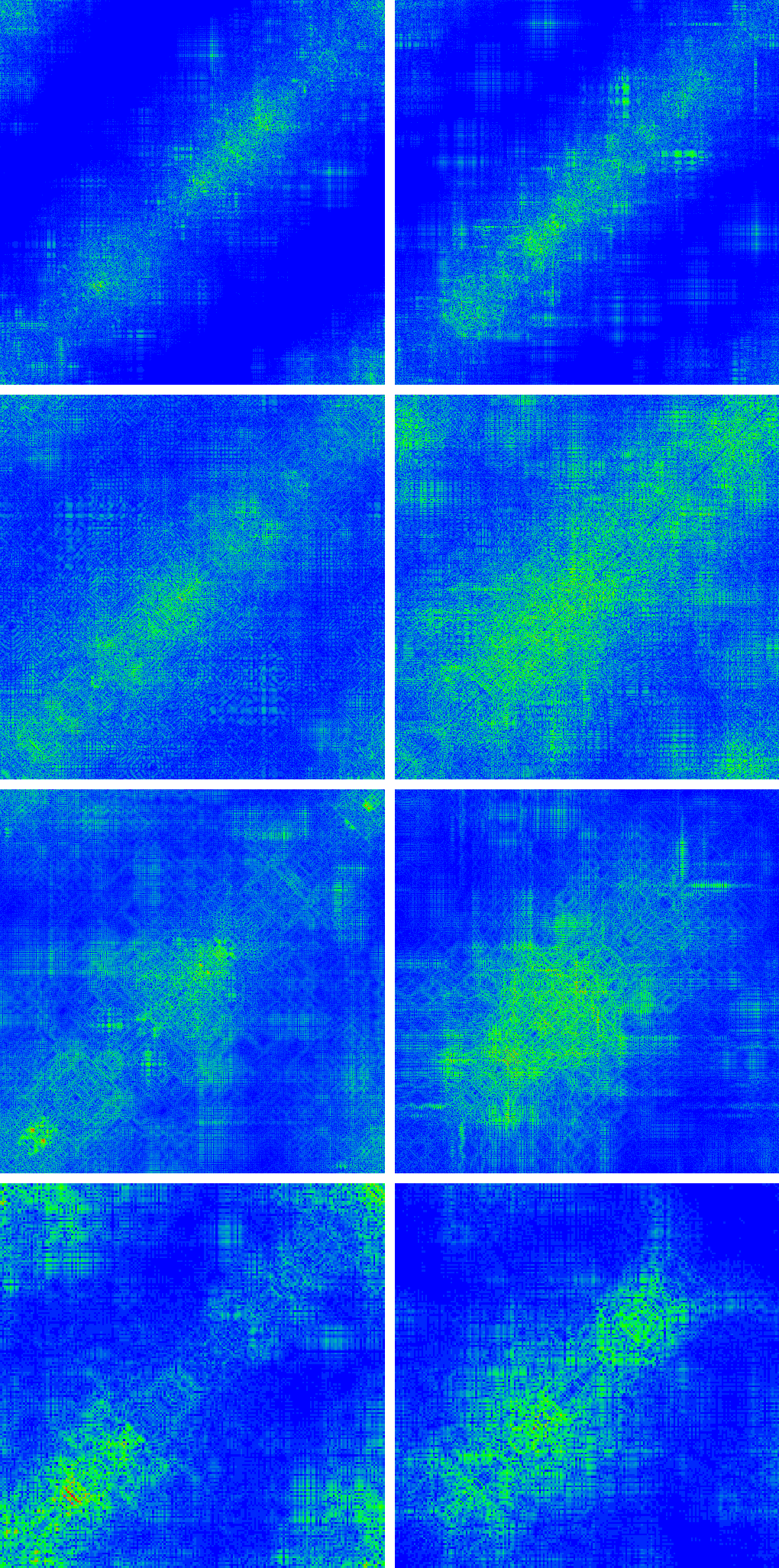}
\caption{Density plot of typical delocalized eigenfunctions of 
(\ref{ham_tot}) for $U=2$, $U_R=1$, energies close to $E=0$ (left) or $E=1$
(right), and 
$W=0.75$, $N=5000$ (first row), \mbox{$W=1$}, $N=1000$ (second row), 
$W=1.5$, $N=500$ (third row), \mbox{$W=2.5$}, $N=200$ (fourth row). 
The values of $\xi_x$, $\xi_{\rm CM}$, $\xi_E$, and $d_{\rm rel}$ are:
{\em first row: left:} $\xi_x=3044.00$, $\xi_{\rm CM}=4993.33$, 
$\xi_E=221.53$, $d_{\rm rel}=288.11$, 
{\em right:} $\xi_x=2977.21$, $\xi_{\rm CM}=6033.04$, 
$\xi_E=167.82$, $d_{\rm rel}=344.77$, 
{\em second row: left:} $\xi_x=773.54$, $\xi_{\rm CM}=1336.25$, 
$\xi_E=100.99$, $d_{\rm rel}=99.65$, 
{\em right:} $\xi_x=804.24$, $\xi_{\rm CM}=1623.51$, $\xi_E=94.32$, 
$d_{\rm rel}=129.13$, 
{\em third row: left:} $\xi_x=305.00$, $\xi_{\rm CM}=505.15$, $\xi_E=55.83$, 
$d_{\rm rel}=55.94$, 
{\em right:} $\xi_x=215.75$, $\xi_{\rm CM}=378.44$, $\xi_E=55.35$, 
$d_{\rm rel}=57.65$, 
{\em fourth row: left:} $\xi_x=79.89$, $\xi_{\rm CM}=108.19$, $\xi_E=38.10$, 
$d_{\rm rel}=19.01$, 
{\em right:} $\xi_x=95.11$, $\xi_{\rm CM}=158.41$, $\xi_E=42.09$, 
$d_{\rm rel}=16.21$.
The two axes correspond to the two particle positions $x_1$ and $x_2$. 
Red/green/blue corresponds to maximum/medium/minimum values 
of $|\psi(x_1,x_2)|$.
}
\label{fig1}
\end{center}
\end{figure}

Figure~\ref{fig1} shows for certain cases with $200\le N\le 5000$ 
color density plots of 
typical rather strongly delocalized two-particle eigenfunctions of the 
TIP Hamiltonian (\ref{ham_tot}) with values of $\xi_x$ and $\xi_{\rm CM}$ 
rather close to the system size $N$ despite the fact that the choice of the 
disorder parameter implies approximately $N\approx 11\,L_1$ (for the cases 
with $N\le 1000$) or even $N\approx 27\,L_1$ (for the case $N=5000$). 
These states are more concentrated close to the diagonal which is 
confirmed by the observation that their values of $d_{\rm rel}$ are 
comparable to $L_1$. Furthermore, the internal structure 
of the eigenfunctions 
is quite complicated with many holes also close to the diagonal. 
Sometimes, especially for $E=1$, one can see certain horizontal and 
vertical structures which indicate a contribution of a non-interacting 
product eigenstate where $L_1$ for one particle is considerably stronger or 
weaker than $L_1$ for the other particle. The values of $\xi_E$ are always 
very clearly above unity, indicating a strong mixing or delocalization, 
We mention that the precise form for other examples of delocalized eigenstates 
varies very strongly, with a rich structure and sometimes even 
the overall cigar-shape along the diagonal is not very clearly visible. 

\begin{figure}
\begin{center}
\includegraphics[width=0.47\textwidth]{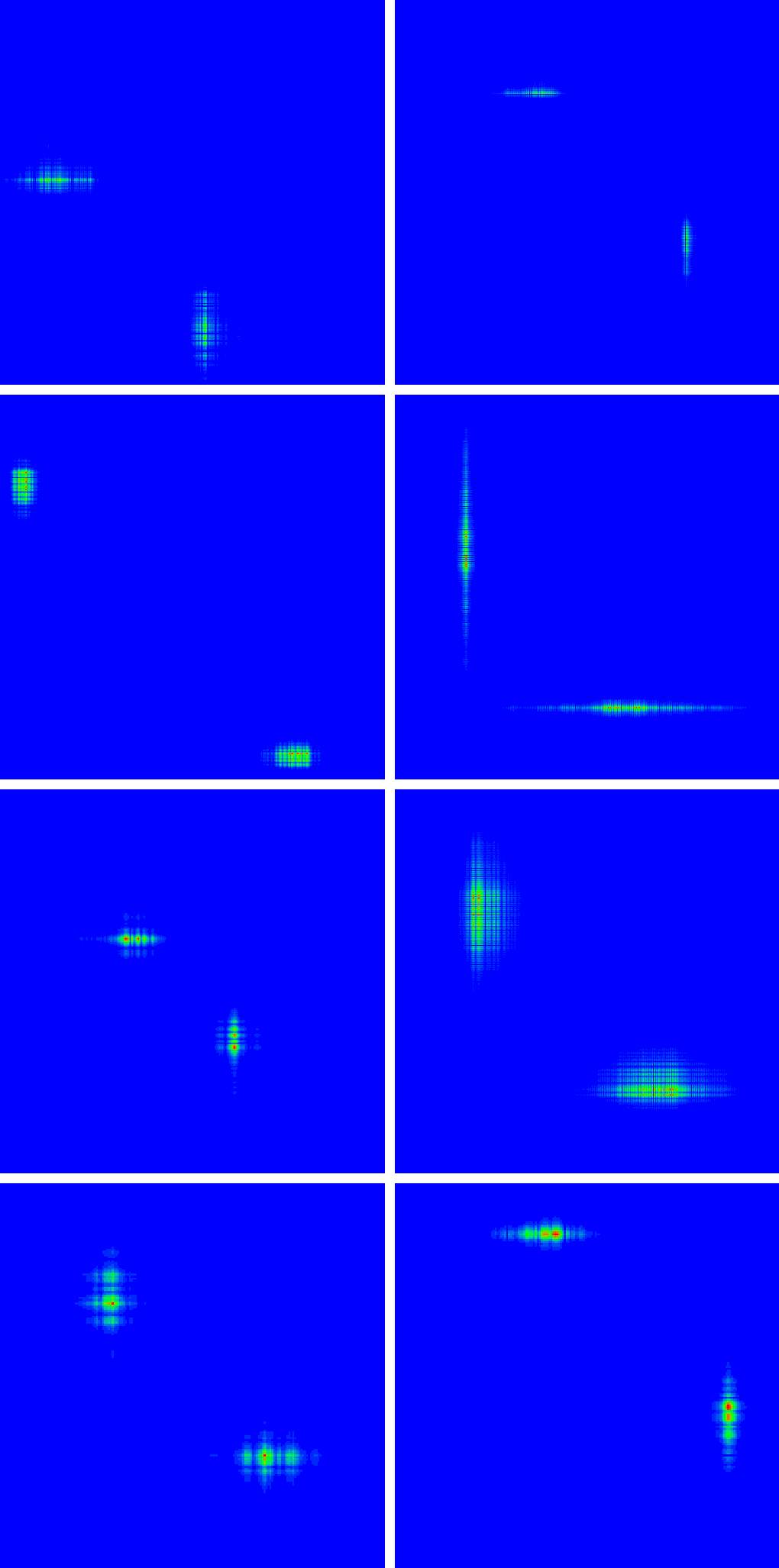}
\caption{Density plot of typical localized product eigenfunctions of 
(\ref{ham_tot}) with $\xi_E=1$ (with a numerical error below $10^{-7}$) 
for $U=2$, $U_R=1$, energies close to $E=0$ (left) or $E=1$
(right), and same values of $W$ and $N$ as in Figure~\ref{fig1} 
for corresponding rows. 
The values of $\xi_x$, $\xi_{\rm CM}$, and $d_{\rm rel}$ are: 
{\em first row: left:} $\xi_x=125.35$, $\xi_{\rm CM}=395.36$, 
$d_{\rm rel}=1947.08$, 
{\em right:} $\xi_x=57.42$, $\xi_{\rm CM}=264.38$, 
$d_{\rm rel}=1898.33$, 
{\em second row: left:} $\xi_x=51.81$, $\xi_{\rm CM}=73.02$, 
$d_{\rm rel}=290.64$, 
{\em right:} $\xi_x=27.80$, $\xi_{\rm CM}=105.41$, 
$d_{\rm rel}=408.16$, 
{\em third row: left:} $\xi_x=18.40$, $\xi_{\rm CM}=23.67$, 
$d_{\rm rel}=133.56$, 
{\em right:} $\xi_x=32.86$, $\xi_{\rm CM}=92.06$, 
$d_{\rm rel}=231.51$, 
{\em fourth row: left:} $\xi_x=8.20$, $\xi_{\rm CM}=9.24$, 
$d_{\rm rel}=80.91$, 
{\em right:} $\xi_x=10.59$, $\xi_{\rm CM}=11.82$, 
$d_{\rm rel}=92.01$.
The two axes correspond to the two particle positions $x_1$ and $x_2$. 
Red/green/blue corresponds to maximum/medium/minimum values 
of $|\psi(x_1,x_2)|$.
}
\label{fig2}
\end{center}
\end{figure}

For comparison we show in Figure~\ref{fig2} for the same parameters as 
in Figure~\ref{fig1} typical localized product eigenstates where $\xi_E$ 
is rather precisely unity and where both particles are localized far away 
such that the interaction does not significantly influence these type 
of states. The Green function Arnoldi method has apparently no problem 
in correctly identifying such states, which form actually the majority 
of found eigenstates for $N\gg L_1$. Their values of $\xi_x$ and 
$\xi_{\rm CM}$ are comparable to $L_1$ while now $d_{\rm rel}$, the 
average particle distance, is significantly larger than $L_1$. 
Sometimes, especially for $E=1$, one can see that the one-particle 
localization length for one particle is considerably larger than 
for the other particle. We mention that for these kind of states 
the values of $\xi_x$ and $\xi_{\rm CM}$ exhibit still quite large 
statistical fluctuations (but still comparable to $L_1$) due to the 
fluctuations of the IPR for the 
one-particle 1d-Anderson model without interaction.

\begin{figure}
\begin{center}
\includegraphics[width=0.47\textwidth]{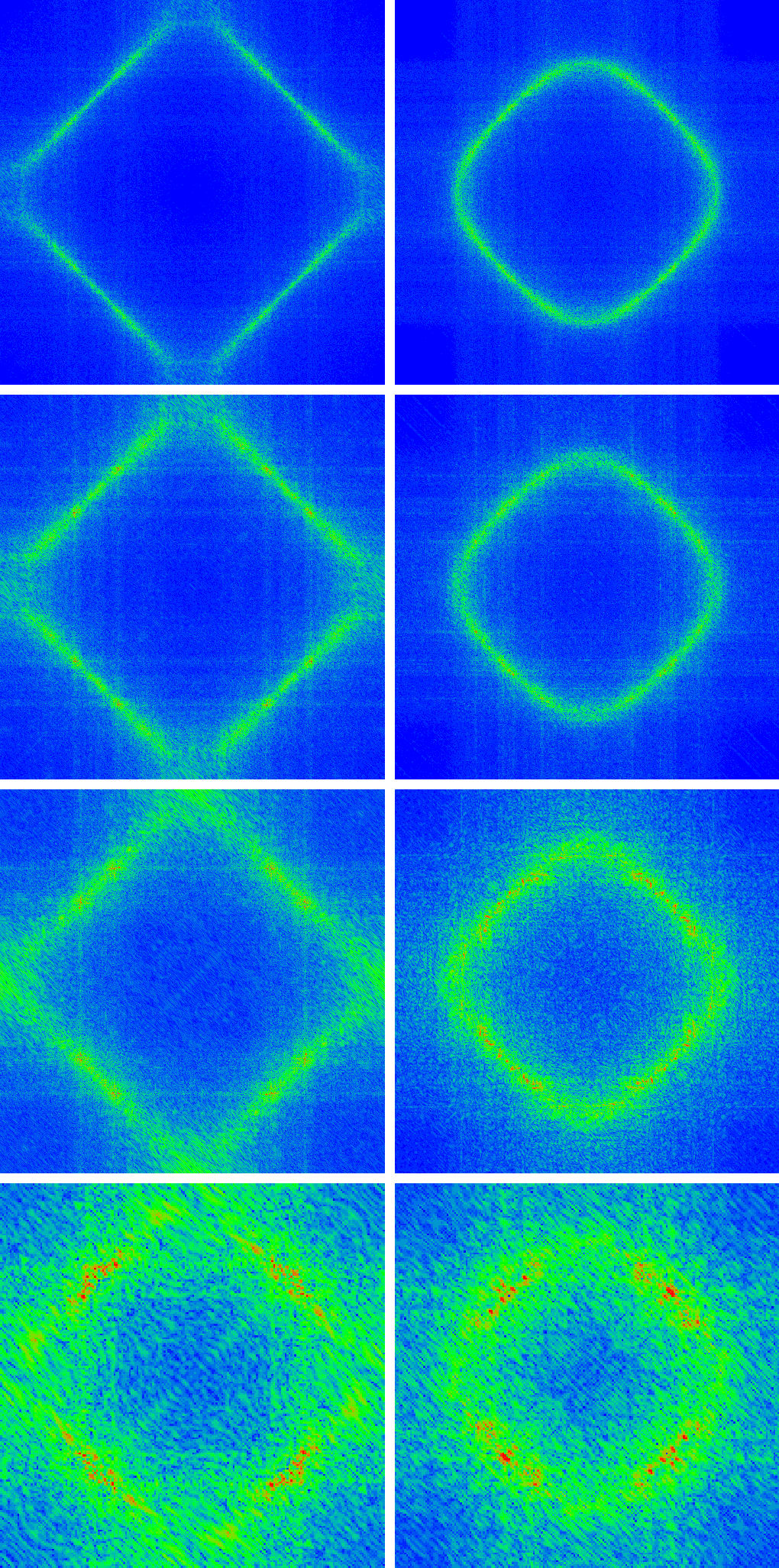}
\caption{Density plot in momentum representation of the 
eigenstates of (\ref{ham_tot}) of Figure~\ref{fig1} (same eigenstates 
for corresponding panels). 
The two axes correspond to the two particle momenta $k_1$ and $k_2$. 
Red/green/blue corresponds to maximum/medium/minimum values 
of $|\bar\psi(k_1,k_2)|$ with $\bar\psi(k_1,k_2)$ being the 
discrete Fourier transform of $\psi(x_1,x_2)$.}
\label{fig3}
\end{center}
\end{figure}

We also determined the wave function $\bar\psi(k_1,k_2)$ 
in momentum representation obtained by a standard 2d-discrete 
Fou\-rier transform 
from $\psi(x_1,x_2)$ and with discrete values $k_j=2\pi p_j/N$, 
$p_j=0,\,\ldots,\,N-1$ for the momenta. Figure~\ref{fig3} shows 
density plots of this quantity for the same eigenstates of Figure~\ref{fig1} 
(for corresponding panels). The amplitudes in momentum representation 
are maximal for momenta close to the Fermi surface of the 2d tight-binding 
model (without disorder/interaction), i.~e. 
$-2\cos(k_1)-2\cos(k_2)\approx E$ for the two cases 
$E=0$ (square form with sides parallel to the diagonals) 
or $E=1$ (a closed curve a bit similar to but still different from a circle). 

To understand this we remind that in the weak disorder limit 
the one-particle eigenfunctions $\phi_\nu(x)$ of the 1d-Anderson model 
(\ref{eq_h1}) have quite well defined momenta $k$ with 
$\epsilon_\nu\approx -2\cos(k)$ and 
the momentum fluctuations due the finite localization length $\sim L_1$ 
are of order $\Delta k\sim 1/L_1$ \cite{ponomarev1} 
implying a one-particle (disorder-induced)
Breit-Wigner width $\Gamma_W\sim 1/L_1$ such that momenta with 
$|\epsilon_\nu+\cos(k)|<\Gamma_W$ contribute to the discrete Fourier expansion 
of $\phi_\nu(x)$. 
Furthermore, in energy representation (\ref{eq_en_rep}) of a 
two-particle eigenstate essentially only non-interacting product eigenstates 
with $|E-(\epsilon_{\nu_1}+\epsilon_{\nu_2})|<\Gamma_U$ contribute 
where $\Gamma_U$ is the (interaction induced) Breit-Wigner width roughly given 
by $\Gamma_U\sim c(U)/L_1$ with a function $c(U)\sim |U|$ for small to 
modest values of $|U|$ \cite{jacquod2,frahm5}. 

In total this implies that in momentum representation momenta obeying 
$|E+2\cos(k_1)+2\cos(k_2)|<\Gamma_{\rm tot}$ contribute to the 
two-particle eigenstate of (\ref{ham_tot}) where 
$\Gamma_{\rm tot}\approx\Gamma_U+\,2\Gamma_W\sim 1/L_1\sim W^2$ is somewhat 
the total momentum Breit-Wigner width. The dependence of this width 
on $L_1$ or $W$ is very clearly visible in Figure~\ref{fig3} with quite 
sharply defined curves for $W=0.75$ (top panels in Figure~\ref{fig3}) 
and quite thick curves for $W=2.5$ (bottom panels in Figure~\ref{fig3}). 
For the case $E=0$ the effective width close to the corners of the square 
(with one momentum close to $\pi$ and the other one close to $0$ or $2\pi$) 
seems strongly enhanced which can be 
understood by the strongly reduced one-particle localization length 
$\tilde L_1\approx L_1\sin^2(k_{1,2})\ll L_1$ for both particles implying 
a strongly enhanced momentum uncertainty and therefore increasing 
the effective value of $\Gamma_W$. 

\begin{figure}
\begin{center}
\includegraphics[width=0.47\textwidth]{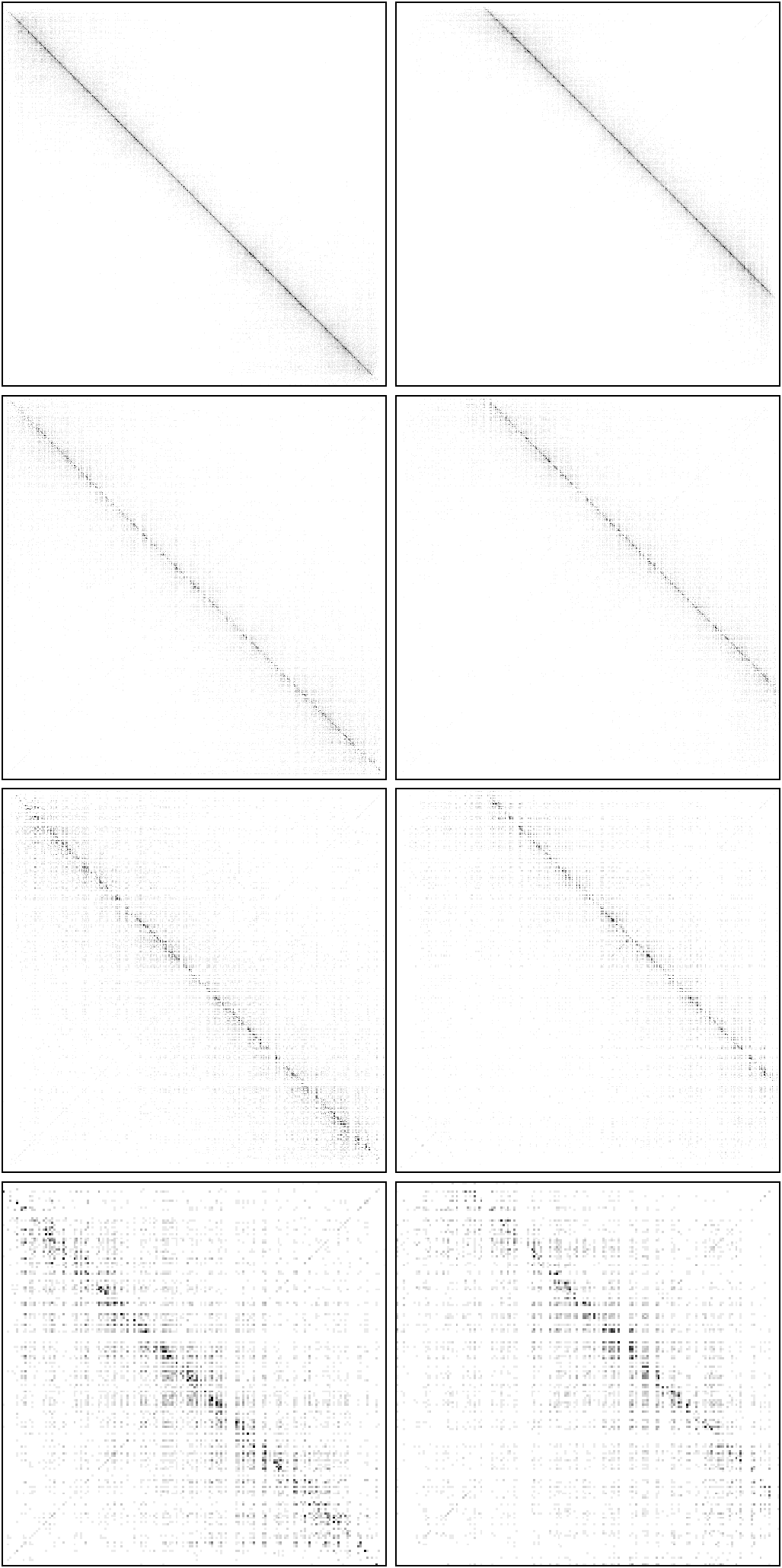}
\caption{Grey scale density plot in one-particle energy representation 
(\ref{eq_en_rep}) of the eigenstates of (\ref{ham_tot}) 
of Figure~\ref{fig1} (same eigenstates for corresponding panels).
Both axes correspond to the (one-particle) energies $\epsilon_{\nu_1}$ 
and $\epsilon_{\nu_2}$ of the two particles. 
The pixel size corresponds to the average level spacing in the band 
center of non-interacting one-particle energies $\epsilon_\nu$ 
(cases with $N\le 1000$, second to fourth rows). This representation implies 
a slight coarse-graining if due to fluctuations several one-particle energies 
correspond to the same cell. 
The pixel size for the case $N=5000$ (top row) corresponds to 5 times 
the average level spacing in the band center 
implying a stronger coarse-graining for a better visibility. 
Black/grey/white corresponds to maximum/medium/minimum values 
of $|\chi_{\nu_1,\nu_2}|$. 
}
\label{fig4}
\end{center}
\end{figure}

To illustrate the effect of the interaction induced Breit-Wigner width 
$\Gamma_U$ we show in Figure~\ref{fig4} density plots of 
the wave function in energy representation (\ref{eq_en_rep}) 
for the same eigenstates of Figure~\ref{fig1} (for corresponding panels). 
The two axes correspond the one-particle energies $\epsilon_{\nu_1}$ 
and $\epsilon_{\nu_2}$ with a pixel size corresponding 
to the average level spacing of $\epsilon_\nu$ in the band-center 
of (\ref{eq_h1}) for the three bottom panels with $N\le 1000$. In this way 
in average a pixel corresponds approximately to one value of $\epsilon_\nu$. 
However, due to fluctuations of the one-particle energies and a reduced 
level spacing at the band edges 
there is a slight coarse-graining, with either some empty 
cells or a few values of $\epsilon_\nu$ for other cells. 
For the case $N=5000$ with such a representation 
the black pixels for maximum values would only be barely visible. 
Therefore we have applied for this case (shown in the top panels) a 
somewhat stronger coarse-graining using a pixel size 
of 5 times the average level spacing in the band center. 

One can clearly see that the maximal contributions in energy representation 
correspond to the lines $E\approx \epsilon_{\nu_1}+\epsilon_{\nu_2}$ 
confirming the expected condition 
$|E-(\epsilon_{\nu_1}+\epsilon_{\nu_2})|<\Gamma_U$ with the 
interaction induced Breit-Wigner width $\Gamma_U$. Furthermore, one 
can also observe that the effective width of the lines increases 
with decreasing values of $L_1$ (or increasing values of $W$ from top to 
bottom panels) which is in qualitative agreement with 
$\Gamma_U\sim 1/L_1\sim W^2$ which is similar to the width 
$\Gamma_{\rm tot}$ visible in Figure~\ref{fig3} but still with 
a considerably smaller numerical prefactor for $\Gamma_U$ as compared 
to $\Gamma_{\rm tot}$. We note that in principle, and without the 
coarse-graining, the quantity $\xi_E$ would correspond the number of 
black pixels in Figure~\ref{fig4}. This figure clearly confirms 
the Breit-Wigner type ``energy space localization'' one can find 
in random band matrix models with a strong diagonal 
\cite{jacquod1,fyodorov1,frahm2} even though the interaction 
dependence of $\Gamma_U$ 
for the TIP problem is different as predicted in such models 
due to the (somewhat incorrect) assumption of random uniform distributions 
of interaction coupling matrix elements for the latter. 

\begin{figure}
\begin{center}
\includegraphics[width=0.48\textwidth]{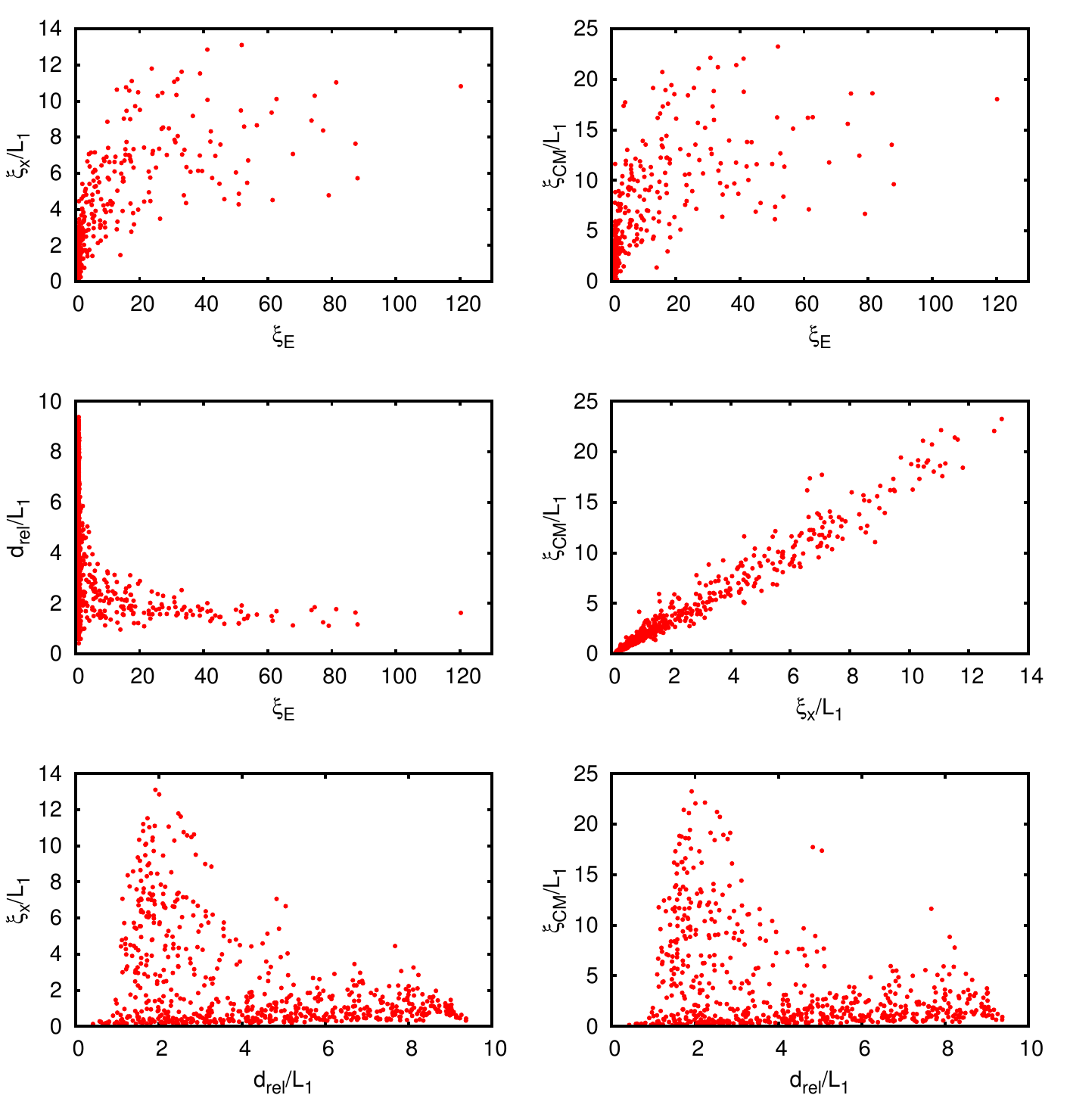}
\caption{Pairwise cross dependence of the four quantities 
$\xi_x/L_1$, $\xi_{\rm CM}/L_1$, $d_{\rm rel}/L_1$ and 
$\xi_E$ computed for the $657$ eigenstates with energies closest to $E=0$ 
for one particular disorder realization of the Hamiltonian (\ref{ham_tot})
for $U=2$, $U_R=1$, $N=2000$ and $W=1$. $L_1$ is the corresponding 
one-particle localization length $L_1=105$. 
Each of the panel shows the dependence of one of these four quantities on 
one of the other three quantities providing 6 combinations. }
\label{fig5}
\end{center}
\end{figure}

The physical picture of the TIP delocalization effect described 
in \cite{dlstip} is that the delocalized TIP eigenstates in position 
representation show 
also a strong delocalization in energy representation and contain 
only non-interacting pair eigenstates where both particles have a typical 
distance $\sim L_1$. Other non-interacting pair eigenstates with 
particles distances $\gg L_1$ are essentially untouched by the interaction 
and are therefore perfectly localized in energy representation with 
$\xi_E=1$. Figure~\ref{fig5} illustrate these points rather clearly 
by showing the cross-dependencies of all combinations between 
two of the four quantities $\xi_x/L_1$, $\xi_{\rm CM}/L_1$, 
$d_{\rm rel}/L_1$ and $\xi_E$ obtained from $657$ eigenstates 
with energies close to $E=0$ for one particular disorder realization of 
the Hamiltonian (\ref{ham_tot}) for $N=2000$ and $W=1$. 

The two quantities $\xi_x$ and $\xi_E$ seem to be loosely correlated 
in the sense that large values of $\xi_E$ imply larger values of 
the ratio $\xi_x/L_1$ but there are statistical fluctuations 
with $\xi_x/L_1$ being large for modest values of $\xi_E$ and vice-versa. 
For example the eigenstate with maximal $\xi_E\approx 120$ corresponds to 
$\xi_x\approx 11 L_1$ while there is another eigenstate with a considerably 
smaller value $\xi_E\approx 25$ and still $\xi_x\approx 12 L_1$. 
Localized pair states with $\xi_E\approx 1$ correspond to small values 
of $\xi_x/L_1$ of order unity but statistical fluctuations of the 
one-particle IPR allow for values up to $4$-$5$ of the latter. The behavior 
for the dependence of $\xi_{\rm CM}$ on $\xi_E$ is rather similar 
with values of $\xi_{\rm CM}$ that are roughly twice the values of $\xi_x$. 

The dependence of the average particle distance $d_{\rm rel}$ on $\xi_E$ is 
rather clear. Large values of $d_{\rm rel}\gg L_1$ are only possible 
for $\xi_E\approx 1$ corresponding to pair localized eigenstates and 
large values of $\xi_E$ imply values of $d_{\rm rel}/L_1$ between $1$-$2$. 

The two quantities $\xi_{\rm CM}$ and $\xi_x$ are rather well correlated 
and the expected behavior $\xi_{\rm CM}\approx 2\xi_x$ is indeed quite 
well verified in average. However, also here we observe some significant 
statistical deviations, probably due to some particular effects of the 
shape of the eigenstate, if it is closer to a cigar form or a more bulky 
shape. 

The dependence of $\xi_x$ ($\xi_{\rm CM}$) on $d_{\rm rel}$ is somewhat 
similar to the dependence of $\xi_E$ on $d_{\rm rel}$, i.~e. large 
values $\xi_x\gg L_1$ ($\xi_{\rm CM}\gg L_1$) require $d_{\rm rel}\sim L_1$ 
and large values $d_{\rm rel}\gg L_1$ correspond to $\xi_x\sim L_1$ 
($\xi_{\rm CM}\sim L_1$). However, the statistical fluctuations with respect to 
these two limits are considerably stronger as compared to the dependence 
of $\xi_E$ on $d_{\rm rel}$. 

The results for exact eigenstates of large TIP systems 
shown in this section illustrate and confirm quite clearly 
many of the physical properties concerning the TIP enhancement of the 
one-particle localization length as described in the early work 
\cite{dlstip,jacquod1,fyodorov1,frahm2} provided that the functional 
dependence of the Breit-Wigner width is corrected taking into account 
realistic distributions of the interaction coupling matrix elements 
\cite{jacquod2,ponomarev1,frahm5}. 

Higher quality gif files for the different panels of Figures~\ref{fig1} 
to \ref{fig4} including color versions for the panels 
of Figure~\ref{fig4} are available for download at 
\cite{webpage}. 

\section{Scaling of IPR}

\label{sec4}

In this section we present and discuss results for the parameter 
dependence on disorder, interaction strength and range of the three 
IPR quantities $\xi_x$, $\xi_{\rm CM}$ and $\xi_E$ obtained from 
effective averages and finite size scaling of several disorder 
realizations. For this we compute appropriate finite size 
(harmonic) averages of these quantities for a selection of 
{\em relevant} TIP eigenstates corresponding to particle distances 
$\sim L_1$ for which the interaction induced enhancement effect is expected 
to be best visible \cite{dlstip}. Explicitely, the relevant eigenstates are 
selected as the fraction $L_1/N$ of eigenstates with maximal values of 
$\xi_E$, the IPR in energy 
representation. This choice seems preferable to us since $\xi_E$ 
measures most directly the interaction induced delocalization effect 
while $\xi_x$ and $\xi_{\rm CM}$ are also influenced by the rather 
considerable statistical fluctuations of the one-particle localization lengths 
of the non-interacting product eigenstates. Actually 
Figure~\ref{fig5} shows that 
correlations of $\xi_x$ (or $\xi_{\rm CM}$) with $\xi_E$ are rather loose 
and therefore the eigenstates with maximal $\xi_E$ are not exactly the same 
as those with maximal $\xi_x$. 

In absence of interaction we have precisely $\xi_E=1$ for all eigenstates and 
in order to be able to determine the set of relevant states for this 
particular case we chose $U=10^{-5}$ and not exactly $U=0$ as reference 
value for ``vanishing interaction strength''. The small interaction 
value does not 
significantly modify the values of the IPR quantities but it ensures small 
differences of $\xi_E$ allowing to distinguish between the relevant 
eigenstates with $\xi_E$ slightly above unity, typically 
$\xi_E-1\sim 10^{-3}$, 
and non-relevant states corresponding to precisely $\xi_E=1$. 

For each parameter set of $N$, $W$, $U$, $U_R$ and $E$ we computed 
about $\sim 650$ two-particle eigenstates (per disorder sample) 
by the Green function Arnoldi method 
using the Arnoldi dimension $n_A=1000$ and for 10 different disorder 
realizations providing $\sim 6500$ eigenstates in total per 
parameter set. For a fixed value of $N$ and different other parameters 
we always chose the same 10 disorder realizations, with the precision that 
``same disorder realization'' for two different disorder values $W$ 
means a uniform scaling factor between the two disorder configurations. 
Then, as already explained, we selected for each sample the fraction of 
$L_1/N$ eigenstates with maximal values of $\xi_E$ as relevant states. 
Using these 
selected states we computed the inverse average (harmonic mean) to obtain 
the (inverse) size dependent ``average values'' for the three quantities 
$\xi_x$, $\xi_{\rm CM}$ and $\xi_E$. The corresponding statistical errors 
are typically between $1$\% and $3$\% and strangely here the relative errors 
are somewhat larger for stronger disorder or smaller interaction values. 

\begin{figure}
\begin{center}
\includegraphics[width=0.48\textwidth]{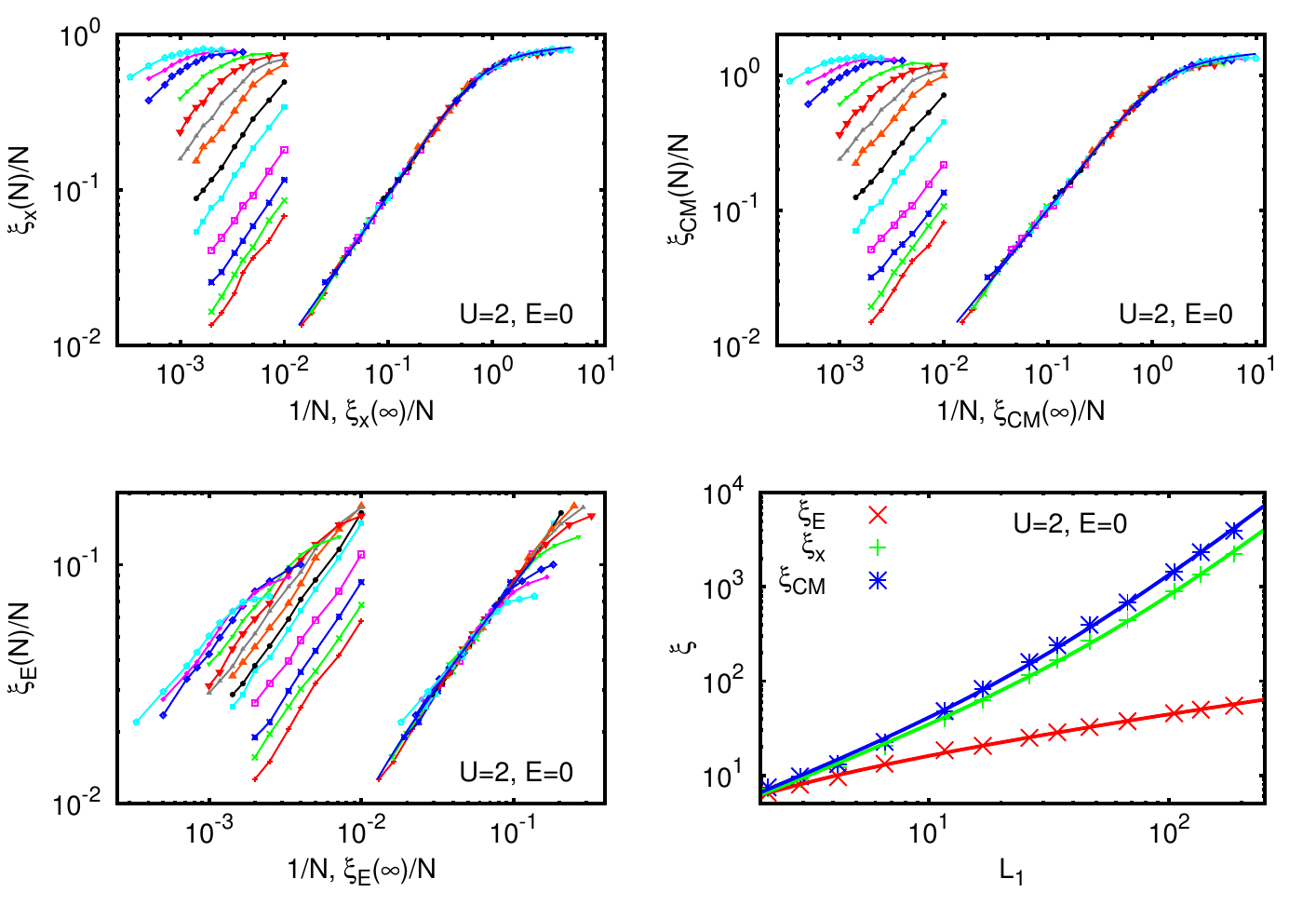}
\caption{Illustration of the scaling procedure to determine the 
three inverse participation ratios $\xi_x$, $\xi_{\rm CM}$ 
and $\xi_E$ at infinite system size. 
Top left panel shows (in a double logarithmic scale) the rescaled 
finite size IPR $\xi_x(N)/N$ versus $1/N$ (left set of curves) 
and versus $\xi_x(\infty)/N$ (right set of curves) 
with $\xi_x(\infty)$ determined such 
that the data points are closest to a single scaling curve (approximately) 
shown by the blue curve corresponding to the fit (\ref{eq_fitscaling}). 
Top right (bottom left) panel shows the scaling 
for $\xi_{\rm CM}$ ($\xi_E$). 
The interaction strength is $U=2$, corresponds to the Hubbard short range 
interaction and the approximate energy is $E=0$. The used disorder values are 
$W=0.75,\,0.875,\, 1,\, 1.25,\, 1.5,\, 1.75,\, 2,
\, 2.5,\, 3,\, 4,\, 5,\, 6,\, 7$ with $W=0.75$ for the top curve and 
$W=7$ for the bottom curve. 
The range of system sizes is 
$100\le N\le N_{\rm max}$ with $N_{\rm max}=500$ 
(for largest values of $W$) or $N_{\rm max}=2000$ 
(for smallest values of $W$). For $W=0.75$ also one data point with 
$N=3000$ has been computed. 
For the scaling procedure only data points with $N\ge 2\,L_1$ have been used 
and are visible in top and bottom left panels. 
Bottom right panel shows (in a double logarithmic scale) 
the three infinite size IPR quantities obtained by 
finite size scaling versus $L_1=105/W^2$ (discrete data points). 
The continuous curves in this panel correspond to the fits: 
$\xi_{\rm CM}=a_2\,L_1+b_2\,L_1^{\gamma_2}$ with $a_2=3.10\pm 0.12$, 
$b_2=0.098\pm 0.038$, $\gamma_2=2.01\pm 0.09$ (top blue curve), 
$\xi_x=a_1\,L_1+b_1\,L_1^{\gamma_1}$ with $a_1=3.00\pm 0.12$, 
$b_1=0.043\pm 0.024$, $\gamma_1=2.04\pm 0.12$ (middle green curve) and
$\xi_E=a_3+b_3\,L_1^{\gamma_3}$ with $a_3=-7.0\pm 1.6$, 
$b_3=10.4\pm 1.4$, $\gamma_3=0.35\pm 0.03$ (lower red curve). 
}
\label{fig6}
\end{center}
\end{figure}

For each set of different values of $E$, $U$, $U_R$, and eventual boson or 
fermion case, we determined the disorder dependent ``infinite size'' IPR 
by the procedure of one-parameter finite size-scaling 
\cite{mackinnon1,mackinnon2,kramer1} 
by fitting the data to a universal scaling function $f$ by~:
\begin{equation}
\label{eq_scale_main}
\frac{\xi(N)}{N}=f\left(\frac{\xi(\infty)}{N}\right)
\end{equation}
where $\xi(N)$ represents one of the three (size and disorder dependent) 
IPR quantities 
($\xi_x$, $\xi_{\rm CM}$ or $\xi_E$) and $\xi(\infty)$ is the 
(disorder dependent) infinite size limit of $\xi(N)$ to be determined by the 
scaling procedure. Details of our implementation of this procedure 
are explained in Appendix~\ref{appA}. 

Concerning $W$ and $N$ we choose values in the range 
$0.75\le W\le 7$ (see caption of Figure~\ref{fig6} for the precise values) 
and $N_{\rm min}=100\le N\le N_{\rm max}$ with $N_{\rm max}=500$ 
(for largest values of $W$) or $N_{\rm max}=2000$ 
(for smallest values of $W$). For $W=0.75$ (and $U=2$, $U_R=1$, $E=0$) 
also one data point with 
$N=3000$ has been computed. For these parameters the scaling procedure 
works actually very well for $\xi_x$ and $\xi_{\rm CM}$ as can be seen 
in the top panels of Figure~\ref{fig6} valid 
for $U=2$, $U_R=1$ and $E=0$, 
and provided we only use data with $N\ge 2L_1$, 
(e.~g. $N\ge 380$ for $W=0.75$) according to the discussion in 
Appendix~\ref{appC} for the validity condition of the scaling approach. 
The scaling curves for these two 
quantities are very nice and we obtain reliable results for $\xi_x(\infty)$ 
and $\xi_{\rm CM}(\infty)$ with relative errors between $6$\% and $8$\% 
for the smallest disorder value $W=0.75$ (see Appendix~\ref{appA} 
for the computation method of these errors). 
We mention that the few data points with $N<2L_1$ (not shown in 
Figure~\ref{fig6}) for the smallest values 
of $W$ and $N$ are clearly below/outside the main scaling curve and 
do not obey one-parameter scaling.

The scaling for the third IPR quantity $\xi_E$ only works approximately 
for still larger values $N\ge 4L_1$ (bottom left panel of 
Figure~\ref{fig6}) but since $\xi_E$ is not defined 
in terms of spatial positions we do not expect the scaling to be perfect. 
However, here the data for larger values of $N$ fall well on the lower 
linear part of the scaling curve where $f(x)\approx x$ for small $x$ 
corresponding to $\xi_E(N)\approx \xi_E(\infty)$. Since 
the scaling procedure optimizes just for this linear region it provides 
therefore correct results for the extrapolated infinite size values 
of $\xi_E$. 

The bottom right panel of Figure~\ref{fig6} shows the dependence of the obtained 
infinite size IPR values on $L_1=105/W^2$ (for $U=2$, $U_R=1$ and $E=0$). 
For the two cases $\xi_x$ and $\xi_{\rm CM}$ 
the power law fit with finite size correction: 
$\xi_{x}=a_1 L_1+b_1L_1^{\gamma_1}$ and 
$\xi_{\rm CM}=a_2L_1+b_2L_1^{\gamma_2}$ 
works very well with both exponents $\gamma_1=\gamma_2=2$ 
within the margin of error (see 
caption of Figure~\ref{fig6} for complete fit results) implying 
the scaling $\xi_{x,{\rm CM}}\sim L_1^2$ for the limit $L_1\to\infty$.
For $\xi_E$ a modified power law fit with a constant term: 
$\xi_E=a_3+b_3L_1^{\gamma_3}$ works very well with $\gamma_3=1/3$ 
(and a negative value of $a_3$) within the margin 
of error implying the scaling $\xi_E\sim L_1^{1/3}$ for 
$L_1\to\infty$. This scaling is clearly below the estimation $\xi_E\sim L_1^2$ 
obtained in \cite{jacquod1,fyodorov1,frahm2} from the simplified 
band matrix model with preferential basis combined with the 
(incorrect) assumption of random and uniform distributions 
for the interaction coupling elements. We will come back to this point 
in Section~\ref{sec6}. 

\begin{figure}
\begin{center}
\includegraphics[width=0.48\textwidth]{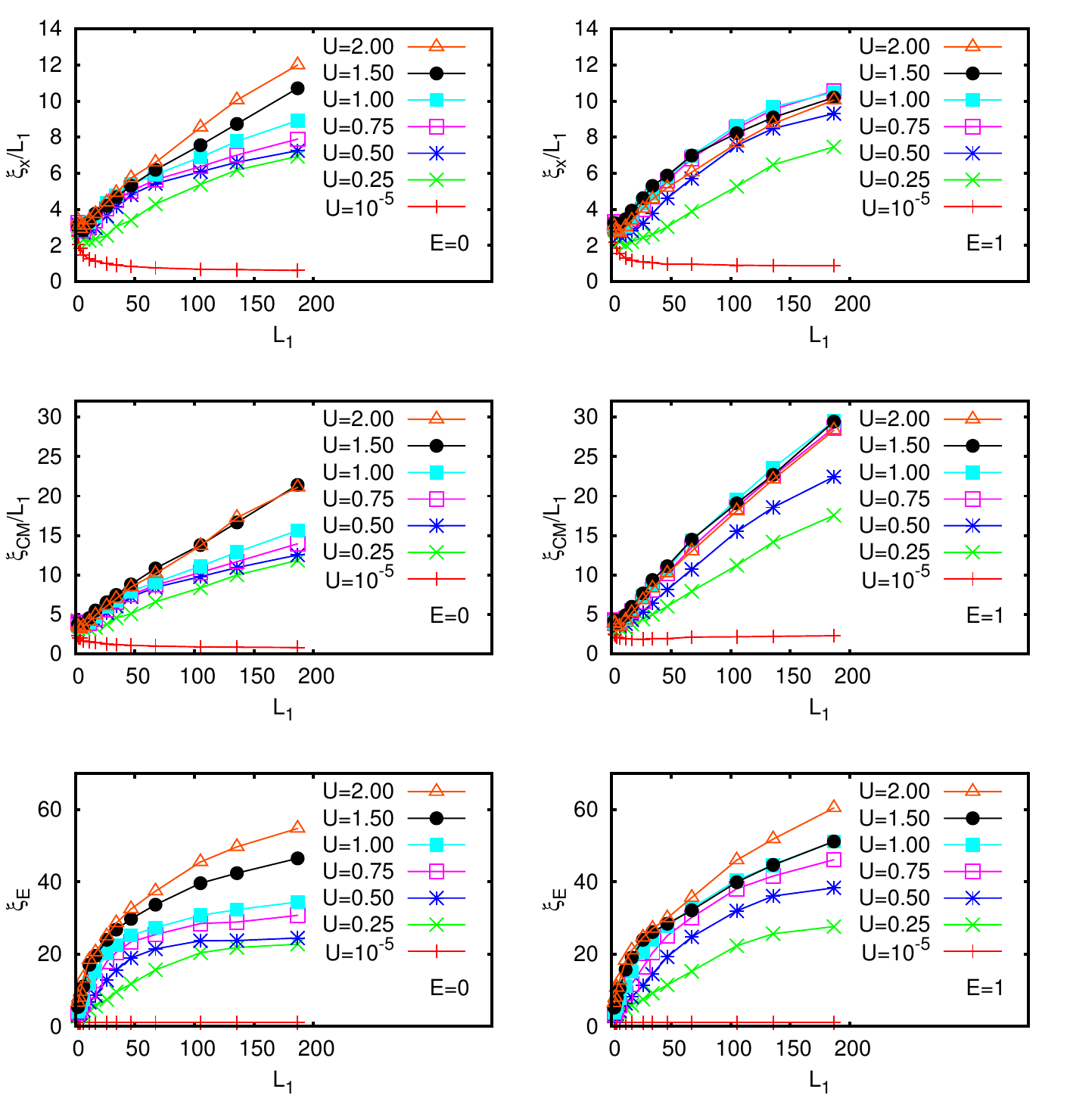}
\caption{Dependence of the three quantities $\xi_x/L_1$ (top panels), 
$\xi_{\rm CM}/L_1$ (center panels) and $\xi_E$ (bottom panels), 
obtained by finite size scaling, on the one-particle localization 
length $L_1=105/W^2$ for the same disorder values used in 
Figure~\ref{fig6}, the interaction values 
$U=10^{-5},\,0.25,\,0.5,\,0.75,\,1,\,1.5,\,2$ for the 
Hubbard interaction and the two energy values $E=0$ (left panels) and 
$E=1$ (right panels). }
\label{fig7}
\end{center}
\end{figure}

These first results are however specific to the case $U=2$, $U_R=1$ and 
$E=0$. Figure~\ref{fig7} shows for $E=0$ and $E=1$ the dependence 
of the enhancement factors $\xi_x/L_1$, $\xi_{\rm CM}/L_1$, and 
of $\xi_E$ (all obtained by finite size scaling) on $L_1$ 
for $U_R=1$ and several values of $0<U\le 2$ including the reference value 
$U=10^{-5}$ for ``vanishing interaction strength''. 
For the energy $E=0$ the IPR values increase with increasing interaction 
strength and the dependence $\xi_{x,{\rm CM}}/L_1\sim L_1+$const. 
only applies to the strongest interaction values 
$U=1.5$ and $U=2$ while for smaller interaction values 
the behavior is sublinear. The behavior of $\xi_E$ is always clearly 
sublinear and for the smallest interaction values one may even observe 
a saturation with increasing $L_1$. For the other energy $E=1$ the situation 
is more complicated. First the dependence of $\xi_{x,{\rm CM}}$ is 
not clearly monotonic for all shown interaction values and a linear behavior 
is only observed for $\xi_{\rm CM}/L_1$ but here for a larger interval 
$0.75\le U\le 2$ of interaction values. Furthermore both $\xi_x/L_1$ 
and $\xi_{\rm CM}/L_1$ seem not to depend strongly on the interaction 
for this interval. For $\xi_E$ the behavior is also sublinear but for 
$U=2$ the exponent $\gamma_3$ (of the power law fit with constant term) 
is close to $1/2$ within the margin of error (instead of $1/3$ for $E=0$). 
The discussion of the particular case $U=10^{-5}\approx 0$ 
(curve closest to the bottom of each panel) is given in Appendix~\ref{appB}. 

\begin{figure}
\begin{center}
\includegraphics[width=0.48\textwidth]{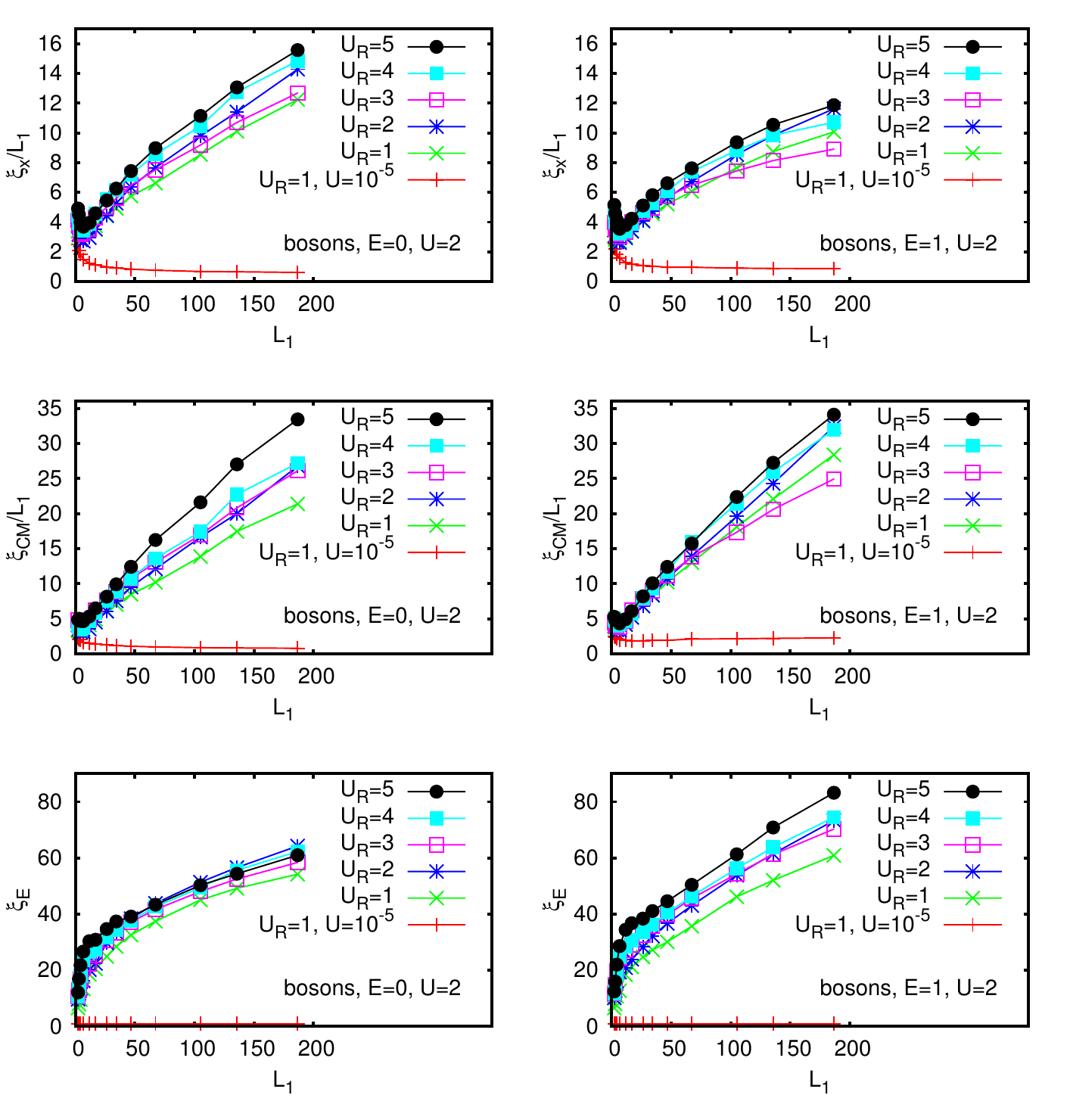}
\caption{Dependence of the three quantities $\xi_x/L_1$ (top panels), 
$\xi_{\rm CM}/L_1$ (center panels) and $\xi_E$ (bottom panels), 
obtained by finite size scaling, on the one-particle localization 
length $L_1=105/W^2$ for the same disorder values used in 
Figure~\ref{fig6}, the interaction value $U=2$ with the boson case and 
interaction ranges $U_R=1,\,2,\,3,\,4,\,5$
and the two energy values $E=0$ (left panels) and 
$E=1$ (right panels). 
For comparison also the data for $U=10^{-5}$ and $U_R=1$ are shown.
}
\label{fig8}
\end{center}
\end{figure}

\begin{figure}
\begin{center}
\includegraphics[width=0.48\textwidth]{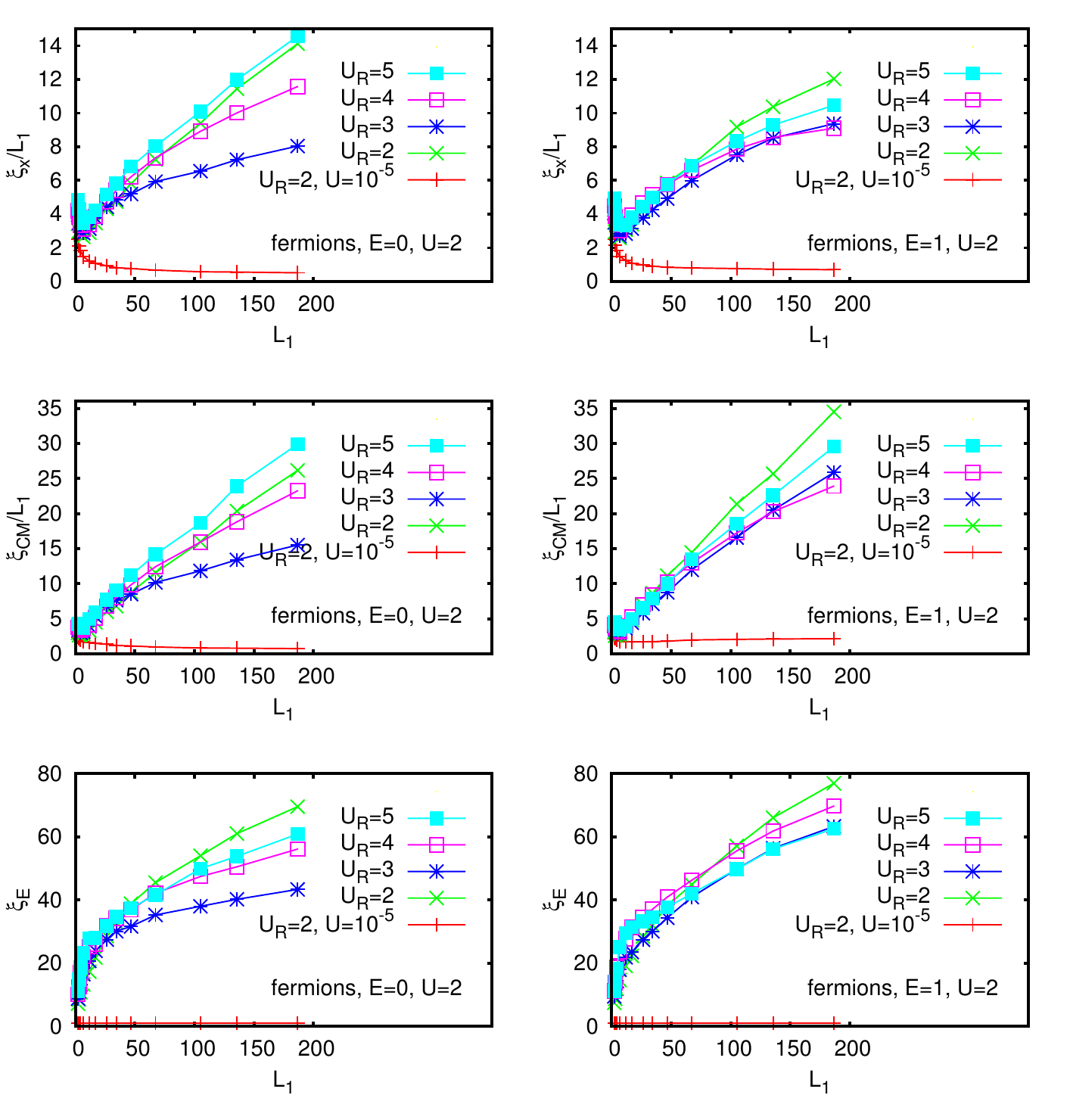}
\caption{Same as Figure~\ref{fig8} but for the fermion case 
with $U_R=2,\,3,\,4,\,5$ and the additional data for $U=10^{-5}$ 
corresponds to $U_R=2$. 
}
\label{fig9}
\end{center}
\end{figure}

We have also studied the case of longer interaction ranges with 
$2\le U_R\le 5$ for a uniform interaction strength $U=2$ 
and both boson and fermion cases with the results shown in 
Figures \ref{fig8} and \ref{fig9}. For comparison both figures also 
show data for the reference value $U=10^{-5}$ with $U_R=1$ (or $U_R=2$) 
for bosons (fermions). 
For bosons the results are rather 
similar to the case $U=2$ of Figure~\ref{fig7} 
with somewhat larger enhancement factors 
$\xi_x/L_1$, $\xi_{\rm CM}/L_1$ for longer interaction ranges 
and a dependence on $L_1$ between sublinear and linear. For $\xi_E$ 
the delocalization effect happens quite abruptly already for quite 
small values of $L_1$ for the largest interaction range $U_R=5$ 
and seems 
to follow a shifted linear dependence. The power law fits with 
constant term, $\xi_E=a_3+b_3\,L_1^{\gamma_3}$, provide for the 
range $20\le L_1\le 200$ and for bosons 
$\gamma_3=0.62\pm 0.05$ ($\gamma_3=0.89\pm 0.05$) for $E=0$ ($E=1$).
For fermions the results are a bit similar to the boson case but with 
the strongest enhancement corresponding either to $U_R=5$ or $U_R=2$. 
Here the same type of fits for $\xi_E$ provide $\gamma_3=0.50\pm 0.07$ 
($\gamma_3=0.70\pm 0.09$) for $E=0$ ($E=1$).

\begin{figure}
\begin{center}
\includegraphics[width=0.48\textwidth]{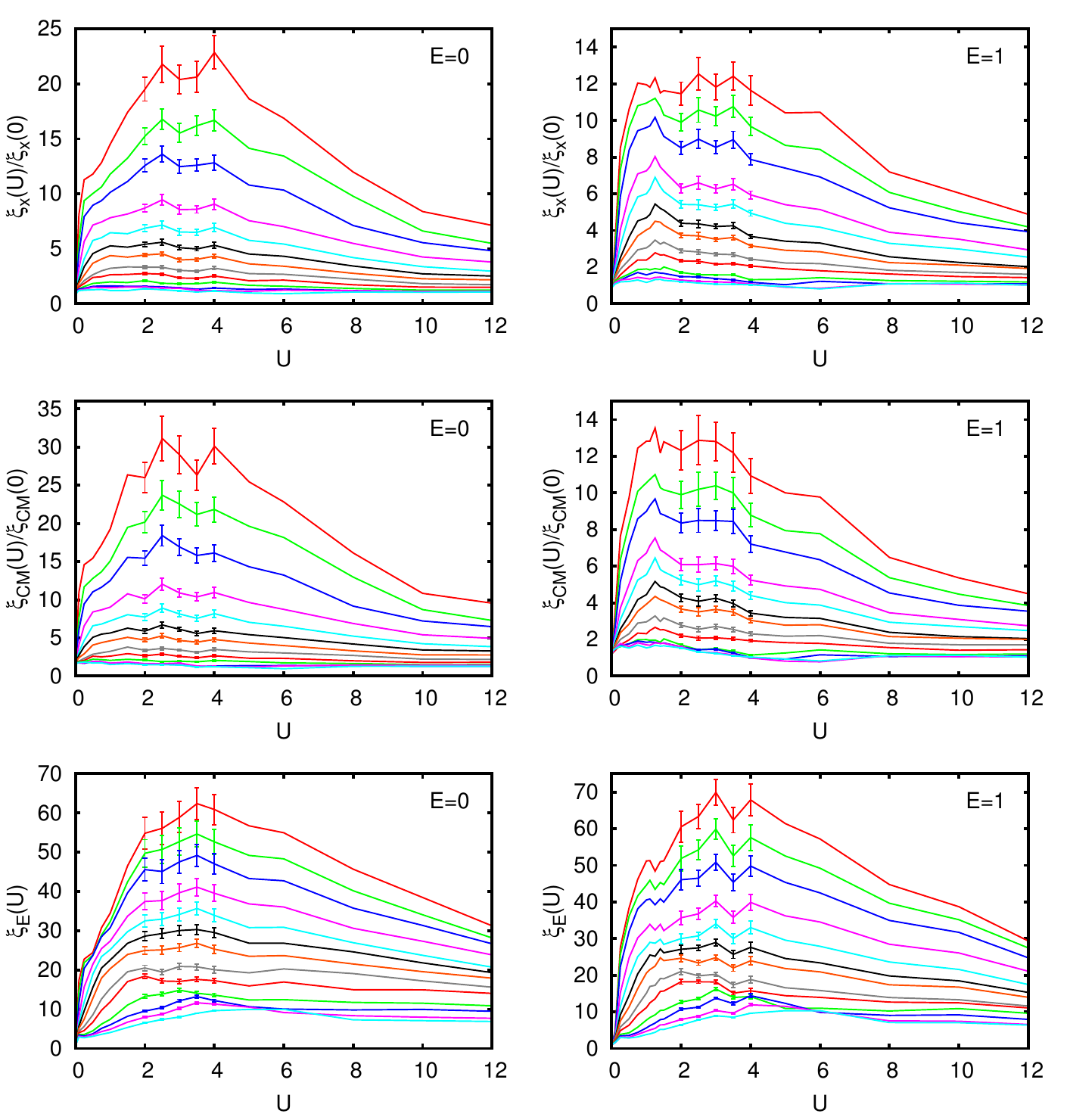}
\caption{Interaction dependence of the three quantities 
$\xi_x(U)/\xi_x(0)$ (top panels), 
$\xi_{\rm CM}(U)/\xi_{\rm CM}(0)$ (center panels) and $\xi_E(U)$ 
(bottom panels) obtained by finite size scaling for $0<U\le 12$, the 
Hubbard interaction and the 
two energy values $E=0$ (left panels) and $E=1$ (right panels). 
The reference values $\xi_x(0)$ and $\xi_{\rm CM}(0)$ have been 
computed using the interaction value $U=10^{-5}\approx 0$. 
Note that the corresponding reference value for $\xi_E$ at $U=10^{-5}$ 
is numerically very close to unity: $|\xi_E(0)-1|<10^{-3}$. 
The different curves correspond to different disorder values 
(same values as in Figure~\ref{fig6}) 
with $W=0.75$ for the top curve and $W=7$ for the bottom curve. 
All panels show error bars for data in the range $2\le U\le 4$. 
}
\label{fig10}
\end{center}
\end{figure}

In Figure~\ref{fig10} the interaction dependence of the 
enhancement factors 
$\xi_x(U)/\xi_x(0)$, $\xi_{\rm CM}(U)/\xi_{\rm CM}(0)$,  and of $\xi_E(U)$ 
for $U_R=1$ and 
several disorder values is shown where $\xi_x(0)$ and $\xi_{\rm CM}(0)$ 
have been computed using the interaction value $U=10^{-5}$ $\approx 0$. 
For $E=0$ the interaction induced enhancement sets already in for $U\ge 0.1$
with enhancement factors between $10$-$15$ and $\xi_E\approx 20$. 
Then there is a region of maximum enhancement $2\le U\le 4$ with 
enhancement factors between $22$-$30$ and $\xi_E\approx 60$. Then 
for $U>4$ the enhancement factors and $\xi_E$ decay (at $U=12$) to values 
between one third and one half of the maximum values at $U=3$-$4$ in 
agreement with the predicted vanishing of the enhancement effect for 
$U\gg 1$ predicted in \cite{ponomarev1}.
For $E=1$ the overall behavior is similar but the effect of a strong 
enhancement already at $U=0.1$ is even stronger and the maximum region 
is extended to $0.5\le U\le 4$. However, the maximum enhancement factors 
are reduced to values $12$-$13$ due to enhanced values of 
$\xi_x(0)$ and $\xi_{\rm CM}(0)$. We remind that, according to 
Figure~\ref{fig7}, the enhancement factor for $E=1$ is comparable or even 
slightly larger as the case $E=0$ when it is measured with respect to $L_1$ 
and not to $\xi_x(0)$ or $\xi_{\rm CM}(0)$. 

We mention that for $U=2$, $U_R=1$ we also computed two set of data points 
at very small disorder $W=0.5$ and $W=0.625$ with 
$100\le N\le N_{\rm max}=2000$. It 
turns out that the scaling for this additional data is very 
problematic and the 
scaling curve for $W=0.5$ does not even overlap (in vertical direction) 
with the previous curves such at best one could try a scaling with an 
extrapolation of the last scaling curve. Due to this we omit these 
data sets and limit ourselves to $W\ge 0.75$ as far the IPR quantities 
are concerned. 

The results of this section clearly that show the interaction induced 
enhancement of the two-particle localization length, measured by $\xi_x/L_1$ 
and $\xi_{\rm CM}/L_1$ using optimal interaction values, behaves as 
$\sim L_1+$const.

\section{Green function localization length}

\label{sec5}

In this section we consider the boson case with the Hubbard interaction 
$U_R=1$ and open boundary conditions (in contrast to the eigenfunction 
calculations of the last two sections with periodic boundary conditions) 
and we study the localization length defined 
by the exponential decay of the projected Green function 
$\bar G(x,y)$ given as $<\!x,x|G|y,y\!>$ between configurations where both 
particles are on the same site $x$ or $y$. 
First, one should note that even though the computational methods for the 
projected Green function used in \cite{song1,leadbeater1} are 
different, less effective than our 
method based on Eq. (\ref{eq_green_project}) (see Section~\ref{sec2}), 
they should provide identical 
results provided that the numerical implementation is stable and sufficiently 
accurate. 

Let us assume that 
we have computed the projected Green function for many different disorder 
realizations of samples of size $N$, for identical other parameters 
($E$, $U$, etc.) and for some values $x$ close to 
one border at $0$ and $y$ being close to the other border at $N-1$. Then we 
define the rather general length scale $L_G$ depending on 
several parameters by
\begin{equation}
\label{eq_L2green_gen}
\frac{1}{L_G(x,y,N,\kappa)}=-\left\langle \frac{1}{y-x}\,
\ln\left(\frac{|\bar G(x,y)|}{|\bar G(x,x)|^\kappa}\right)\right\rangle_W
\end{equation}
where $\langle (\cdots)\rangle_W$ represents the ensemble average with 
respect to different disorder realizations. The parameter $\kappa$ is 
chosen either $1$ or $0$ depending if we want to take into account or not a 
finite size correction by the extra contribution of $|\bar G(x,x)|$ 
in the denominator. Furthermore, let $L^{-1}_{\rm av}(x,N,\kappa)$ be defined 
as the average of (\ref{eq_L2green_gen}) with respect to 10\% of $y$-values 
close to the second border $N-1$, i.~e.: $N-N/10\le y<N$. The hope 
behind this average in $y$ is to reduce short range fluctuations in 
the projected Green function due to the ballistic behavior for small length 
scales and small disorder values. In \cite{frahm5} we used 
the quantity 
\begin{equation}
\label{eq_L2green}
L_2(N)=L_{\rm av}(N/20,N,1)
\end{equation}
using the average for $y$, the position $x=N/20$ and the choice 
with denominator $|\bar G(x,x)|$ to define ``the'' finite size two-particle 
Green function localization length called $L_2$. In \cite{song1} the 
quantity $L_G(0,N-1,N,0)$ was used, i.~e. using the choice $x=0$, 
$y=N-1$ and without the denominator $|\bar G(x,x)|$ while in 
\cite{leadbeater1} apparently the 
quantity $L_G(p,N-1-p,N,0)$ (or similar) 
was used where both positions $x$ and $y$ are taken slightly inside 
the sample (at $x=p$ and $y=N-1-p$ for some suitable small value of $p$) 
to reduce possible boundary effects. 

In the limit of samples in the strongly localized regime, with $N$ 
being much larger than the two-particle localization length, and assuming 
only small particular boundary effects (a problematic assumption as we will 
see) one would expect that $L_G$ and 
$L_{\rm av}$ provide identical localization lengths for reasonable parameter 
choices for the two positions $x$, $y$ and the parameter $\kappa=0$ or $1$. 
However, in realistic situations, when trying to compute the infinite size 
localization length by finite size scaling and for small disorder values, 
the size $N$ is comparable or even quite smaller than 
the two-particle localization length. In this regime the precise choice of 
parameters $x$, $y$ and $\kappa$ may indeed have an important impact on the 
results. 

To test the effect of this we have therefore simultaneously computed eight
quantities $L_G(0,N-1,N,\kappa)$ (both 
positions at the boundary), $L_G(N/20,N-1-N/20,N,\kappa)$ (both positions 
5\% inside the boundary), $L_{\rm av}(0,N,\kappa)$ ($x$ at the boundary and 
10\% average for $y$ at the other boundary), and 
$L_{\rm av}(N/20,N,\kappa)$ ($x$ 5\% inside the boundary and 
10\% average for $y$ at the other boundary),
for both values $\kappa=0$ and $\kappa=1$, several interaction values, 
and $E=0$ or $E=1$. 

Then we have applied finite size scaling, using the automatic procedure 
described in appendix \ref{appA}, to the raw 
data to determine the associated infinite size localization lengths for 
each quantity. For this we used 15 disorder values in the 
range $0.5\le W\le 7$ (see caption of Figure~\ref{fig11} for precise values) 
and system sizes in the range $N_{\rm min}=50\le N\le N_{\rm max}$ 
with $N_{\rm max}=215$ (for largest values of $W$) and 
$N_{\rm max}=2000$ (for smallest values of $W$). The density of 
$N$-values corresponds to an approximate factor of $1.2$ between two neighbor 
values of $N$. For the scaling procedure we also limited ourselves 
to data points with $N\ge 1.5L_1$ since according to the discussion 
of Appendix~\ref{appC} $N$ must be larger than $L_1$ for the validity of the 
one-parameter scaling hypothesis. The average over different disorder 
realization has been performed up to a 
precision of 1\% or better for six interaction values 
$U\in\{0.25,\,0.5,\,0.75,\,1,\,1.5,\,2\}$ which requires 20 samples 
for $W=7$ at $N=215$, $E=0$ (minimum number) and 
\mbox{$\approx 1.3\times 10^5$}
samples for $W=0.5$, $N=50$, $E=1$ (maximum number). 

For $E=0$ we find that for all eight cases the scaling procedure works 
very well 
with well defined scaling curves. The 4 cases with one or two positions 
exactly at the boundary produce (for smallest values of $W$) rather 
considerable variations of the infinite size localization length while the 
values for the 4 cases with one or two positions 5\% inside the boundary 
are somewhat smaller but also closer together. Furthermore, the cases 
$\kappa=1$ [i.e. ``with'' the denominator $|\bar G(x,y)|$ in 
(\ref{eq_L2green_gen})] produce at same system size $N$ larger values 
as the cases with $\kappa=0$ which is not a problem as such 
if after finite size scaling the results are coherent. However, due to this 
for $\kappa=1$ the scaling curves for small $W$ are a bit lower (not in the 
flat regime of the scaling curve) with stronger slopes such the 
scaling is more reliable. 
For $E=1$ the situation is somewhat similar but here the two cases 
without average for the $y$-position and with both positions at the 
boundary do not scale correctly and the individual curves cannot be matched 
to one scaling function (for $U=2$). 
The other six cases with either average or 
positions 5\% inside the boundary produce rather nice scaling curves but 
here the condition $N\ge 1.5L_1$ is indeed important, actually somewhat 
more important than for the case $E=0$.
Furthermore, for $U=0$ we have the impression that 
the cases with average produce 
a smaller variation for the dependence on $L_1$ which seems more 
reasonable to us. Therefore, in summary we choose for this work 
(and except the particular cases studied in Appendix~\ref{appC}) 
the case with average, with $\kappa=1$ and the $x$-position 5\% inside the 
boundary, i.~e. we stick to our initial choice \cite{frahm5} with 
$L_2$ given by (\ref{eq_L2green}). 

\begin{figure}
\begin{center}
\includegraphics[width=0.48\textwidth]{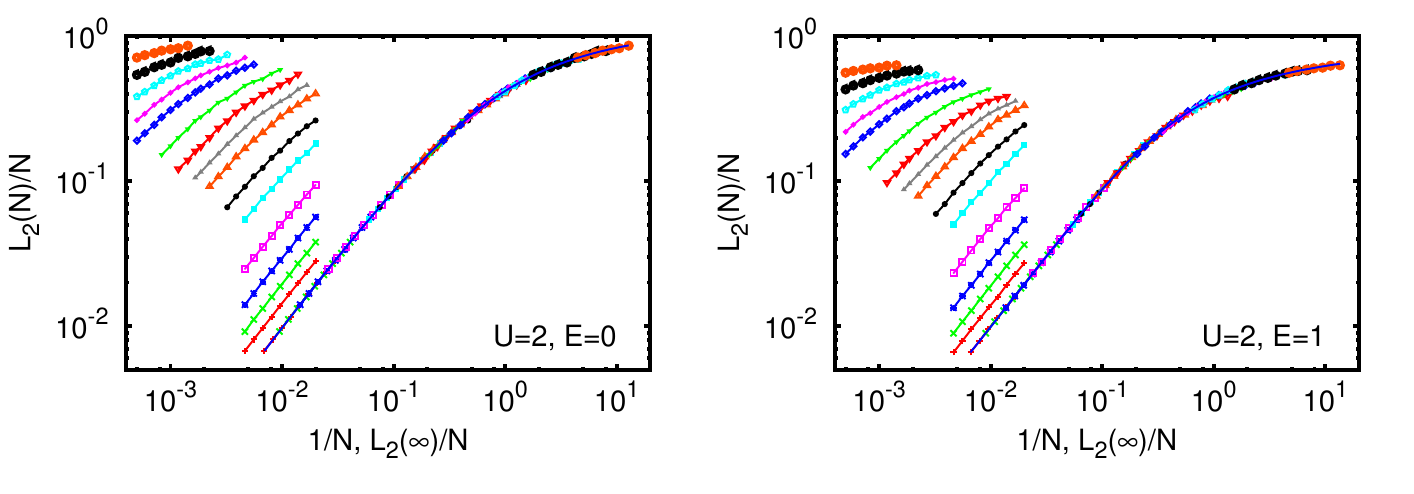}
\caption{Illustration of the scaling procedure to determine the 
Green function two-particle localization length $L_2$ at infinite system size 
for $U=2$ and energy values $E=0$ (left panel) 
and $E=1$ (right panel). 
Both panels show (in a double logarithmic scale) $L_2(N)/N$ versus $1/N$ 
(left set of curves) and versus $L_2(\infty)/N$ (right set of curves) 
with $L_2(\infty)$ determined such that the data points are closest to a 
single scaling curve (approximately) 
shown by the blue curve corresponding to the fit (\ref{eq_fitscaling}). 
$L_2(N)$ has been computed from (\ref{eq_L2green}) as explained in 
the text. The used disorder values are 
$W=0.5,\,0.625,\,0.75,\,0.875,\, 1,\, 1.25,\, 1.5,\, 1.75,\, 2,
\, 2.5,\, 3,\, 4,\, 5,\, 6,\, 7$ with $W=0.5$ for the top curve and 
$W=7$ for the bottom curve. The used values for the system size $N$ are 
between $N_{\rm min}=50$ and $N_{\rm max}=215$ 
(for largest values of $W$) and $N_{\rm max}=2000$ 
(for smallest values of $W$). 
Only data points with $N\ge 1.5\,L_1$ have been used for the 
scaling procedure and are shown in the figure. }
\label{fig11}
\end{center}
\end{figure}

Figure~\ref{fig11} illustrates the scaling procedure for this quantity 
and the case $U=2$ and both energies $E=0$ and $E=1$. The quality of the two 
scaling curves is very impressive and appears even better than the quality 
of the scaling curves of the IPR quantities shown in Figure~\ref{fig6}. 
The two-particle localization lengths $L_2$ for infinite system size 
obtained from this coincide (for the case 
$E=0$) within the margin of error with our previous results 
\cite{frahm5} for disorder values $1\le W\le 7$ and 
obtained by finite size extrapolation using data with $100\le N\le 1400$. 
However, our results deviate considerably from those 
of \cite{song1,leadbeater1} which we attribute to the limited system sizes 
$N\lesssim 250$ used in these two works not respecting the condition $N>L_1$ 
of the one-parameter scaling approach \cite{kramer1} for smaller 
disorder values. A detailed analysis of this point by simulating different 
scaling scenarios for limited system size and other parameters used 
in (\ref{eq_L2green_gen}) is given in Appendix~\ref{appC}. In this appendix 
also discrepancies between \cite{song1} and \cite{leadbeater1} are 
explained by another scaling related problem. 

\begin{figure}
\begin{center}
\includegraphics[width=0.48\textwidth]{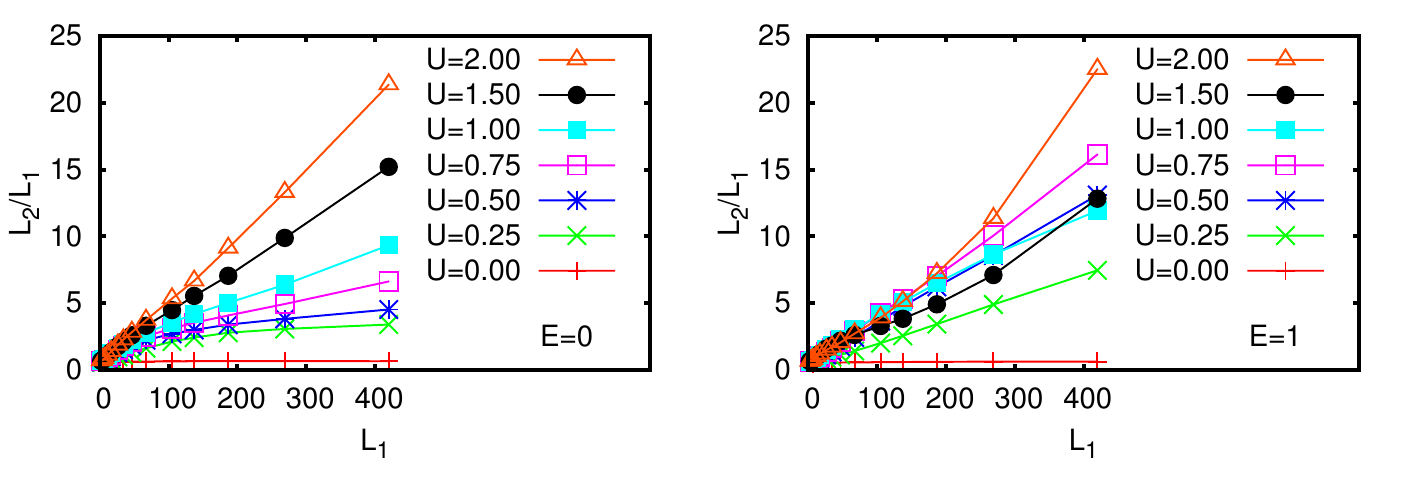}
\caption{
Enhancement factor 
$L_2/L_1$ versus $L_1=105/W^2$ for 
$U=0,\,0.25,\,0.5,\,0.75,\,1,\,1.5,\,2$ and 
energy $E=0$ ($E=1$) for left (right) panel. 
The used disorder values $W$ are as in Figure~\ref{fig11}.
}
\label{fig12}
\end{center}
\end{figure}

In the following (and except Appendix \ref{appC}) 
the quantity $L_2$ always denotes the 
infinite size localization length obtained by finite size scaling 
from $L_2(N)$ defined in (\ref{eq_L2green}). 
Figure~\ref{fig12} shows the dependence of the 
enhancement factor $L_2/L_1$ on $L_1$ for certain selected interaction values 
in a similar way as in Figure~\ref{fig7}. For $E=0$ we see a linear behavior 
for larger interaction values and sublinear form for smaller interaction 
strengths and the overall dependence on $U$ is clearly monotonic.
We mention that the approximate formula 
$L_2\approx L_1/2+c(U)L_1^2$ with $c(U)\approx 0.074\,|U|/(1+|U|)$ 
suggested in \cite{frahm5} works rather well for disorder values 
$W>1$ and $|U|\le 2$ 
corresponding to the available data of \cite{frahm5}. 
However for smaller disorder values, there are significant deviations 
due to the cases of sublinear behavior. 
For $E=1$ the situation is more complicated with even stronger than linear 
behavior for certain interaction values and the $U$-dependence is not 
monotonic. In particular the enhancement factor is quite reduced for 
$U=1$ and $U=1.5$ if compared to $U=2$ and $U=0.75$. This strange 
behavior will be better clarified below in the discussion of 
Figure~\ref{fig15}. 
As for the IPR quantities 
the discussion of the particular case $U=0$ 
(curve closest to the bottom of each panel) is given in Appendix~\ref{appB}. 

\begin{figure}
\begin{center}
\includegraphics[width=0.48\textwidth]{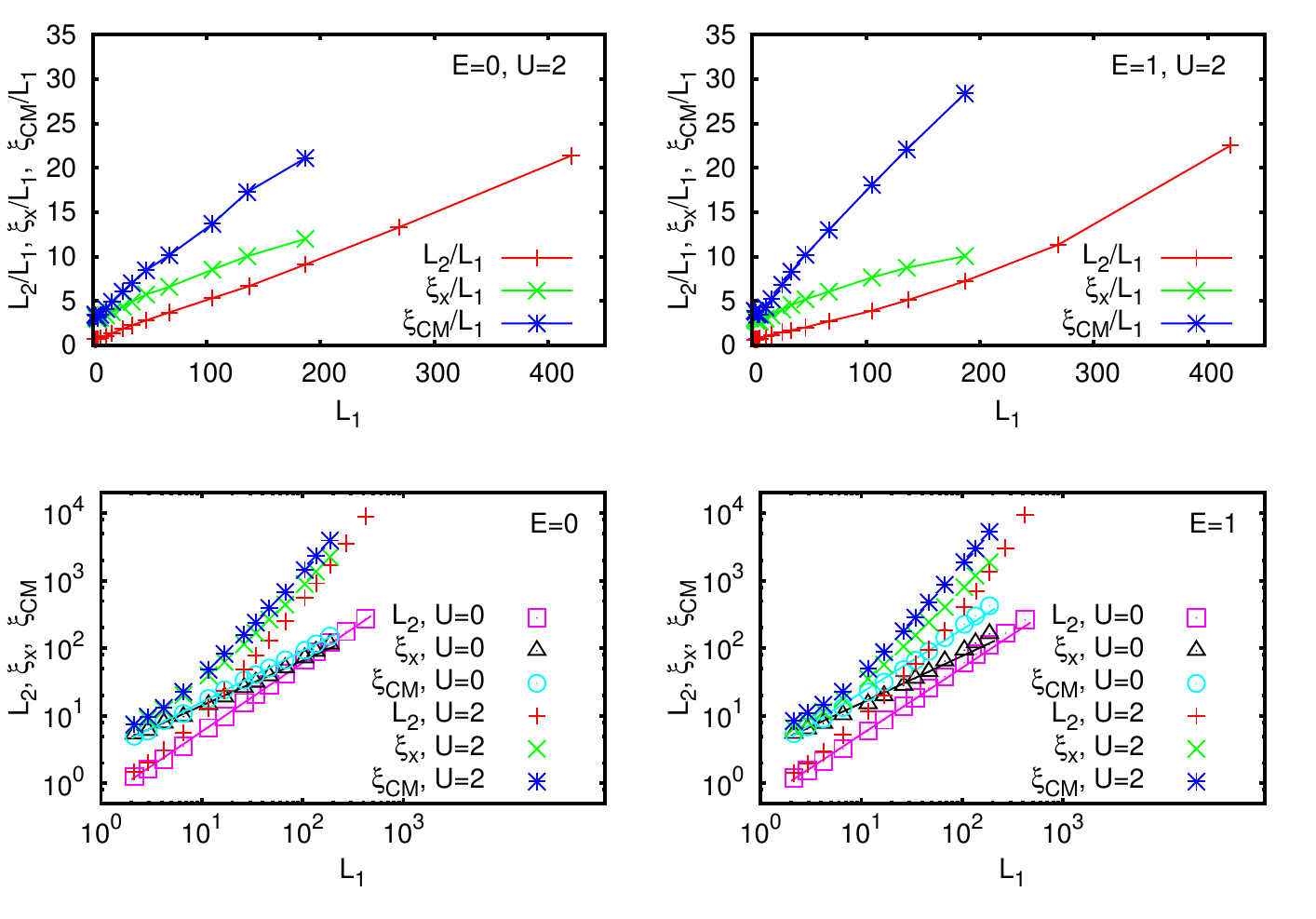}
\caption{{\em Top panels:} 
Comparison of the dependence of $L_2/L_1$, $\xi_x/L_1$ and $\xi_{\rm CM}/L_1$
on $L_1$ for $U=2$. 
{\em Bottom panels:} 
$L_2$, $\xi_x$ and $\xi_{\rm CM}$
versus $L_1$ in double logarithmic scale for $U=0$ and $U=2$. 
The full lines for the data with $U=0$ show the power law fits 
$C\,L_1^\gamma$ with the fit results:
for $L_2$, $E=0$: $C=0.56\pm 0.01$, $\gamma=1.02\pm 0.01$, 
for $\xi_x$, $E=0$: $C=3.05\pm 0.10$, $\gamma=0.67\pm 0.01$, 
for $\xi_{\rm CM}$, $E=0$: $C=2.76\pm 0.04$, $\gamma=0.76\pm 0.01$, 
for $L_2$, $E=1$: $C=0.55\pm 0.01$, $\gamma=0.99\pm 0.02$, 
for $\xi_x$, $E=1$: $C=3.11\pm 0.11$, $\gamma=0.70\pm 0.02$, 
for $\xi_{\rm CM}$, $E=1$: $C=2.53\pm 0.11$, $\gamma=0.94\pm 0.02$, 
{\em All panels:} The used disorder values $W$ are as in 
Figure~\ref{fig6} ($\xi_x$ and $\xi_{\rm CM}$) or in Figure~\ref{fig11} 
($L_2$).
The energy is either $E=0$ (left panels) or $E=1$ (right panels).
For $\xi_x$ and $\xi_{\rm CM}$ the reference values for vanishing 
interaction in bottom panels were computed using the value $U=10^{-5}$.
}
\label{fig13}
\end{center}
\end{figure}

The top panels of Figure~\ref{fig13} compare the dependence 
of $L_2/L_1$, $\xi_x/L_1$ and $\xi_{\rm CM}/L_1$
on $L_1$ for $U=2$. We see a linear or slightly 
stronger than linear behavior (for $E=1$ and $L_2/L_1$) with 
a slope for $\xi_{\rm CM}/L_1$ being larger than for the other cases, 
roughly by factor $\approx 2$ for $E=0$ and a factor $\approx 3$ for $E=1$. 
We attribute this difference between the two energies to fact that 
for $E=1$ the contributing non-interacting pair eigenstates to a 
full two-particle eigenstates are more likely to have two very different 
one-particle localization lengths and for the center of mass IPR it is the 
larger of the two who dominates (contrary to $\xi_x$ where the smaller of 
the two dominates; see also the discussion in Appendix~\ref{appB}). 
The slopes for $\xi_x/L_1$ and $L_2/L_1$ 
are comparable but there is rather constant shift between these quantities 
with $\xi_x/L_1\approx L_2/L_1+3$ which can be understood by the fact that 
the IPR $\xi_x$ measures the localization length in the main maximal 
part of an eigenstate while $L_2$ measures the exponential decay length 
of eigenstates far away from the maximal part. The eigenfunction 
structure close to the main part is indeed very complicated with strong 
fluctuations enhancing somewhat $\xi_x$ (see 
Figure~\ref{fig1} and corresponding discussion).

The bottom row of panels of Figure~\ref{fig13} show the dependence of 
$L_2$, $\xi_x$ and $\xi_{\rm CM}$ on $L_1$ for $U=0$ and $U=2$ in 
a double logarithmic scale confirming the above observations. 
The case of vanishing interaction $U=0$, including the results 
of the power law fits for this case shown in Figure~\ref{fig13}, 
is discussed in Appendix~\ref{appB}. 

\begin{figure}
\begin{center}
\includegraphics[width=0.48\textwidth]{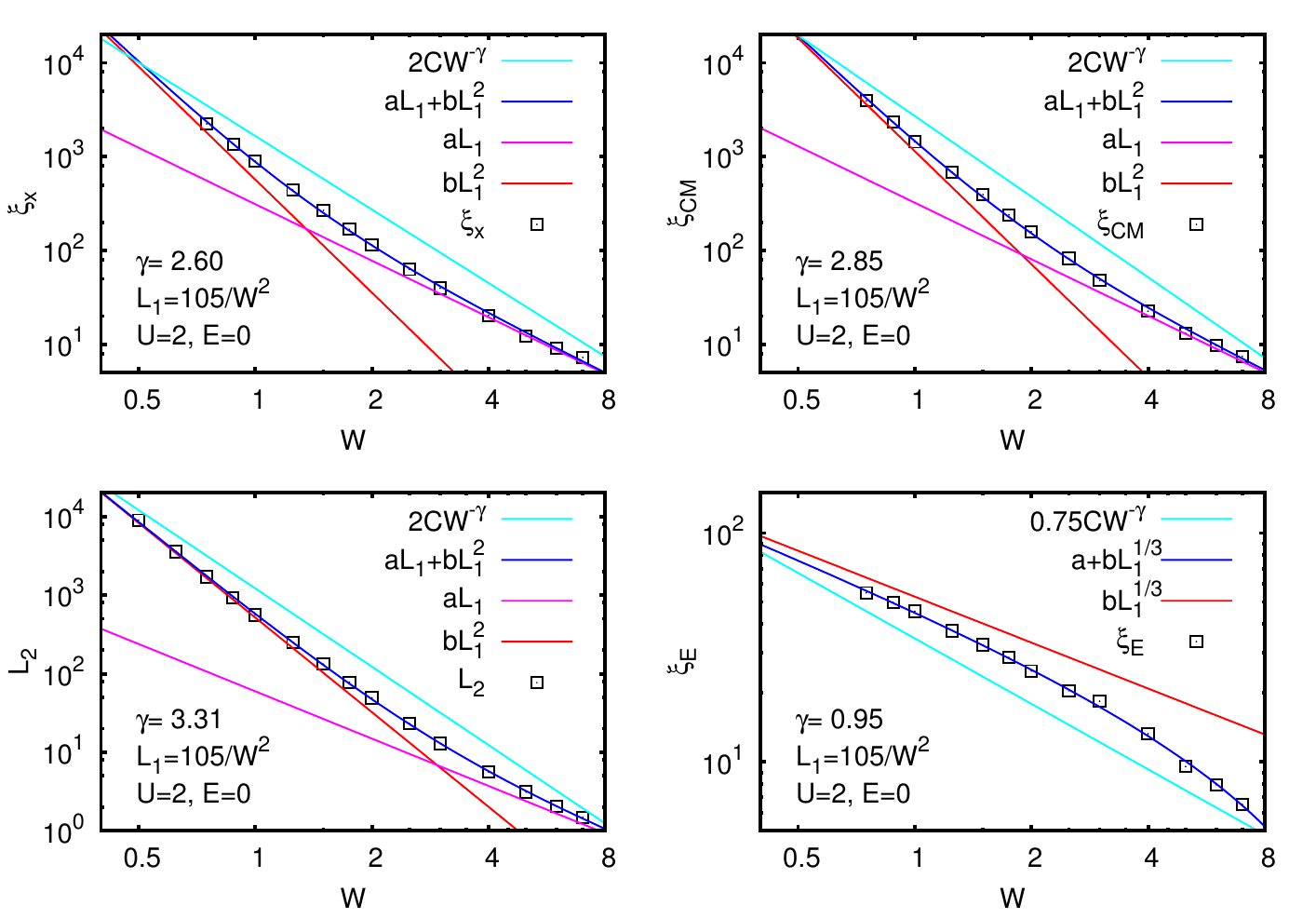}
\caption{Dependence of the four quantities 
$\xi_x$, $\xi_{\rm CM}$, $\xi_E$ and $L_2$ on disorder $W$ 
in a double logarithmic scale for $U_R=1$, $U=2$ and 
$E=0$ (black squares). 
Furthermore for the cases $\xi_x$, $\xi_{\rm CM}$ and $L_2$ 
also the fit $aL_1+bL_1^2$ with $L_1=105/W^2$ (blue curve), 
the asymptotic formulas $aL_1$ (pink line), $bL_1^2$ (red line) 
and the power law fit $CW^{-\gamma}$ (light blue line, shifted by 
a factor of 2) are shown. 
For the case $\xi_E$ also the fit $a+bL_1^{1/3}$ (blue curve), 
the asymptotic formula $bL_1^{1/3}$ (red line) 
and the power law fit $CW^{-\gamma}$ (light blue line, shifted by 
a factor of 0.75) are shown. 
The fit results are 
for $\xi_E$~: 
$a=-8.0\pm 0.3$, $b=11.2\pm 0.2$, $C=46.2\pm 1.4$, $\gamma=0.95\pm 0.03$, 
for $\xi_x$~: 
$a=2.97\pm 0.08$, $b=0.052\pm 0.004$, $C=830\pm 60$, $\gamma=2.60\pm 0.07$, 
for $\xi_{\rm CM}$~: 
$a=3.08\pm 0.07$, $b=0.104\pm 0.006$, $C=1360\pm 110$, $\gamma=2.85\pm 0.07$, 
and for $L_2$~: 
$a=0.569\pm 0.008$, $b=0.047\pm 0.001$, $C=610\pm 50$, $\gamma=3.31\pm 0.07$. 
}
\label{fig14}
\end{center}
\end{figure}

In previous numerical works (e.~g. \cite{song1,leadbeater1}) but 
also more recently in \cite{flach2011}, a lot of effort 
was devoted to characterize the enhancement effect (or ``absence'' of it) 
by a simple power law fit $L_2=CW^{-\gamma}$ which typically provides 
some exponent $\gamma$ somewhat larger than $2$ (behavior for absence of 
interaction) but still clearly below $4$ (behavior expected if 
$L_2\sim L_1^2$). As already discussed in \cite{frahm5} 
one must be very careful with such a fit which is not really justified if 
there are finite size corrections corresponding to a different 
behavior such as 
$L_2=aL_1+bL_1^2$ resulting actually in $L_2\sim W^{-4}$ when 
taking the formal limit $W\to 0$ or $L_1=105/W^2\to \infty$. However, 
in numerical computations such a limit may be difficult to access, 
especially if the constant $b$ is rather small as compared to $a$, and in 
order to distinguish between the two scenarios one must carefully analyze the 
dependence of $L_2$ on $W$, especially the curvature in double logarithmic 
scale. 

In Figure~\ref{fig14} we show for $U=2$ and $E=0$ (cases with 
linear behavior of the enhancement factor in Figures~\ref{fig7} and 
\ref{fig12}) the dependence of the four quantities 
$\xi_x$, $\xi_{\rm CM}$, $L_2$, and $\xi_E$ on disorder $W$ 
in a double logarithmic scale. The case of $\xi_E$ is somewhat particular. 
For the other three quantities we compare the simple power law fit 
$CW^{-\gamma}$ with the square polynomial fit $aL_1+bL_1^2$ and show for 
the latter also the asymptotic limits $aL_1$ and $bL_1^2$ 
for small or large values of $L_1$. 
For all three cases there is a clear and significant non-vanishing curvature 
and the square polynomial fit works very well with $a\approx 3$ for 
$\xi_x$ and $\xi_{\rm CM}$ and $a\approx 0.6$ for $L_2$ while 
$b\approx 0.05$ for $\xi_x$ and $L_2$ and $b\approx 0.1$ for $\xi_{\rm CM}$ 
also confirming the observations of Figure~\ref{fig13} 
(see caption of Figure~\ref{fig14} for precise fit results). 

The overall power law fit for these cases provide exponents 
$\gamma=2.61$ for $\xi_x$, $\gamma=2.86$ for $\xi_{\rm CM}$ and 
$\gamma=3.31$ for $L_2$. At first sight these fits appear indeed rather 
close to the data points (in double logarithmic scale) but the deviations are 
systematic and not random. Furthermore, when the lines obtained by power 
law fits are slightly shifted up one sees very clearly that the deviations 
are due to the non-vanishing curvature of the data. 
However, if the interval of available data values of $W$ is reduced (e.~g. 
for $2\le W\le 5$) or if the data for small $W$ are simply invalid 
(for example when using finite size scaling for too small system sizes in 
the raw data) one may get the wrong impression that the simple power 
law appears justified. 

For $\xi_E$ the dependence on $W$ is quite different. Motivated by the fit 
result of Figure~\ref{fig6} with $\gamma_3=0.35$ we use here the 
fit $\xi_E=a+bL_1^{1/3}$ which gives $a=-7.9$ and $b=11.2$ and is very 
accurate while the power law fit $\xi_E\propto W^{-\gamma}$ gives the 
exponent $\gamma=0.94$ corresponding $\xi_E\sim L_1^{0.47}$ and 
shows quite significant deviations. 
In all four cases one sees that the slope of the simple power law fit 
is quite different from the slope of the asymptotic behavior 
for large $L_1$ and that the former is not sufficiently accurate 
for the full interval of considered disorder values. 

Even tough the square polynomial fit in $L_1$ (for the first three 
quantities) does not apply to all interaction values according 
to Figures~\ref{fig7} and \ref{fig12} the analysis shown in 
Figure~\ref{fig14} illustrates clearly the problems and limitations 
associated to the simple power law fit for the disorder dependence of 
the different types of two-particle localization lengths. 

\begin{figure}
\begin{center}
\includegraphics[width=0.48\textwidth]{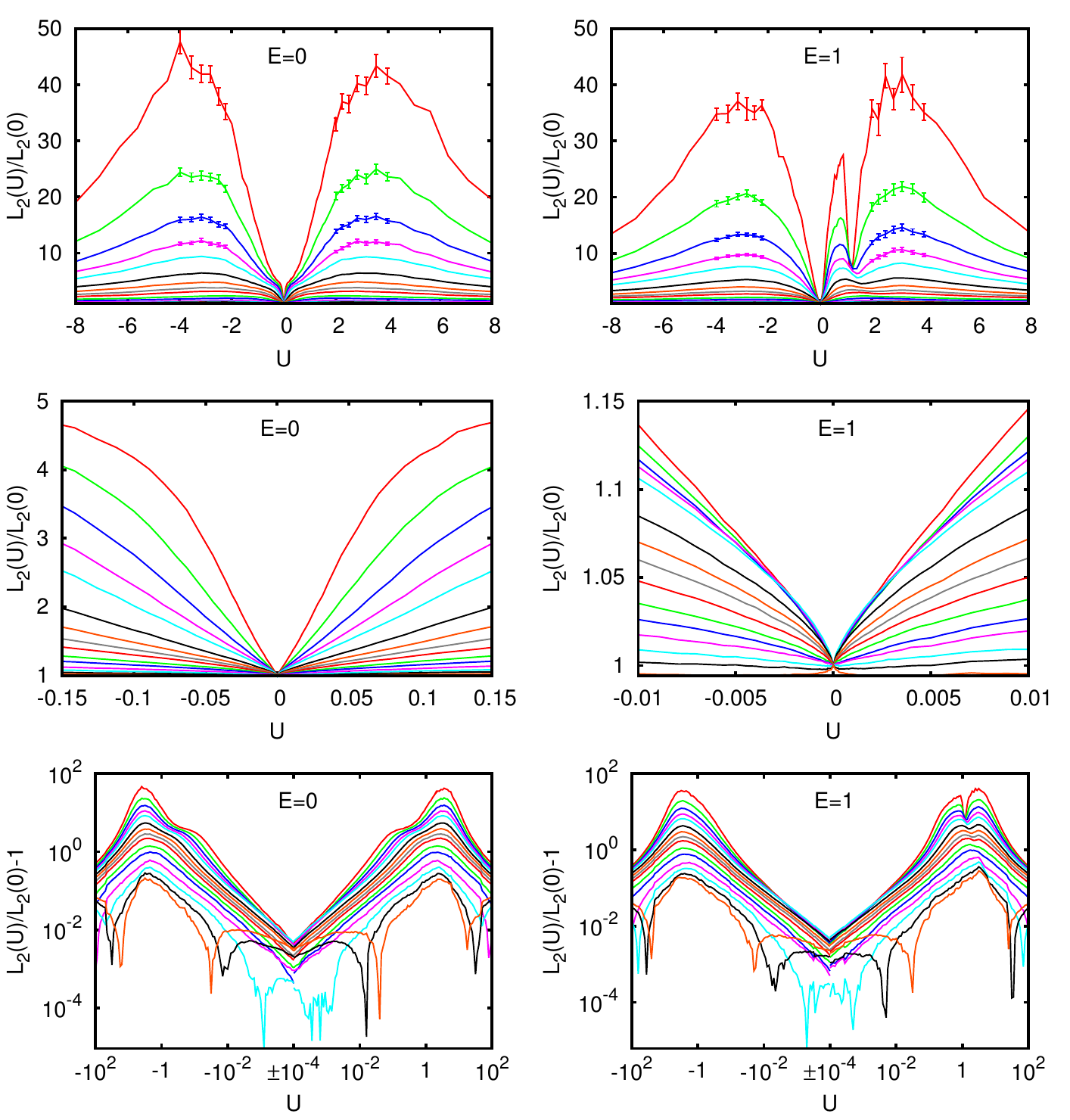}
\caption{Dependence of $L_2(U)/L_2(0)$ on $U$ 
for $E=0$ (left panels) and $E=1$ (right panels) 
and the disorder values $W$ used in Figure~\ref{fig11} with 
$W=0.5$ for the top red curves and increasing values of $W$ corresponding 
to decreasing curves. Here $L_2(U)$ represents the infinite size localization 
length obtained by finite size scaling for the interaction strength $U$. 
Top panels show the curves on normal scale 
for $-8\le U\le 8$ with error bars for data in the range 
$2\le |U|\le 4$ and top four curves. Error bars for other curves are 
below line size. 
Center panels show the zoomed range $-0.15\le U\le 0.15$ 
($-0.01\le U\le 0.01$) for $E=0$ ($E=1$) on normal scale. 
Bottom panels show $L_2(U)/L_2(0)-1$ in logarithmic 
scale versus $U$ for $10^{-4}\le |U|\le 10^2$ in a logarithmic scale 
with sign corresponding to $U\le -10^{-4}$ for the left half and $U\ge 10^{-4}$
for the right half of the $U$-axis. 
}
\label{fig15}
\end{center}
\end{figure}

We also studied the dependence of the infinite size localization length 
$L_2(U)$ on the interaction strength $U$ (with $L_2(U)$ 
obtained by finite size scaling 
and not to be confused with the finite size quantity $L_2(N)$ used in 
Figures~\ref{fig11} and \ref{fig18}). Exploiting the optimization of 
the Green function expression (\ref{eq_green_project3}) we computed 
simultaneously $L_2(U)$ with nearly no additional effort for a 
large number of interaction values which are the 7 
reference values used in Figure~\ref{fig12}, 
121 values in the range $10^{-4}\le U\le 10^2$ and further 
121 values in the range $-10^{2}\le U\le -10^{-4}$ giving 249 different 
values for $U$. The latter two groups are uniformly distributed in 
logarithmic scale for $|U|$, i.~e. with a constant factor 
$10^{1/20}\approx 1.122$ between two neighbor values of $|U|$. 
According to the discussion at the end of Section~\ref{sec2}, we used 
the optimized expression (\ref{eq_green_project3}) for the most 
difficult cases of smallest disorder and $N\le 20L_1$ while for 
some less difficult cases for large disorder and $N>20L_1$ we directly 
used the more expensive matrix inversion in (\ref{eq_green_project}) 
for reasons of numerical stability. 

The dependence of $L_2(U)/L_2(0)$ on $U$ for the two energies $E=0$ and 
$E=1$ is shown in Figure~\ref{fig15}. The top panels 
show this quantity in normal scale for $-8\le U\le 8$ and with 
error bars for the top 4 curves (for smallest values of $W$) 
in the ranges $2\le |U|\le 4$ close to the maxima. Since the behavior 
for very small interaction values appears to be very 
particular, we also show (in center panels) the zoomed region 
$|U|\le 0.15$ ($|U|\le 0.01$) for $E=0$ ($E=1$). In bottom panels the quantity 
$L_2(U)/L_2(0)-1$ is shown in logarithmic scale versus a 
logarithmic scale with sign for $U$, i.~e. the two regions 
of positive and negative values of $U$ are both presented in logarithmic 
scale of $|U|$ and they are joined together at $U=\pm 10^{-4}$. 

For $E=0$ the first observation is that the dependence of $L_2(U)/L_2(0)$ 
on $U$ is an even function in average but that there are small statistical 
fluctuations within the margin of statistical error that do not respect this 
symmetry. This behavior is easily understood theoretically since a change 
of sign of $U$ can be taken into account by replacing the disorder potential 
according to $V(x)\to -V(x)$, which corresponds to a {\em different} 
statistical sample, and by the transformation 
$\psi(x_1,x_2)\to (-1)^{x_1+x_2}\psi(x_1,x_2)$ which accounts for the 
change of sign of the hopping matrix element in (\ref{ham_tot}). 
Furthermore we observe roughly a linear behavior $\sim |U|$ for small 
$|U|$ and a decay $\sim |U|^{-1}$ for large $|U|$, which is also somehow 
suggested by the analytic form of the projected Green function 
(\ref{eq_green_project}) in terms of $U$. We therefore confirm our above 
observation of Figure~\ref{fig10} that the enhancement effect indeed vanishes 
for $U\gg 1$ not only for $\xi_x$, $\xi_{\rm CM}$ and $\xi_E$ but 
also for $L_2$ in agreement with the theoretical predictions 
of \cite{ponomarev1}. The curves are maximal 
in the region $3\le |U|\le 4$, at least for the smallest disorder values 
(top curves) where the maxima are rather clearly visible. 
There is a tendency that the 
maximum positions are slightly moving closer to 0 with increasing disorder.
The curves increase from $U=0$ to $|U|\approx 0.1$ quite abruptly 
with values up to $4$-$5$ for the two smallest disorder 
values. Actually, the double logarithmic scale of bottom panels shows that 
there are two different linear regimes for small and medium values 
with two different slopes. For example for $W=0.5$ the fit 
$L_2(U)/L_2(0)=a+b|U|$ in the range $|U|\le 0.1$ provides 
$a=0.999\pm 0.003$ and $b=35.7\pm 0.2$ while for larger values 
$0.8\le|U|\le 2$ it gives $a=-3.9\pm 0.4$ and $b=18.1\pm 0.3$ 
corresponding to roughly a factor of two between the two slopes. 

For the other energy $E=1$ one does not expect a symmetry between positive 
and negative values of $U$ and indeed for positive $U$ there is 
for small disorder values (top curves) a well pronounced local minimum 
close to $U\approx 1.2$ which is completely absent for negative values of 
$U$. Motivated by this finding we have also computed a few additional 
data points for $1<U<1.5$ and $E=1$ for the IPR quantities. These 
data points were included in Figure~\ref{fig10} where one can see a slight 
reduction for a similar $U$ value and smallest disorder 
but this reduction is also of the order of statistical 
fluctuations. The three IPR quantities do not show the clear local 
minimum 
as the Green function localization length $L_2$ but the region of maximum 
values in Figure~\ref{fig10} for $E=1$ is quite large which is coherent with 
a scenario that the minimum is somehow smoothed out for the IPR quantities. 
Apart from this the overall form of the curves in Figure~\ref{fig10}
 for $E=1$, 
with an enlarged $U$-regime for high values, is coherent 
with Figure~\ref{fig15}. Furthermore in Figure~\ref{fig15} for $E=1$ 
the behavior $\sim |U|^{-1}$ for large $|U|$ appears to be similar as 
to $E=0$. 
For the region $|U|\le 0.01$ there is for $E=1$ 
a slight sublinear behavior and 
the power law fit $L_2(U)/L_2(0)-1=C|U|^\gamma$ provides for 
$0<U\le 0.01$ and $W=0.5$ 
the values $C=6.62\pm 0.10$ and $\gamma=0.830\pm 0.003$
and for $-0.01\ge U>0$ the values $C=5.65\pm 0.10$ and 
$\gamma=0.812\pm 0.003$. The differences between the two cases are due 
to a slight asymmetry. Using all positive and negative values in 
the range $|U|\le 0.01$ for the fit one obtains $C=6.1\pm 0.2$ 
and $\gamma=0.822\pm 0.006$. 

\begin{figure}
\begin{center}
\includegraphics[width=0.48\textwidth]{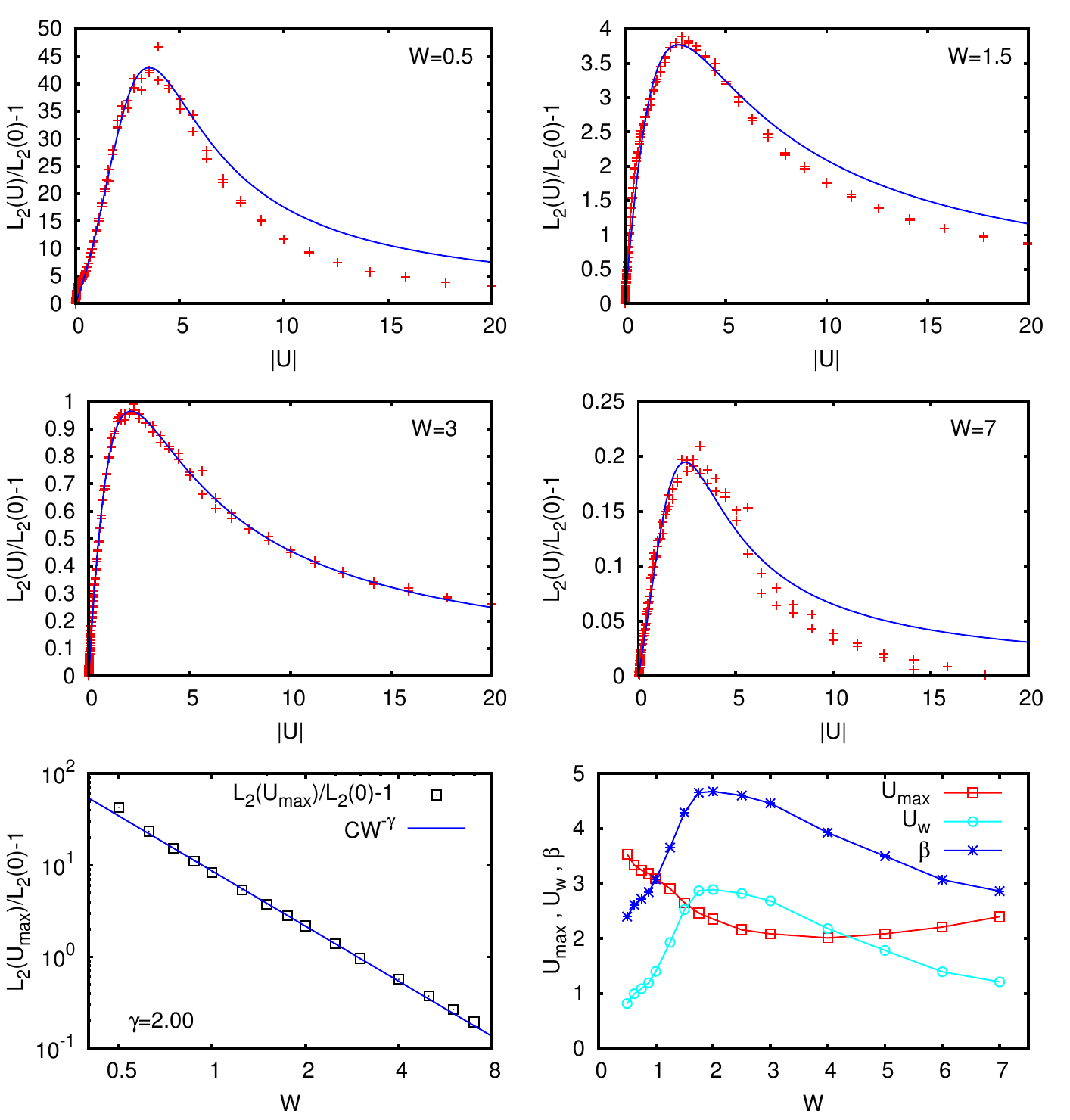}
\caption{{\em Top and center panels:} Comparison of the 
interaction dependence of $L_2(U)/L_2(0)-1$ 
(same data as in Figure~\ref{fig15}) 
with the fit (\ref{eq_L2L1Ufit}) for $E=0$ and the 
disorder values $W=0.5,\,1.5,\,3,\,7$. Note that due to the symmetry 
of $F(U)$ data points for $U<0$ are mapped to the positive $U$-axis. 
{\em Bottom left:} Dependence of $F_{\rm max}=L_2(U_{\rm max})/L_2(0)-1$ 
on the disorder strength $W$ (black squares). The full blue line shows 
the power law fit $L_2(U_{\rm max})/L_2(0)-1=C\,W^{-\gamma}$ 
with $C=8.66\pm 0.10$ and $\gamma=1,999\pm 0.017$. 
{\em Bottom right:} Dependence of the fit parameters 
$U_{\rm max}$ (red squares), $U_w$ (light blue circles) and the quantity 
$\beta$ (blue stars) given by 
(\ref{eq_beta_Uw}) on the disorder strength $W$. 
}
\label{fig16}
\end{center}
\end{figure}

Since for the case $E=0$ and each value of $W$ 
the curve $L_2(U)/L_2(0)-1$ is an even function 
in $U$ and due to the above observation that it obeys the 
limits $L_2(U)/L_2(0)-1\sim |U|$ for $|U|\ll 1$ and 
$L_2(U)/L_2(0)-1\sim |U|^{-1}$ for $|U|\gg 1$ one can try 
(for each disorder value $W$) the fit 
\begin{equation}
\label{eq_L2L1Ufit0}
L_2(U)/L_2(0)-1=F(U)
\end{equation}
with 
\begin{equation}
\label{eq_L2L1Ufit}
F(U)=F_{\rm max}\,\frac{|U|\,U_w/U_{\rm max}}{
(1-|U|/U_{\rm max})^2+|U|\,U_w/U_{\rm max}}
\end{equation}
being a rational function in $|U|$ and 
where the positive quantities $F_{\rm max}$, $U_{\rm max}$ and $U_w$ 
represent the three (disorder dependent) 
fit parameters. One verifies directly that 
this function has its maxima at $U=\pm U_{\rm max}$ with the value 
$F(\pm U_{\rm max})=F_{\rm max}$. The quantity $U_w$ represents 
somehow the (square of the) decay width around the maxima for 
the dimensionless quantity $|U|/U_{\rm max}$. Furthermore, the ansatz 
(\ref{eq_L2L1Ufit}) obeys both limits for small and large $|U|$-values, and 
the duality relation $F(U)=F(U_{\rm max}^2/U)$. 
Let us introduce the quantity $\beta>1$ such that 
$F(\beta U_{\rm max})=F(\beta^{-1}U_{\rm max})=F(U_{\rm max})/2$, i.~e.~: 
$\beta^{-1}U_{\rm max}$ and $\beta U_{\rm max}$ are the two values on the 
positive $U$-axis where the value of $F(U)$ is reduced by a factor $1/2$ 
with respect to its maximum value. From (\ref{eq_L2L1Ufit}) one finds that 
$\beta$ is related to $U_w$ by:
\begin{equation}
\label{eq_beta_Uw}
\beta=1+U_w/2+\sqrt{U_w+U_w^2/4}\ .
\end{equation}
Increasing values of $U_w$ and $\beta$ indicate a larger width around the 
maxima of $F(U)$. 
Performing the fit with the 
ansatz (\ref{eq_L2L1Ufit}), using the data of Figure~\ref{fig15}, 
we determined for each disorder 
value the three parameters $F_{\rm max}$, $U_{\rm max}$, $U_w$, and 
via (\ref{eq_beta_Uw}) the related quantity $\beta$. The non-linear fit 
is a bit tricky and we used stronger weights for larger data values closer 
to the maximum. The particular region $U<0.1$ and the limit $U\gg 1$ are 
not very precisely captured but the data close to the maximum 
are quite accurately represented by the fit as can be seen in 
Figure~\ref{fig16}. Furthermore, the bottom panels of Figure~\ref{fig16} 
show the disorder dependence of the fit parameters. 
The quantity $F_{\rm max}=L_2(U_{\rm max})/L_2(0)-1$ obeys a nearly 
perfect power law $L_2(U_{\rm max})/L_2(0)-1\approx 8.7\,W^{-2.00}$ 
and using the behavior 
$L_2(0)\approx 0.56\,L_1^{1.02}\approx 59\,W^{-2.04}$ (see Figure~\ref{fig13}) 
we find the expression 
$L_2(U_{\rm max})\approx 59\,W^{-2.04}+513\,W^{-4.04}$ which 
is indeed very accurate. 
A more direct fit with integer exponents gives a very similar expression: 
$L_2(U_{\rm max})\approx 54\,W^{-2}+586\,W^{-4}$. 
The finite size correction $\sim L_1\sim W^{-2}$ is quite important and a 
(too) simple power law fit without this correction would provide 
$L_2(U_{\rm max})\approx 622\,W^{-3.49}$ with rather 
strong systematic deviations 
due to a non-vanishing curvature (in double logarithmic scale) 
in a similar way as in Figure~\ref{fig14} for $L_2(U=2)$.
The values of $U_{\rm max}$ and the width parameters $U_w$ or $\beta$ 
are not constant with respect to the disorder strength and for smaller 
values of $W$ close to $0.5$ the width of the curve $F(U)$ around its 
maxima is considerably reduced. This point explains that $L_2(U)/L_1$ 
is below the behavior $a+bL_1$ for $U\le 1.0$ in the 
left panel of Figure~\ref{fig12} since for smaller values of $W$ 
(larger values of $L_1$) the interaction values $U\le 1$ are 
already out of resonance with respect to their optimal value $U_{\rm max}$. 
However, for $U=2$, which is closer to $U_{\rm max}$, 
the behavior $L_2(U)/L_1\approx a+b\,L_1$ is clearly 
valid for all considered disorder values $W\ge 0.5$.

Another interesting point concerns the duality with respect 
to $U\to \sqrt{24}/U$ predicted in \cite{waintal1a}. The fit function 
$F(U)$ verifies such a duality relation provided 
$U_{\rm max}^2=\sqrt{24}$ corresponding to $U_{\rm max}\approx 2.2$ 
which is approximately valid for $W\ge 2$ according to 
Figure~\ref{fig16}. Therefore we can approximately confirm this duality 
for such disorder values but not for the region $0.5\le W<2$. 
However, in general we have a modified duality relation $U\to U_{\rm max}^2/U$ 
with $U_{\rm max}$ depending on $W$ according to Figure~\ref{fig16}. 
Actually a more general duality relation $U\to {\rm const.}/U$ 
was suggested in \cite{ponomarev1}. 
Furthermore, the duality relation does not precisely extend to the extreme 
regions $|U|\le 0.1$ or $|U|\ge 50$. 

We have also applied the fit (\ref{eq_L2L1Ufit}) to the data of 
Figure~\ref{fig10} 
for $E=0$ concerning the interaction dependence of the three 
IPR quantities $\xi_x$, $\xi_{\rm CM}$ and $\xi_E$. Due to less 
available data points the fits are more difficult. We mention only that 
we find the following power laws~: $\xi_E(U_{\rm max})-1\sim W^{-0.88}$, 
$\xi_x(U_{\rm max})-\xi_x(0)\sim W^{-3.24}$, and 
$\xi_{\rm CM}(U_{\rm max})-\xi_{\rm CM}(0)\sim W^{-3.40}$. 
Here the last two exponents for $\xi_x$ and $\xi_{\rm CM}$ are quite 
different from $-4$ found for $L_2$. To understand this we first note 
that the exponents of the reference values at $U=0$ of 
$\xi_x(0)\sim W^{-1.33}$ and $\xi_{\rm CM}(0)\sim W^{-1.52}$ 
are different from $-2$ (see the fits for $E=0$ and $U=0$ 
of Figure~\ref{fig13}). 
Furthermore for $\xi_x$ and $\xi_{\rm CM}$ 
the maximum position $U_{\rm max}$ moves to quite small values 
well below $1$ for larger disorder values which changes the functional 
dependence of $\xi_{x,{\rm CM}}(U_{\rm max})$ on $W$ since at 
small $U$ and large $W$ the two IPR quantities are relatively enhanced 
as compared to $L_2$. 

In summary, in this section we have established the behavior 
$L_2=aL_1+bL_1^2$ for optimal interaction values, clarified that 
a simple power law fit $L_2\sim W^{-\gamma}$ is not well justified 
and how to understand exponents $\gamma$ below $4$ obtained by such fits. 
We have also 
obtained new and interesting results for the precise interaction dependence, 
such as a special regime for very small interaction values or a 
well pronounced local minimum at a finite value $U>0$ for $E=1$. 
Furthermore the discussion in Appendix \ref{appC} 
shows that the finite size scaling procedure requires 
a careful treatment of the condition $N>L_1$ on used data points, implying 
that previous results \cite{song1,leadbeater1} 
obtained for $N\le 250$ and $W\le 1.5$ are simply 
invalid. Also the use of constant offsets (independent of sample size) 
for the reference positions when 
measuring the localization length by the exponential decay of the 
Green function must be avoided since they imply a non-trivial 
transformation on the raw-data tainting completely the results of the 
scaling procedure. 

\section{Internal eigenfunction structure inside the localization domain}

\label{sec6}

A typical TIP localized eigenstate of length $L_2$ in the center of mass 
coordinate and width $L_1$ in the relative coordinate extends to 
a domain of potential $L_1L_2\sim L_1^3$ non-interacting product eigenstates 
in energy representation. Our results of Section~\ref{sec4} 
clearly indicate that the number of such states that really contribute 
to a TIP eigenstate, which is roughly $\xi_E$ the IPR in energy 
representation, is far below the size of this domain providing 
therefore a non-trivial internal eigenfunction structure. 

Theoretically it was first expected 
that only non-inter\-acting product eigenstates in an energy 
interval $\sim\Gamma$ are mixed where $\Gamma$ is the Breit-Wigner width 
implying the estimate $\xi_E\sim L_1 L_2\Gamma/B_W$ where 
$B_W\sim 1$ is the total band width of two-particle energies 
\cite{jacquod1,fyodorov1,frahm2}. The first scenario proposed in the 
initial work \cite{dlstip} assumed that a typical interaction matrix element 
(for $U_R=1$), 
\begin{equation}
\label{eq_int_matrix_element}
U_{\nu_1\nu_2,\,\nu_3\nu_4}=U\sum_x\phi_{\nu_1}(x)\,\phi_{\nu_2}(x)\,
\phi_{\nu_3}(x)\,\phi_{\nu_4(x)},
\end{equation}
behaves as $\sim U/L_1^{3/2}$ if all one-particle wave functions 
are localized at roughly the same position with amplitudes 
$\sim L_1^{-1/2}$ and random phases inside the localization 
domain providing an additional factor $\sim L_1^{1/2}$ due to the sum of 
$L_1$ random numbers. Using this assumption the Breit Wigner width is 
estimated as $\Gamma\sim U^2/L_1$ (assuming a unit coupling element $t=1$ 
in the initial Anderson model) leading to the estimates 
$L_2\sim U^2 L_1^2$ and $\xi_E\sim U^4 L_1^2$ 
\cite{jacquod1,fyodorov1,frahm2}. These two estimates disagree both 
with the numerical results of Figures~\ref{fig10} and \ref{fig15} concerning 
the interaction dependence and the expression for $\xi_E$ also disagrees 
strongly for the dependence on $L_1$ 
with the results of Figures~\ref{fig6} (lowest curve in bottom right panel) 
or \ref{fig7}-\ref{fig9} (bottom panels) predicting a power law 
$\xi_E\approx a+bL_1^\gamma$ with constant term and $\gamma$ clearly below 1 
(e.~g. $\gamma=1/3$ for $U=2$, $E=0$, $U_R=1$). 

The main reason of this discrepancy 
is that the phases of the localized one-particle wave functions are (for 
small disorder) quite strongly correlated due a plane wave structure 
with rather well defined momenta. Therefore the interaction matrix 
elements strongly fluctuate with maximum values $\sim U/L_1$ due 
to an approximate momentum conservation with uncertainty $\sim 1/L_1$ and 
much smaller values for non-conserved momenta \cite{ponomarev1}. 

The analytical calculation (to all orders in $U$) 
of the Breit-Wigner width for the case 
of vanishing disorder \cite{jacquod2} and the extension in 
\cite{frahm5}  provide indeed a modified dependence $\Gamma\sim c(U)/L_1$ 
with $c(U)\sim |U|$ for $|U|\lesssim 1$ 
resulting in $L_2\sim c(U) L_1^2$. This behavior is closer to the 
numerical results but at first sight the modification of the above estimate 
of $\xi_E$ would provide $\xi_E \sim c(U)^2 L_1^2\sim U^2 L_1^2$ 
which still clearly contradicts our numerical data. 

We attribute this to the fact that the Breit-Wigner width actually depends 
strongly on the quasi-momenta $k_{\nu_1}$ and $k_{\nu_2}$ of 
the initial non-interacting product eigenstate for which it is computed 
as can be clearly seen 
in the calculations of \cite{jacquod2,frahm5}, i.~e. the estimate 
$\Gamma\sim c(U)/L_1$ corresponds to the average 
$\Gamma=\langle \Gamma^{(1)}_{\nu_1\nu_2}\rangle_{\nu_1,\nu_2}$ with 
respect to these momenta with $\Gamma^{(1)}_{\nu_1\nu_2}$ for example given 
by equation (22) of \cite{frahm5}. The variations of 
$\Gamma^{(1)}_{\nu_1\nu_2}$ are also visible in Figure~\ref{fig4} 
due to the non-uniform structure of the energy line 
$E\approx \epsilon_{\nu_1}+\epsilon_{\nu_2}$. 
When determining $L_2\sim \Gamma L_1^3$ the average Breit-Wigner width 
seems to produce rather reasonable dependencies of $L_2$ on 
$U$ and $L_1$ (even 
though a more accurate theory is still lacking) but for $\xi_E$, 
requiring an harmonic average 
$\xi_E\sim L_2 L_1\langle(B_W/\Gamma^{(1)}_{\nu_1\nu_2})\rangle_{\nu_1,\nu_2}^{-1}$, 
the strong fluctuations of $\Gamma^{(1)}_{\nu_1\nu_2}$ with possible quite 
small values will considerably reduce $\xi_E$ thus explaining the 
lower exponents $\gamma$ clearly below unity. 

It is interesting to note that for the random matrix ensembles proposed 
by Ponomarev et al. \cite{ponomarev1}, which are modelized by carefully 
taking into account the strong fluctuations of the interaction matrix 
elements as well as the approximate momentum conservation for best 
coupled sta\-tes, the power law $\xi_E\sim L_1^{\gamma_{\rm ipr}}$ is expected 
with typical values of $\gamma_{\rm ipr}$ clearly below unity 
according to Figure~1 of \cite{ponomarev1} for at least one variant of the 
modified random matrix ensembles studied in \cite{ponomarev1}. 

The physical picture of strongly fluctuating interaction coupling matrix 
elements and Breit-Wigner widths depending on the initial state corresponds 
to the situation where among $\sim L_1L_2$ potential non-interacting product 
eigenstates in the localization domain 
one has to select first states fulfilling the condition 
\begin{equation}
\label{eq_select1}
|E-\epsilon_{\nu_1}-\epsilon_{\nu_2}-U_{\rm diag}+
\Gamma^{(0)}_{\nu_1\nu_2}/2|
\lesssim \Gamma^{(1)}_{\nu_1\nu_2}
\end{equation}
where 
$U_{\rm diag}=<\!\phi_{\nu_1}\,\phi_{\nu_2}|\hat U
|\phi_{\nu_1}\,\phi_{\nu_2}\!>\,\, \sim U/L_1$ is a 
diagonal interaction matrix element with the same sign as $U$ but with 
considerable fluctuations. $\Gamma^{(0)}_{\nu_1\nu_2}$ is 
the real part of the self-energy while the Breit-Wigner width 
$\Gamma^{(1)}_{\nu_1\nu_2}$ is the imaginary part as given 
in equations (14) and (15) of 
\cite{frahm5}. The approximate momentum conservation 
implies an additional selection criterion 
\begin{equation}
\label{eq_select2}
|k_{\nu_1}+k_{\nu_2}-k_{\nu_3}-k_{\nu_4}|\lesssim 1/L_1
\end{equation}
(or similar with modified signs for the different momenta) for two 
strongly coupled non-interacting product eigenstates \cite{ponomarev1}. 
For the energy $E=0$ at the band center, with an approximate Fermi surface 
with linear borders in momentum space (see left panels of Figure~\ref{fig3}), 
both selection criteria (\ref{eq_select1}) and (\ref{eq_select2}) 
seem to select rather similar states 
but the situation is more complicated due 
to the shifts from $U_{\rm diag}$ and $\Gamma^{(0)}_{\nu_1\nu_2}/2$ 
and of course due to the strong fluctuations of $\Gamma^{(1)}_{\nu_1\nu_2}$. 
For the energy $E=1$ outside the band center the form of 
the approximate Fermi surface is different (see right panels of 
Figure~\ref{fig3}) and the overlap for both criteria appears 
to be somewhat reduced. However, this effect does not seem to reduce 
$\xi_E$ for $E=1$ if compared to $E=0$ according to results 
shown in Figures \ref{fig7}-\ref{fig10}. 
Combining the effects of the strong fluctuations of the Breit-Wigner width, 
the shifts due to the diagonal interaction matrix elements and the 
real part of the self-energy, and the additional approximate momentum 
conservation it finally appears 
that typical TIP eigenstates select only a rather modest number 
$\sim\xi_E\sim L_1^\gamma\ll L_1^2$ of 
non-interacting product eigenstates with $\gamma<1$. However, due to a 
complicated spatial distributions of such states and their large 
individual values of $\Gamma^{(1)}_{\nu_1\nu_2}$ they still produce 
an overall localization length $L_2\sim L_1^2$ (for certain optimal 
interaction values).

\section{Discussion}

\label{sec7}

In this work numerous new numerical results for various quantities 
characterizing the localization and other properties of TIP eigenfunctions 
for the one-dimensional Anderson model have been obtained. The dependence of 
the three types of localization lengths $\xi_x$, $\xi_{\rm CM}$ and $L_2$ on 
$L_1$ can be well fitted by $aL_1+bL_1^\gamma$ for a large range of disorder 
and with $\gamma\approx 2$ for a considerable interval of optimal 
interaction values. 

However, for the interaction dependence 
the behavior $L_2/L_1\approx c(U)L_1+$const. 
as suggested in \cite{vonoppen,jacquod2,frahm5} with various propositions 
for the coefficient $c(U)$ is not completely confirmed by our new results 
for the full interval of considered disorder values. In the band center $E=0$ 
the fit (\ref{eq_L2L1Ufit}) of the quantity $L_2(U)/L_2(0)-1$ 
(see Figure~\ref{fig15}) 
provides for each disorder value roughly a similar form with an 
approximate behavior $\sim |U|$ ($\sim |U|^{-1}$) for $|U|\ll 1$ ($|U|\gg 1$) 
and the maximal amplitude scales very precisely as 
$L_2(U_{\rm max})/L_2(0)-1\sim L_1$. 
However, the maximum position $U_{\rm max}$ and the effective width parameter 
of (\ref{eq_L2L1Ufit}) 
depend on disorder according to Figure~\ref{fig16}. In 
particular the width of these curves decreases considerably for the smallest 
disorder values thus explaining the sublinear behavior of 
$L_2(U)/L_1$ in $L_1$ for interaction values 
not sufficiently close to $U_{\rm max}$. Furthermore, 
the duality with respect to $U\to {\rm const.}/U$ predicted in 
\cite{ponomarev1,waintal1a} is roughly confirmed 
by our data. 

The new claims of \cite{flach2011,flach2014} concerning a strongly reduced 
TIP enhancement are based on numerical data for limited parameters in 
system size ($N\le 234$) and disorder ($W\ge 2$) for TIP eigenfunctions 
and without use of finite size scaling. Furthermore in 
\cite{flach2011} the oversimplified power law 
fit without finite size correction was used (see our above discussion of 
Figure~\ref{fig14}). Our numerical results obtained for 
very large system sizes and by careful finite size scaling, especially with 
respect to the condition $N>L_1$ neglected in previous work 
\cite{song1,leadbeater1}, refute very clearly the new claims 
of \cite{flach2011,flach2014}. 

We have also considered 
the inverse participation ratio in energy representation (of non-interacting 
product eigenstates) which clearly demonstrates the interaction 
induced delocalization by values $\xi_E\gg 1$ such as $\xi_E\approx 222$ 
or $\xi_E\approx 168$ for the two example eigenstates with $N=5000$ shown in 
Figure~\ref{fig1}. This quantity obeys a different dependence on $L_1$ as 
$\xi_E\approx a+bL_1^\gamma$ with $\gamma\approx 1/3$ 
(for $U=2$, $E=0$ and the Hubbard interaction case) and somewhat larger values 
with $0.6\le \gamma\le 0.9$ for the interaction range $U_R=5$. This 
behavior is indeed unexpected if compared to the early results based 
on the random band matrix model with preferential basis 
that suggested $\xi_E\sim L_1^2$ \cite{jacquod1,fyodorov1,frahm2}. As 
explained in Section~\ref{sec6} this estimate was obtained from an incorrect 
hypothesis about uncorrelated phases inside the localization domain of 
non-interacting one-particle eigenfunctions. An accurate quantitative 
analytical theory for this quantity, beyond the random band matrix 
model of \cite{jacquod1,fyodorov1,frahm2}, is still missing but qualitatively 
it seems that the scaling $\xi_E\sim L_1^\gamma$ with $\gamma<1$ is related 
to very strong fluctuations 
of the Breit-Wigner width depending strongly on the unperturbed initial 
state for which it is computed. 

We have also established a particular regime of rather strong 
enhancement for quite low interaction values and obtained 
new very interesting results for an energy $E=1$ outside the band center 
such as a strong local minimum in the interaction dependence 
of $L_2(U)/L_2(0)$ at the value $U\approx 1.2$ (see Figure~\ref{fig15}). 
Also for these new results a precise analytical theory is still missing. 

The very efficient numerical methods used in this work allowed to 
considerably extend the range of parameters in system size, low disorder 
values, very small and large interaction values for which results for 
various quantities were obtained. These methods are potentially 
also applicable for TIP in higher dimensions (see e.~g. \cite{ortuno1})
even though the efficiency gain will be more moderate, especially for 
the computation of the projected Green function $\bar G_0$ at vanishing 
interaction which is the basic step for both methods. 

The author thanks Dima Shepelyansky for numerous fruitful 
discussions and  Gabriel Lemari\'e for his insight on the 
numerical scaling procedure. The author is particularly 
grateful for the access to the HPC resources of 
CALMIP (Toulouse) under the allocation 2015-P0110 during this work. 

\appendix

\section{Computation of the scaling curve}

\label{appA}

In this appendix we provide some details on the explicit 
procedure we used to compute the scaling curve and 
to determine the infinite size localization lengths for the TIP problem. 
Our approach is somewhat different from the scaling fit procedure given 
in \cite{mackinnon2} and in particular it also allows 
to take into account (and to compute) the statistical errors of the 
finite (infinite) size localization lengths in a quite simple way. 

Let us assume that we have 
numerical raw data of finite size localization lengths $\xi_j(N)$ 
dependent on system size $N$ and on an index $j=0,\,1,\,2,\,\ldots$ 
representing several data sets for different disorder values 
(with increasing values of $j$ corresponding to decreasing of values of 
the disorder strength). The ``localization length'' 
$\xi$ may correspond to one of the three IPR-localization lengths 
$\xi_x$, $\xi_{\rm CM}$ or $\xi_E$ or to the Green function localization 
length $L_2$. Typically these localization lengths are obtained by 
some kind of averaging procedure which also provides data for the 
statistical errors $\delta \xi_j(N)$. Furthermore we assume 
that all data sets correspond to the same values of other physical parameters 
such as energy, interaction strength, interaction range, 
boson- or fermion case etc. 

Assuming the validity of one-parameter scaling we need to find a 
universal scaling function $f(x)$ and ``infinite size localization lengths'' 
$\xi_j(\infty)$ for each data set 
such that the data are well fitted by \cite{mackinnon1,mackinnon2,kramer1}:
\begin{equation}
\label{eq_scale1}
\frac{\xi_j(N)}{N}=f\left(\frac{\xi_j(\infty)}{N}\right)\ .
\end{equation}
The scaling function has the obvious limit $f(x)=x$ for $x\ll 1$. 
Furthermore it turns out that $f(x)$ is monotonically increasing and 
tends to some limit $f_\infty$ for $x\to\infty$ corresponding to the 
behavior $\xi_j(N)\approx f_\infty\,N\sim N$ in the regime 
$N\ll\xi_j(\infty)$. We mention that for the TIP problem 
this point requires to exclude data with $N<L_1$ where $L_1=105/W^2$ is the 
one-particle localization length and as discussed in Appendix~\ref{appC} 
the validity of one-parameter scaling indeed requires $N$ to be larger than 
all other length scales \cite{mackinnon2,kramer1}, especially 
$L_1$ which plays in our case somehow the role of the mean free path. 

Usually, to perform the scaling procedure 
in a graphical way one draws for each 
data set the logarithm of $\xi_j(N)/N$ versus the logarithm of $1/N$ and 
applies a horizontal shift on each curve such that all data fall on the 
same universal scaling curve where the horizontal shift provides 
the (logarithm of the) infinite size localization length $\xi_j(\infty)$ (see 
Figures~\ref{fig6} and \ref{fig11}). 
Below we describe an explicit, efficient and reliable implementation 
of this procedure suitable to treat automatically 
a large amount of different data sets for many 
different cases and to obtain the best possible accuracy together 
with an estimate of the statistical error of $\xi_j(\infty)$. 

Explicitely, let us define for each available value of $N=N_l$ and 
disorder index 
$j$~: $x_{j,l}=1/N_l$ and $y_{j,l}=\xi_j(N_l)/N_l$. Furthermore, let 
$y_{j,{\rm max}}=\max_l\,(y_{j,l})$, 
$y_{j,{\rm min}}=\min_l\,(y_{j,l})$, and $\bar x_j$ the $x$-value associated 
to $y_{j,{\rm max}}$. First we remove all data points with 
$x_{j,l}>\bar x_j$ and $y_{j,l}<y_{j,{\rm max}}$ corresponding to data 
on the right side of (and below) the maximum in $y$ which are 
problematic since they clearly violate the monotonicity between $x$ and $y$ 
and produce a partly double valued curve. 
Such data points are either in the problematic 
flat regime of the scaling curve for $x\gg 1$ or indicate that for them 
the one-parameter scaling hypothesis is not verified, e.~g. if the ratio 
$N_l/L_1$ is not yet sufficiently large (this may especially happen 
for the cases of $\xi_x$ and $\xi_{\rm CM}$). 
Let us denote by $h(y)$ the inverse of the scaling 
function $f(x)$ such that $h(y)=x$ is equivalent to $y=f(x)$. 
Furthermore we translate the known statistical errors 
$\delta y_{j,l}=\delta \xi_j(N_l)/N_l$ for $y_{j,l}$ to errors 
$\delta x_{j,l}\approx (\Delta x/\Delta y)\,\delta y_{j,l}$ for $x_{j,l}$ 
where $\Delta x/\Delta y$ is the local slope of the $(y,x)$ curve (at 
given value $j$) obtained from the neighbor values of $x$ and $y$ on this 
curve.

Let us assume that we have for a given value of $j$ already determined the 
inverse scaling functions $h_i(y)$ for $i=0,\,1,\,\ldots,\,j-1$ of the 
previous data sets (see below for details on this). 
Then the horizontal shift for the data set $j$ 
can be determined from the average~:
\begin{equation}
\label{eq_average}
\ln[\xi_j(\infty)]=C\sum_{i=0}^{j-1}\sum_{l\in S_i} w_{j,l}
\Bigl(\ln[h_i(y_{j,l})]-\ln(x_{j,l})\Bigr)
\end{equation}
with the weights $w_{j,l}= (\delta x_{j,l}/x_{j,l})^{-2}$ and the $l$-sum runs 
over the set $S_i$ containing the values of $l$ such that 
$y_{i,{\rm min}}\le y_{j,l}\le y_{i,{\rm max}}$ because $h_i(y)$ is not 
defined for $y>y_{i,{\rm max}}$ or $y<y_{i,{\rm min}}$. 
The normalization constant $C$ is determined by the inverse sum of the weights 
over exactly the same contributions in $i$ and $l$. It is easy to verify 
that this average corresponds to the least square (one-parameter) fit 
to minimize the horizontal differences in logarithmic scale 
of the data curve $j$ to the previous already rescaled curves and 
using the weights $w_{j,l}$. We note that the procedure of 
\cite{mackinnon2} is quite different since it performs a single fit to 
determine {\em simultaneously all} shifts and furthermore 
it corresponds to the choice of constant weights. 
Our choice of weights proportional to the inverse square of the 
statistical errors of $\ln(x_{j,l})$ 
is quite standard in the context of least square fits and has the advantage 
that it automatically takes into account the quality of individual 
data points. In particular in the flat region of the scaling curve the slope 
$\Delta x/\Delta y$ is typically quite large thus strongly increasing the 
errors in $x$ and reducing the weight of less reliable data in this 
problematic region. 

We also compute the statistical error $\delta\varepsilon_j$ 
of the average (\ref{eq_average}) in the standard way 
as the square root of the averaged squared deviations of the shifts 
multiplied by $\sim 1/\sqrt{d-1}$ where $d$ is the number of independent 
data points used for the fit. We have chosen 
for $d$ the number of different $l$ values which contribute 
in the sum of (\ref{eq_average}) even though this
sum runs over more terms due to the sum over $i$ of previous scaling curves 
which is only an artificial effect of the notation. 
In particular for the case where only one single data point 
($d=1$) contribute this implies an ``infinite error'', a case which actually 
appears in Figure~\ref{fig19}. 
Furthermore, for $j>0$ we increase the error by 
$\delta \varepsilon_j\to\sqrt{\delta \varepsilon_j^2+\delta 
\varepsilon_{j-1}^2}$ using the error $\delta \varepsilon_{j-1}$ of 
the previous data set $j-1$. This takes 
into account that also the previous inverse scaling curves 
$h_i(y)$ are subject to an error of their respective scaling factors with 
the most important contribution arising form the last previous data set $j-1$. 
Finally the error of $\xi_j(\infty)$ is obtained as 
$\delta\xi_j(\infty)=\delta\varepsilon_j\,\xi_j(\infty)$. 

Once $\xi_j(\infty)$ is known we replace 
$x_{j,l}\to \xi_j(\infty)\,x_{j,l}$ and determine the next piece $h_j(x)$ of 
the inverse scaling function either by linear interpolation [in $\ln(x)$ and 
$\ln(y)$] or suitable fits of the rescaled discrete data set $j$ (see below). 

At the beginning of this procedure at $j=0$, the case with strongest 
disorder ($W=7$) or smallest localization length ($\xi\approx 1.5-4$), 
we choose in (\ref{eq_average}) the limit $h(y)=y$ for $y\ll 1$ 
as initial scaling curve in order to set approximately 
the global scale. To improve the precision of the 
global scale we choose at the end 
the first data set $j$ containing values with $\xi_j(N)\ge 10$, which 
typically happens at $W=3$ and still provides data points with 
rather small ratios $\xi_i(N)/N \ll 1$, use the 
fit $\xi_j(N)^{-1}=A+B\,N^{-1}$ 
to determine accurately $\xi_j(\infty)=A^{-1}$ for this case, 
and apply the appropriate correction factor to the other infinite size 
localization lengths. It turns out that the correction of the global scale 
due to this improvement is actually quite small and in particular the 
condition $y\ll 1$ is well fulfilled for the first data set with $W=7$. 

This scaling procedure depends on the way the rescaled discrete data points 
of the set $j$ are translated into a continuous inverse 
scaling function $h_j(y)$ used in the average (\ref{eq_average}). A simple 
and quite effective choice is linear interpolation using the logarithmic 
quantities $X=\ln(x)$ and $Y=\ln(y)$ which already produces very nice 
scaling curves provided that the raw data are of reasonable quality 
(good statistical precision of the finite size localization lengths) and 
that the validity of one-parameter scaling is fulfilled. However, if 
fluctuations of individual data points are somewhat stronger it is 
more reasonable to determine $h_j(y)$ by a fit such as a polynomial of 
degree 3 for the dependence of $X$ on $Y$~:
\begin{equation}
\label{eq_fitpoly}
X=\ln[h(e^Y)]=a_0+a_1\,Y+a_2\,Y^2+a_3\,Y^3\ .
\end{equation}
The degree 3 appears to be a good compromise between being able to modelize 
a reasonable curvature and to avoid artificial minima and maxima when using 
data sets of typically 10-20 points. Even though the fit (\ref{eq_fitpoly}) 
also provides good scaling curves it can still be improved by incorporating 
the singularity of $h(y)$ when $y\to f_\infty$, with $f_\infty$ being 
the limit of the scaling function $f(x)$ for $x\to\infty$, using a modified 
fit:
\begin{equation}
\label{eq_fitscaling}
X=\ln[h(e^Y)]=a_0+a_1\,Y+a_2\,Y^2+\frac{a_3}{F_\infty-Y}
\end{equation}
with $F_\infty=\ln(f_\infty)$ and $f_\infty$ 
is determined in advance from the last data set (with 
smallest disorder) by the fit $y^{-1}=f_\infty^{-1}+A\,x^{-1}$. 
We emphasize that during the scaling procedure both fits 
(\ref{eq_fitpoly}) and (\ref{eq_fitscaling}) are used only 
on the individual data sets and not on the complete scaling curve obtained 
so far. Furthermore the value of $F_\infty$ is fixed in advance such that both 
fits are linear in the parameters $a_j$, $j=0,\,1,\,2,\,3$ assuring a unique, 
well defined fit optimum. It turns out that the 
scaling results are not very sensible to the precise 
choice of $F_\infty$ and are still reasonable if one chooses a rather 
large value, e.~g. $F_\infty=\ln(2)$ in case if the automatic 
determination using the last data set fails for some reason. 

One might have the idea to use $F_\infty$ as an additional non-linear 
fit parameter but this works quite badly for the first data sets that 
do not allow to extract the information concerning the singularity. 
However, once the scaling procedure using (\ref{eq_fitscaling}) 
with a fixed value of $F_\infty$ is finished we performed such a non-linear 
fit on all rescaled data to optimize the value $F_\infty$ in order to 
have a full analytical scaling curve for the purpose of illustration 
(see full blue lines shown in the scaling curves of Figures~\ref{fig6}, 
\ref{fig11} and panel (a) of Figure~\ref{fig18}). 

We have also tested modified variants of the scaling procedure 
where the least square procedure to determine $\xi_j(\infty)$ and 
the linear interpolation or fits to determine $h_j(y)$ are not done in 
logarithmic scale for $X$ and $Y$ but rather in the original variables 
$x$ and $y$. In this case we have to 
minimize the weighted sum of squares of the quantity $\alpha_j h_i(y)-x$
where $\alpha_j=1/\xi_j(\infty)$ is a (linear) fit parameter. It is 
straightforward to work out the details of the modified 
fit procedure, and without going into the details, we mention 
that it is possible to determine $\xi_j(\infty)$ as well as its statistical 
error $\delta \xi_j(\infty)$ by closed formulas. 
The choice between $y$ and $Y$ does not 
affect the least square procedure but it modifies the 
linear interpolation or fits to determine $h_j(y)$. 
In total we have implemented many different variants of the scaling procedure, 
four for each pairing of $x$ or $X$ with $y$ or $Y$ combined 
with several cases to obtain $h_j(y)$ either by linear interpolation or 
the two fits (\ref{eq_fitpoly}) and (\ref{eq_fitscaling}). 
It turns that for good quality raw data all these variants provide coherent 
results with localization lengths which coincide well within the bounds 
of the statistical errors obtained from the scaling procedure. 
However, for bad data (e.~g. when trying the very small 
disorder values $W=0.5$ and $W=0.625$ also for $\xi_x$ and $\xi_{\rm CM}$ 
at $U=2$) the different variants may provide different results or even 
partly fail. Therefore, the comparison of the different variants 
gives a good indication on the quality of the data.

Actually, in some cases of ``very bad'' raw data one may even have a data set 
$j$ such 
that all values of $y_{j,l}$ are above all $y$-values of the previous data 
sets, i.~e. where the curves we want to rescale do not even overlap 
in $y$-direction. In this case one might still try to apply 
(\ref{eq_average}) simply by extending the validity of the fits for 
$h_i(y)$ beyond the maximal value of the corresponding curve 
(``extrapolation mode'') and here the second fit (\ref{eq_fitscaling})
is certainly superior than the polynomial fit (\ref{eq_fitpoly}). 
Even though this might still provide some estimation of the scaling factors, 
we did not push this point very far. 
In particular, all curves and data shown in this work 
(except for some of the particular and problematic cases shown in 
Figures~\ref{fig18} and \ref{fig19}) correspond to good quality raw data
where all of the above variants of the method as well as 
the precise fit procedure of \cite{mackinnon2}, which we also tested, 
produce rather precisely the same results (within the margin of errors).

As default method for the scaling procedure we 
used the variant with both logarithmic variables $X$ and $Y$ and 
the second fit (\ref{eq_fitscaling}). 

\section{Two-particle localization lengths at vanishing interaction}

\label{appB}

In this appendix we discuss our numerical results for the different 
two-particle localization lengths at vanishing 
interaction $U=0$ (or very small interaction $U=10^{-5}\approx 0$ for 
the three IPR quantities). The behavior $\xi_E\approx 1$ 
(lowest curve in bottom panels of Figure~\ref{fig7}) is rather precisely 
confirmed as expected. For the other three localization lengths scales 
the ratios $\xi_x/L_1$, $\xi_{\rm CM}/L_1$ 
(lowest curve of top and center panels of Figure~\ref{fig7}) 
and $L_2/L_1$ (lowest curve of both panels of Figure~\ref{fig12}) 
are of order unity $\sim 1$ but there are still slight variations 
in the dependence on $L_1$ due to the effective average in one-particle 
energy when measuring these quantities in two-particle space. Also 
the Green function localization length at $U=0$ contains implicitly such 
an average as can be seen from the expression (\ref{eq_proj_resolv000}) 
for $\bar G_0$ (see also \cite{leadbeater1} for a simplified analytical 
estimation).

\begin{figure}
\begin{center}
\includegraphics[width=0.48\textwidth]{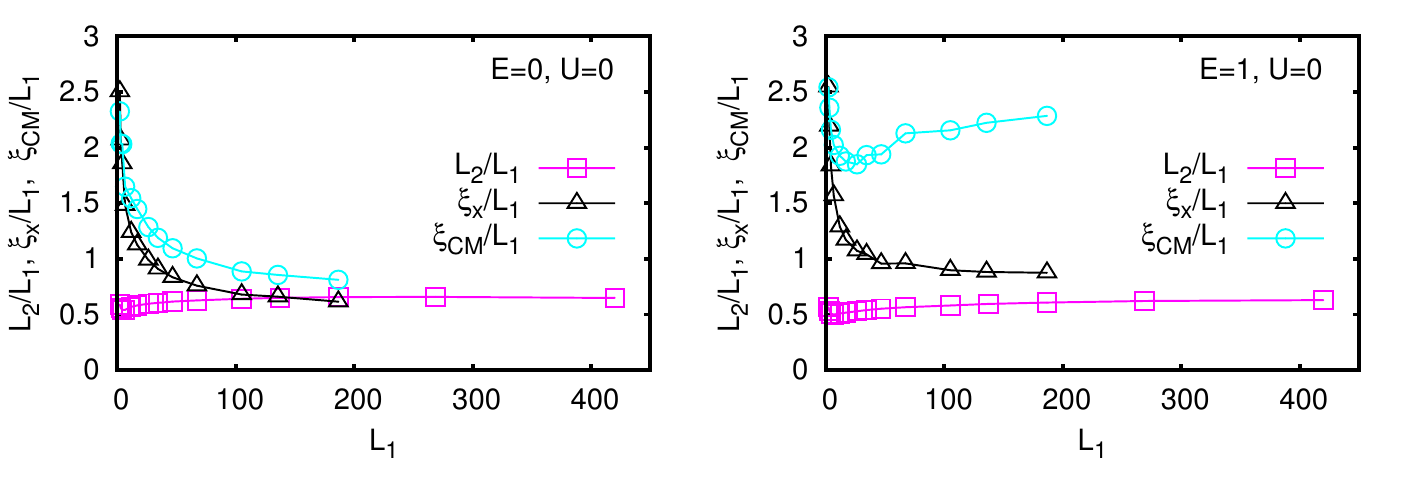}
\caption{Comparison of the dependence of $L_2/L_1$, $\xi_x/L_1$ and 
$\xi_{\rm CM}/L_1$
on $L_1$ for $U=0$ and energy $E=0$ ($E=1$) for left (right) panel. 
The used disorder values $W$ are as in Figure~\ref{fig6} ($\xi_x$ and 
$\xi_{\rm CM}$) or in Figure~\ref{fig11} ($L_2$).
For $\xi_x$ and $\xi_{\rm CM}$ the value $U=10^{-5}\approx 0$ was used.
}
\label{fig17}
\end{center}
\end{figure}

In Figure~\ref{fig17} these three ratios are shown again as a function 
of $L_1$ but using an increased scale. For $\xi_{x,{\rm CM}}/L_1$ 
we observe a slight decreasing behavior with $L_1$, i.e. with strongest values 
$\approx 2$ for smallest values of $L_1$ (largest disorder values)
and values slightly below unity for largest values of $L_1$ 
concerning $\xi_{x}/L_1$ (both energies) 
and $\xi_{\rm CM}/L_1$ (for $E=0$). For $E=1$ the ratio $\xi_{\rm CM}/L_1$ is 
rather constant with values $\approx 2.2$ and with a very modest 
local minimum at $L_1\approx 25$ and value  $\approx 1.9$. 
To understand this we note that for (nearly) vanishing interaction 
$\xi_{x,{\rm CM}}$ is computed from a harmonic mean of non-interacting product 
eigenstates with approximate momenta (or energies) close to the curves 
(or lines) shown in Figure~\ref{fig3} (or Figure~\ref{fig4}). Let $\xi_1$ and 
$\xi_2$ be the two IPR values of the two one-particle eigenstates 
contributing to such a product state. Then we have 
for the case $\xi_2\gg \xi_1$:  
$\xi_x=2/(\xi_1^{-1}+\xi_2^{-1})\approx 2\xi_1$ and 
$\xi_{\rm CM}\approx 2\xi_2$. Furthermore, we may assume that roughly 
$\xi_j\sim L_1\sin^2(k_j)$ for $|k_j|>L_1^{-1/3}$ 
where $k_1$ and $k_2$ are the approximate momenta 
of the two one-particle eigenstates. The harmonic average is dominated 
(reduced) by contributions from small values of $\xi_j$, i.~e. $k_j$ close 
to $0$, $\pi$, or $2\pi$ (with uncertainty $\sim L_1^{-1/3}$) explaining 
the slight decrease of $\xi_{x,{\rm CM}}/L_1$ with increasing $L_1$. 
However for $E=1$ and $\xi_{\rm CM}/L_1$ it is more difficult to have 
product states where both approximate momenta produce simultaneously 
small values of $\xi_1$ and $\xi_2$ and here the larger of the two 
contribute stronger thus explaining the absence of decrease 
of $\xi_{\rm CM}/L_1$ for large $L_1$. 

$L_2/L_1$ is for $U=0$ rather constant with values close to $0.5$ for 
small $L_1$ and to $0.65$ for large $L_1$. Note that in 
\cite{song1,leadbeater1} somewhat larger values closer to $1$ 
for large $L_1$ and $U=0$ where found which is due to a different choice 
of the parameters used in (\ref{eq_L2green_gen}). 

The power law fits $\propto L_1^\gamma$ of the three localization lengths 
for $U=0$ shown in the bottom panels of Figure~\ref{fig13} 
provide exponents close to $1$ for both cases of $L_2$. For 
$\xi_x$ we have $\gamma=0.66$ ($\gamma=0.70$) and for 
$\xi_{\rm CM}$ we have $\gamma=0.76$ ($\gamma=0.94$) both for $E=0$ ($E=1$) 
(see caption of Figure~\ref{fig13} for precise fit results). 
The reduced exponents for $\xi_x$ and $\xi_{\rm CM}$ are due to 
quite enhanced values at $U=0$ for small $L_1$ (stronger $W$) with 
values close to $2$-$2.5$. 

\section{Effects of scaling with insufficient system size}

\label{appC}

The validity of one-parameter scaling theory 
very clearly requires \cite{mackinnon2,kramer1} that the 
system size $N$ is significantly larger than all other typical length scales 
in the system, especially the mean free path for the usual case of the 
Anderson model of non-interacting particles. In the case of TIP the 
one-particle 
localization length $L_1$ is certainly such a length scale and 
actually plays somehow the role of a mean free path since the dynamics is 
ballistic for $N<L_1$ and diffusive with coherent pair propagation for 
$L_1<N<L_2$. Therefore one expects the scaling theory for TIP only to be 
valid for $N>L_1$. Also the physical argumentation 
in terms of the Thouless scaling block picture 
\cite{imry,thouless1} shows that it is simply not reasonable 
to consider values of $N$ below $L_1$ in the context of scaling. 

The results of \cite{song1,leadbeater1} obtained by 
finite size scaling for small disorder values, where $N<L_1$ for 
many or even all data points (for the data sets with lowest disorder) 
are therefore very problematic in this context. 
Our numerical results obtained by finite size scaling of the quantity $L_2$, 
defined by (\ref{eq_L2green}) and shown in the figures of Section~\ref{sec5},  
are for small disorder indeed quite different from the results of 
\cite{song1,leadbeater1}. 

For example for $E=0$, 
$U=1$ and $W=0.5$ we obtain $L_2=3922\pm 102$ while from Figure~2(c) 
of \cite{song1} the corresponding data point provides $L_2\approx 1900$ 
and from Figure~11 of \cite{leadbeater1} we get $L_2\approx 6500$ (both 
values were graphically determined from these figures with 
the best possible accuracy). We attribute the discrepancy of more than 
a factor of three between the two values of \cite{song1,leadbeater1} 
to a particular problem concerning the precise definition of $L_2$ 
used in \cite{leadbeater1} and which will be discussed below in more detail. 
Apart from this in \cite{song1,leadbeater1} 
there are simply no (or too few) valid data points for the smallest disorder 
values and we find indeed significant 
deviations to our results at least for values up to $W\le 1.5$. 
On the other hand, as can be seen in Figure~\ref{fig11}, our data set for 
$W=0.5$ still contains seven valid data points satisfying the 
condition $1.5L_1=630\le N\le 2000$ which allow for a reliable scaling 
of $L_2$ even for this disorder value. 

\begin{figure}
\begin{center}
\includegraphics[width=0.48\textwidth]{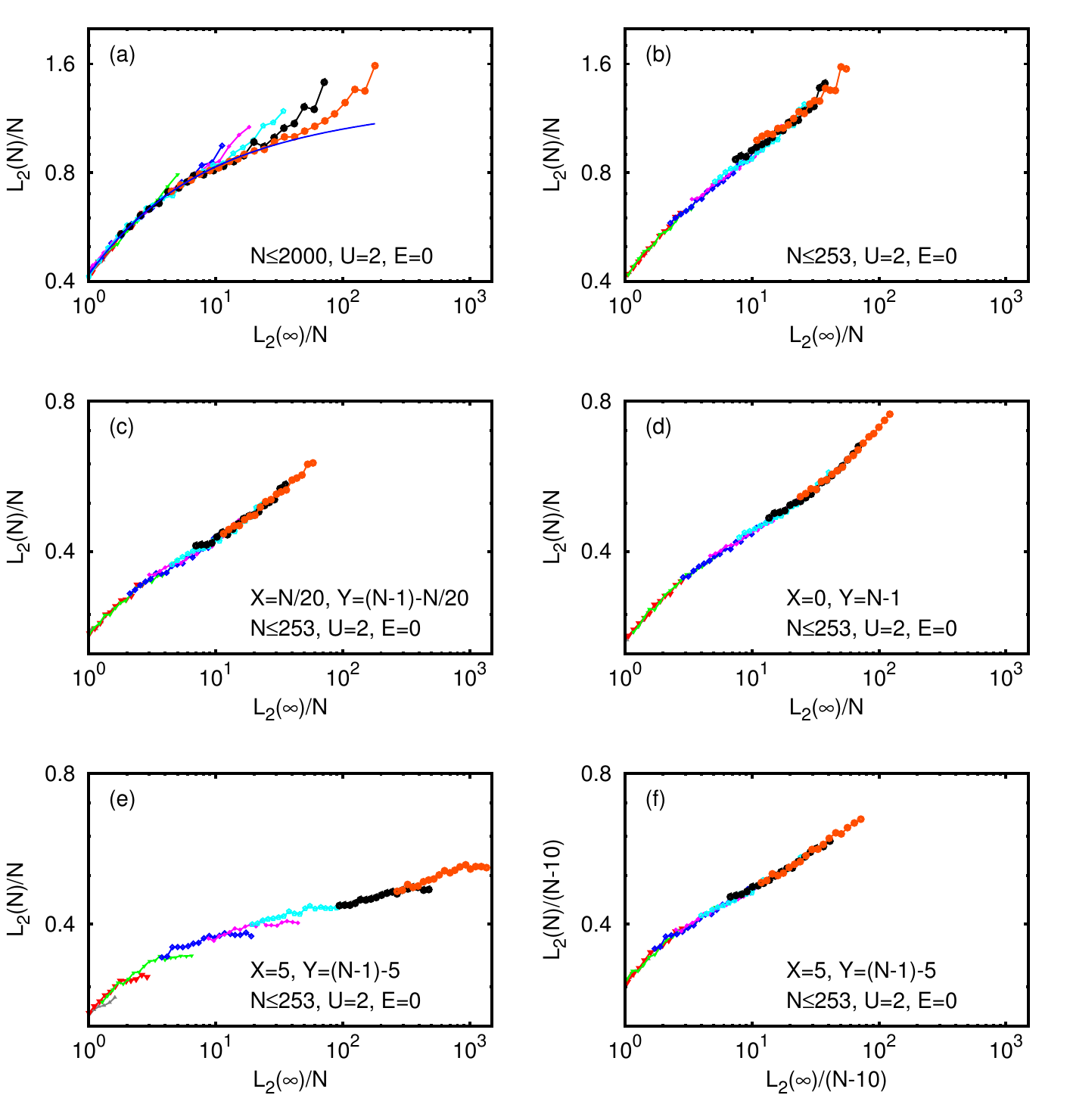}
\caption{Comparison of scaling curves shown in the zoomed region
$1\le L_2(\infty)/N\le 1500$ for $U=2$ and $E=0$ for different cases 
concerning the range of used system sizes 
and the determination of $L_2(N)$ from the two-particle Green function. 
Panel (a) shows the same scaling curve as in the left panel of 
Figure~\ref{fig11}, 
i.~e. $L_2(N)$ is given by (\ref{eq_L2green}) and $L_2(\infty)$ has 
been determined from data points with $50\le N\le 2000$ and 
$N\ge 1.5\,L_1$. However, in panel (a) also 
the non-used data points with $50\le N<1.5\,L_1$ are shown. The blue full 
curve in panel (a) corresponds to the same fit of the scaling curve 
shown in the left panel of Figure~\ref{fig11}. 
In panels (b)-(f) a (denser) set of data points for smaller 
system sizes $50\le N\le 253$ is used. 
In panel (b) $L_2(N)$ is  given by (\ref{eq_L2green}) as in panel (a) while 
in panels (c)-(f) $L_2(N)=L_G(X,Y,N,0)$ as defined 
in (\ref{eq_L2green_gen})
with $X=N/20$ and $Y=(N-1)-N/20$ [panel (c)], 
$X=0$ and $Y=N-1$ [panel (d)], 
$X=5$ and $Y=(N-1)-5$ using $N$ as system size for the 
scaling [panel (e)] and 
$X=5$ and $Y=(N-1)-5$ using the reduced effective system size 
$N-10$ for the scaling [panel (f)]. The used disorder values and 
corresponding colors are as in Figure~\ref{fig11} but due to the zoomed 
range only the data points for the 7-8 smallest disorder values are visible. 
}
\label{fig18}
\end{center}
\end{figure}

To investigate in more detail the issues of limited system size 
and also of the precise choice for the definition of the two-particle 
localization length in terms of the projected Green function, 
we computed (with 1\% accuracy) special data sets 
of smaller system sizes $50\le N\le 253$, with an approximate ratio $1.1$ 
between two neighbor values of $N$. To be more precise 
we determined for these limited system sizes 
our variant (\ref{eq_L2green}) of the two-particle 
localization length $L_2(N)$ 
and of some other variants with different parameters in 
(\ref{eq_L2green_gen}), including the two cases of 
\cite{song1,leadbeater1}. The results of this 
comparative study of ``scaling in the invalid regime'' (with 
$N\le 253<L_1=420$ for $W=0.5$) are shown in Figure~\ref{fig18} 
for the case $U=2$ and $E=0$. 

First we present again in panel (a) of Figure~\ref{fig18} 
the same ``correct'' scaling curve of the left panel of Figure~\ref{fig11} 
in a zoomed presentation and 
computed with data for $L_2(N)$, given by (\ref{eq_L2green}), with 
$50\le N\le 2000$ and $N\ge 1.5L_1$. 
However in addition also the non-used data points 
with $N<1.5L_1$ are shown for illustration but these 
data points were not used for the computation of the scaling curve. 
We indeed observe that they are clearly outside (above) the 
scaling curve thus confirming the condition $N>L_1$ for the validity 
of the one-parameter scaling approach. 

In panel (b) the scaling curve of the same quantity (\ref{eq_L2green}) but 
using the data set for smaller system sizes $N\le 253$ is visible. 
The quality of the scaling curve appears to be rather satisfactory but its 
form is very different from the scaling curve of panel (a) or 
Figure~\ref{fig11}. In particular it does not have a finite limit 
for large $L_2(\infty)/N$ which is indeed quite unusual if compared to 
the standard form of the scaling curves for the non-interacting 
Anderson model in 2 dimensions. The curve (b) can be obtained 
from (a) by removing in the latter the ``good'' data points (with $N>253$) 
in the lower parts of the individual curves and then using the ``bad'' 
data points (those with $N<L_1$) to recalculate the scaling parameters. 
This changes considerably the form of the scaling curve and reduces the 
values of infinite size localizations lengths for small disorder values. 

Panel (c) shows, also for the case $N\le 253$, 
the scaling curve of a different quantity 
$L_2(N)=L_G(N/20,N-1-N/20,N,0)$ 
corresponding to two reference positions 5\% inside the sample 
and absence of the denominator $|\bar G(x,x)|$ in (\ref{eq_L2green_gen}). 
The scaling curve of panel (d) is similar but for the quantity 
$L_2(N)=L_G(0,N-1,N,0)$ with two reference positions exactly at the boundary 
(and also absence of the denominator $|\bar G(x,x)|$) 
which corresponds exactly to the choice of \cite{song1}. 
We note that the scaling curves of (b), (c), and (d) have a rather similar 
form and slightly increasing values for large $L_2(\infty)/N$. 

In panel (e) we consider the case of $L_2(N)=L_G(p,N-1-p,N,0)$ for 
$p=5$ corresponding to reference positions 5 sites inside the sample 
which is similar (or even identical) to the choice of 
\cite{leadbeater1}. The scaling curve has a very different form than 
the three previous cases. The matching of the different data points is not 
very good and one observes small but clearly significant and 
systematic deviations from a perfect scaling. More importantly the 
obtained infinite size localization lengths at smallest disorder 
are strongly increased as compared to all previous cases. 
The question arises why 
we observe this strong difference, especially between the two scenarios 
(e) and (c) where for both cases 
the two reference positions are slightly inside the sample ? 
The reason is that for (e) the distance between the two reference positions 
is effectively $N_{\rm eff}=N-10$ instead of $N$ and therefore the 
scaling procedure should be done with this effective system size 
which provides indeed a non-trivial complicated transformation as can be 
seen in panel (f) where the same data of (e) is submitted to a scaling 
procedure using $N_{\rm eff}=N-10$ as effective system size. The scaling 
curve of (f) is very different from (e) but quite similar in form to the 
cases (b), (c) and (d). 

For the case (c) we have also a reduced effective system $N_{\rm eff}=0.9N$ 
but here the situation is very different since this transformation 
corresponds in double logarithmic scale to a simple small shift up and 
to the right of all data and the scaling curve. Actually this 
transformation maps somehow the case of (c) to the case of (d) (with 
reference points precisely at the boundaries). 
Therefore the resulting infinite size 
localization lengths are not modified when using a scaling procedure 
with $N_{\rm eff}=0.9N$ instead of $N$. However, the transformation 
$N_{\rm eff}=N-10$ affects seriously in a non-trivial way 
the form of the individual data curves (also in double logarithmic scale) 
and modifies strongly the obtained infinite size localization 
lengths. Therefore  
it is indeed necessary to be careful when using offset parameters by 
choosing reference positions 
slightly inside the sample. A constant choice of them {\em 
independent of $N$} will produce tainted scaling results and one must 
instead choose them proportional to $N$, for example 5\% inside the sample.

\begin{figure}
\begin{center}
\includegraphics[width=0.48\textwidth]{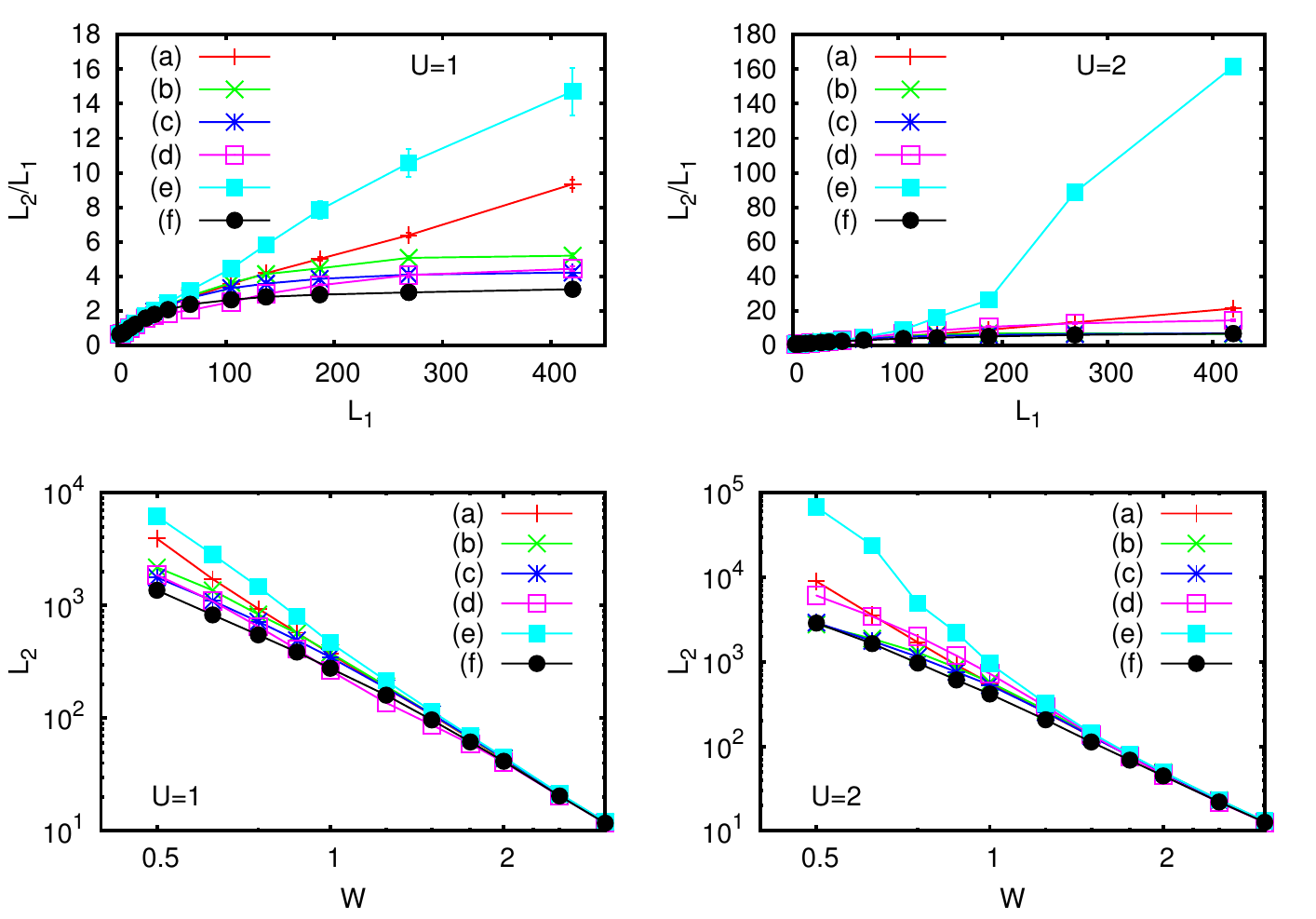}
\caption{Comparison of $L_2$ obtained by the six different 
cases (a)-(f) of finite size scaling illustrated in Figure~\ref{fig18} for 
$E=0$ (all panels), and $U=1$ 
(left panels) or $U=2$ (right panels). Top panels show the 
enhancement factor $L_2/L_1$ versus 
$L_1=105/W^2$ and bottom panels show $L_2$ versus $W$ on a double logarithmic 
scale. Top panels also show error bars obtained by the scaling procedure 
and the absence or non-visibility of them indicates errors below 
point size except in top right panel for the two data points of curve (e) 
with largest values of $L_1$ and $L_2/L_1$ which have 
infinite errors. In the bottom panels errors are 
always below point size except in bottom right panel for the two points 
of curve (e) with largest values of $L_1$ and $L_2$ corresponding to 
infinite errors. 
}
\label{fig19}
\end{center}
\end{figure}

Figure~\ref{fig19} shows the dependence of the 
infinite size localization lengths on $L_1$ or $W$ 
(for $E=0$ and two interaction values $U=1$ and $U=2$) 
obtained by the different scaling scenarios (a)-(f) 
of Figure~\ref{fig18}. 
The cases (b), (c), (d) and (f), all obtained for small system sizes 
$N\le 253$, seem to form one group with roughly 
comparable results and only slight deviations among them indicating 
that the precise choices of $L_2(N)$ either by (\ref{eq_L2green}) 
or (\ref{eq_L2green_gen}) with different parameters is not very 
important in this context. However, they all produce significantly 
smaller values of infinite size localization lengths 
for large $L_1$ (small $W$) than the scenario (a) which is 
the only case with large system sizes up to $N=2000$ ensuring that the 
important condition $N\ge 1.5L_1$ is verified for a sufficient number of 
data points. 
The scenario (e), also valid for small system sizes, suffers additionally 
from the constant offset of the reference points in the definition of 
$L_2(N)$ which seriously taints the scaling results and produce for large 
$L_1$ considerably larger values of infinite size 
localization lengths than the case (a). Actually, for $U=2$ there 
is only one (usable) data point in the second last curve having an overlap 
with the third last curve and resulting formally in an infinite margin 
of error for the two last localization lengths which are by a factor 
$\sim 10$ larger than for $U=1$. 

We mention that using the procedure (d) for $U=1$, $E=0$, $W=0.5$ 
we find the value $L_2=1863\pm 62$ which coincides very well 
with the value $L_2\approx 1900$ of \cite{song1}. 
Furthermore, using the procedure (e) for the same parameters we obtain 
$L_2=6179\pm 576$ which also agrees with the value 
$L_2\approx 6500$ of \cite{leadbeater1} (within margin of error). These 
agreements confirm that the raw data for the projected Green function 
of \cite{song1,leadbeater1} are in agreement with our raw data 
(obtained by a different and more efficient but equivalent 
method to compute the projected Green function). Of course both of these 
values are incorrect since there were obtained by scaling procedures 
which suffer from one or two serious flaws as explained above and the 
correct value is $L_1=3922\pm 102$ obtained by the procedure (a) using 
large system sizes up to $N=2000$ and the expression (\ref{eq_L2green}) 
for the definition of $L_2(N)$. In all figures and discussions 
of Section~\ref{sec5} and of Appendix~\ref{appB} 
the quantity $L_2$ represents always the infinite size value obtained 
by finite size scaling from $L_2(N)$ for the case (a). 

We mention that also for the two IPR length scales $\xi_x$ and $\xi_{\rm CM}$ 
considered in Section~\ref{sec4} the data with $N<2L_1$ 
do not obey one-parameter scaling. In this case these data produce for 
small disorder a maximum in the individual curves and quite strongly decaying 
values right to this maximum. This behavior is somewhat 
different as compared to the increasing curves visible in panel (a) of 
Figure~\ref{fig18} but it provides further confirmation 
that the application of one-parameter scaling theory to TIP indeed 
requires $N$ being sufficiently larger than $L_1$.


\begin{thebibliography}{99}
\bibitem{lee1} Patrick A. Lee and T. V. Ramakrishnan, Rev. Mod. Phys. 
  {\bf 57}, 287 (1985). 

\bibitem{mackinnon1} A. MacKinnon, and B. Kramer, Phys. Rev. Lett. {\bf 47}, 
  1546 (1981).

\bibitem{mackinnon2} A. MacKinnon, and B. Kramer, Z. Phys. B {\bf 53}, 
  1 (1983).

\bibitem{kramer1} B. Kramer and A. MacKinnon, Rep. Prog.
        Phys. {\bf 56}, 1469 (1993). 

\bibitem{efetov1} K. B. Efetov, {\it Supersymmetry in Disorder and Chaos}, 
        Cambridge University Press (1997).

\bibitem{guhr1} T. Guhr, A. M\"uller-Groeling, and H. A. Weidenm\"uller,
	Phys. Rep. {\bf 299}, 189 (1998).

\bibitem{been_rev1} C. W. J. Beenakker, Rev. Mod. Phys. {\bf 69}, 731 (1997). 

\bibitem{brouwer1} P. W. Brouwer and K. Frahm, Phys. Rev. B {\bf 53}, 
  1490 (1996).

\bibitem{dorokhov1} O. N. Dorokhov, Zh. Eksp. Teor. Fiz. {\bf 98}, 646
	(1990) [Sov. Phys. JETP {\bf 71}, 360 (1990)].

\bibitem{dlstip}  D. L. Shepelyansky, Phys. Rev. Lett. {\bf 73}, 2607 (1994).

\bibitem{bloch} M.~Screiber, S.S.~Hodgman, P.~Bordia, H.~L\"uschen,
                 M.H.~Fischer, R.~Vosk, E.~Altman,
                 U.~Schneider, and I.~Bloch,
                 Science {\bf 349}, 842 (2015).

\bibitem{imry} Y.Imry, Europhys. Lett. {\bf 30}, 405 (1995).

\bibitem{thouless1} D. C. Thouless, Phys.\ Rev.\ Lett.\ {\bf 39}, 1167 (1977).

\bibitem{frahm1} K. Frahm, A. M\"uller--Groeling, J.-L. Pichard, and
	D.~Weinmann, Europhys. Lett. {\bf 31}, 169 (1995).

\bibitem{weinmann} D.~Weinmann, A. M\"uller--Groeling, J.-L.~Pichard, 
  and K.~Frahm, Phys. Rev. Lett. {\bf 75}, 1598 (1995).

\bibitem{jacquod1} P. Jacquod and D. L. Shepelyansky, Phys. Rev. Lett. 
        {\bf 75}, 3501 (1995). 

\bibitem{fyodorov1} Y. V. Fyodorov and A. D. Mirlin, Phys. Rev. B 
	{\bf 52}, R11580 (1995).

\bibitem{frahm2} K. Frahm and A. M\"uller--Groeling, Europhys. Lett. 
        {\bf 32}, 385 (1995). 

\bibitem{borgonovi1} F. Borgonovi and D. L. Shepelyansky, Nonlinearity 
	{\bf 8}, 877 (1995).

\bibitem{borgonovi1a} F. Borgonovi and D. L. Shepelyansky, 
  J. Phys. I France {\bf 6}, 287 (1996).

\bibitem{halfpap1} O. Halfpap, A. MacKinnon and B. Kramer, 
  Solid State Communications {\bf 107}, 379 (1998).


\bibitem{vonoppen} F. von Oppen, T.~Wetting, and J.~M\"uller,
                Phys. Rev. Lett. {\bf 76}, 491 (1996).

\bibitem{jacquod2} Ph. Jacquod, D. L. Shepelyansky, and O. P. Sushkov, 
	Phys. Rev. Lett. {\bf 78}, 923 (1997).

\bibitem{shep2} D. L. Shepelyansky, Proceedings of les Rencontres de 
	Moriond 1996 on ``Correlated Fermions and Transport in 
	Mesoscopic Systems'', edited by T. Martin, G. Montambaux and 
	J. Tr\^an Thanh V\^an, 201 (1996). 

\bibitem{ponomarev1} I. V. Ponomarev and P. G. Silvestrov, Phys. Rev. 
	B {\bf 56}, 3742 (1997).

\bibitem{roemer4} R. A. Roemer, M. Schreiber and T. Vojta, 
  phys. stat. sol. (b) {\bf 211}, 681 (1999).

\bibitem{frahm3} K. Frahm, A. M\"uller--Groeling, and J.- L. Pichard, 
	Phys. Rev. Lett. {\bf 76}, 1509 (1996). 

\bibitem{frahm3a} K. Frahm, A. M\"uller--Groeling, and J.- L. Pichard, 
	Z. Phys. B {\bf 102}, 261 (1997).

\bibitem{richert1} J. Richert and H. A. Weidenm\"uller, 
  J. Phys. A: Math. Gen. {\bf 36}, 3281 (2003).

\bibitem{richert1a} J. Richert and H. A. Weidenm\"uller, 
  Annals of Physics {\bf 306}, 96 (2003).


\bibitem{wein2} D. Weinmann, J.-L. Pichard, Phys. Rev. Lett. 
	{\bf 77}, 1556 (1996).

\bibitem{akkermans} E. Akkermans and J.-L. Pichard, 
   Eur. Phys. J. B {\bf 1}, 223 (1998). 

\bibitem{waintal1} X. Waintal, and J.-L. Pichard, Eur. Phys. J. B {\bf 6}, 
	117 (1998).

\bibitem{waintal1a} X. Waintal, D. Weinmann, and J.-L. Pichard, 
	Eur. Phys. J. B {\bf 7}, 451 (1999). 

\bibitem{waintal1b} S. De Toro Arias, X. Waintal, 
	and J.-L. Pichard, Eur. Phys. J. B {\bf 10}, 149 (1999). 

\bibitem{roemer1} R.A. R\"omer and M. Schreiber, Phys. Rev. Lett. 
	{\bf 78}, 515 (1997). 

\bibitem{roemer1a} R.A. R\"omer and M. Schreiber, Phys. Rev. Lett. 
	{\bf 78}, 4890 (1997). 

\bibitem{frahm4} K. M. Frahm, A. M\"uller-Groeling, J.-L. Pichard and 
        D. Weinmann, Phys. Rev. Lett. {\bf 78}, 4889 (1997). 

\bibitem{song2} P. H. Song and F. v. Oppen, Phys. Rev. B {\bf 59}, 
	46 (1999). 

\bibitem{song1} P. H. Song and Doochul Kim, Phys. Rev. B {\bf 56}, 
	12217 (1997). 

\bibitem{leadbeater1} M. Leadbeater, R. A. R\"omer, and M. Schreiber,
	Eur. Phys. J. B {\bf 8}, 643 (1999). 

\bibitem{frahm5} K.~M.~Frahm, Eur. Phys. J. B, {\bf 10}, 371 (1999).

\bibitem{flach2011} D.~O.~Krimer, R.~Khomeriki, S.~Flach, 
  JETP Letters {\bf 94}, 406 (2011) 
               
\bibitem{flach2014} D.~O.~Krimer, S.~Flach, Phys. Rev. B {\bf 91}, 
  100201(R) (2015). 

\bibitem{frahm6} K. M. Frahm and D. L. Shepelyansky, 
  Eur. Phys. J. B {\bf 88}, 337 (2015). 

\bibitem{flach2012} S.~Flach, M.~Ivanchenko, and R.~Khomeriki,
               Europhys. Lett. {\bf 98}, 66002 (2012).

\bibitem{stewart} G.~W.~Stewart,  
         {\em Matrix Algorithms Volume II: Eigensystems}, 
         SIAM (2001).

\bibitem{frahmulam} K.~M.~Frahm, and D. L.~Shepelyansky,
        Eur. Phys. J. B {\bf 76},  57 (2010).

\bibitem{schwartz1} H. R. Schwartz, {\em Numerische Mathematik}, 
	B. G. Teubner, Stuttgart (1986). 

\bibitem{webpage} K.~M.~Frahm, 
  \url{http://www.quantware.ups-tlse.fr/QWLIB/tipdisorder1d}, 
  Accessed February (2016).

\bibitem{ortuno1} M. Ortuno and E. Cuevas, Europhys. Lett., {\bf 46}, 
  224 (1999). 

\end{thebibliography}
\end{document}